%% file: ms.tex
\documentclass[11pt,a4paper,twoside,openright]{article}
\usepackage{authblk}

\usepackage[utf8]{inputenc}
\usepackage[russian, francais, english]{babel}
\usepackage[T1]{fontenc}
\usepackage[left=3cm,right=3cm,top=3cm,bottom=3cm]{geometry}
\usepackage{amsmath}
\allowdisplaybreaks %allow page break in 'align' (\\* to disallow)
\usepackage{amsfonts}
\usepackage[numbers]{natbib}
\usepackage[globalcitecopy, labelstoglobalaux]{bibunits}
\usepackage{bibentry}
\nobibliography* %for bibentry
\usepackage{titletoc} % pour liste des figures partielle. Commandes : startlist printlist stoplist
\usepackage{appendix}%
\usepackage[tabbotcap, TABBOTCAP]{subfigure}
\usepackage[hidelinks]{hyperref}
\usepackage{import}
\usepackage{amssymb}
\usepackage{graphicx}
  \usepackage[export]{adjustbox}%
\usepackage{setspace}
\usepackage{placeins}
\usepackage{changepage}
\usepackage{caption}
\usepackage{numprint}
\npstyleenglish
\npthousandsep{}
\usepackage{multicol}
\usepackage{multirow, bigdelim}
\usepackage{longtable}
\usepackage{indentfirst}
\usepackage{url}
\usepackage[outdir=fig/]{epstopdf}
\usepackage{color}
\usepackage[usenames,dvipsnames,svgnames,table]{xcolor}
\usepackage{arydshln}%for \hdashline
\usepackage{afterpage}
\usepackage{float}
\usepackage{listings}
\usepackage{xpatch}

\usepackage{titlesec}

\titleformat{\paragraph}
{\normalfont\normalsize\bfseries}{\theparagraph}{1em}{}
\titlespacing*{\paragraph}
{0pt}{3.25ex plus 1ex minus .2ex}{1.5ex plus .2ex}

\title{\Large \textbf{A priori tests of subgrid-scale models in an anisothermal turbulent channel flow at low Mach number}}
\def\runtitle{A priori tests of subgrid-scale models}

\def\runauthor{D. Dupuy, A. Toutant and F. Bataille}
\author[1]{Dorian Dupuy}
\author[1]{Adrien Toutant\thanks{Corresponding author : adrien.toutant@univ-perp.fr}}
\author[1]{Fran\c coise Bataille}
\affil[1]{PROMES CNRS, Universit\'{e} de Perpignan Via Domitia, Rambla de la thermodynamique, Tecnosud, 66100 Perpignan, France}

\date{\vspace{-4ex}\itshape\small (Published version: International Journal of Thermal Sciences 145, 105999 (2019); https://doi.org/10.1016/j.ijthermalsci.2019.105999)\vspace{-4ex}}

\usepackage{fancyhdr}
\usepackage{scalerel,stackengine}
\stackMath
\DeclareRobustCommand\reallywidehat[1]{%
\savestack{\tmpbox}{\stretchto{%
  \scaleto{%
    \scalerel*[\widthof{\ensuremath{#1}}]{\kern-.6pt\bigwedge\kern-.6pt}%
    {\rule[-\textheight/2]{1ex}{\textheight}}%WIDTH-LIMITED BIG WEDGE
  }{\textheight}% 
}{0.5ex}}%
\stackon[1pt]{\displaystyle #1}{\tmpbox}%
}
\stackMath
\DeclareRobustCommand\reallywidetilde[1]{%
\savestack{\tmpbox}{\stretchto{%
  \scaleto{%
    \scalerel*[\widthof{\ensuremath{#1}}]{\kern-1.0pt\sim\kern-1.0pt}%
    {\rule[-\textheight/2]{1ex}{\textheight}}%WIDTH-LIMITED BIG WEDGE
  }{\textheight}% 
}{0.5ex}}%
\stackon[1pt]{\displaystyle #1}{\tmpbox}%
}
\newcommand{\overbar}[1]{\mkern 1.5mu\overline{\mkern-1.5mu#1}}

\DeclareMathOperator{\tr}{tr}

\let\OLDthebibliography\thebibliography
\renewcommand\thebibliography[1]{
  \OLDthebibliography{#1}
  \setlength{\parskip}{0pt}
  \setlength{\itemsep}{0pt plus 0.3ex}
}

\begin{document}

\hyphenation{aniso-thermal}

\newcommand{\der}[2]{\frac{\partial #1}{\partial #2}}
\newcommand{\dd}[2]{\frac{\partial^2 #1}{\partial #2^2}}
\newcommand{\dtot}[2]{\frac{D #1}{D #2}}
\newcommand{\ov}[1]{\overline{#1}}
\renewcommand{\f}[1]{\overbar{#1}}
\newcommand{\fa}[1]{\widetilde{#1}}
\newcommand{\FA}[1]{\reallywidetilde{#1}}
\newcommand{\ff}[1]{\widehat{#1}}
\newcommand{\FF}[1]{\reallywidehat{#1}}
\newcommand{\m}[1]{\overbar{#1}}
\newcommand{\ma}[1]{\widetilde{#1}}
\newcommand{\MA}[1]{\reallywidetilde{#1}}
\newcommand{\vv}[1]{\boldsymbol{#1}}
\newcommand{\vt}[1]{\boldsymbol{#1}}
\newcommand{\w}[1]{\widehat{#1}}
\newcommand{\W}[1]{\reallywidehat{#1}}

\newcommand{\refv}[1]{#1^{b}}
\newcommand{\adim}[1]{#1^{\circ}}

\newcommand{\upi}[0]{\pi}

\newcommand\Real{\mbox{Re}}
\newcommand\Imag{\mbox{Im}}
\newcommand\Rey{\mbox{\textit{Re}}}
\newcommand\Pran{\mbox{\textit{Pr}}}
\newcommand\Pen{\mbox{\textit{Pe}}}
\newcommand\Ai{\mbox{Ai}}
\newcommand\Bi{\mbox{Bi}}

\newcommand{\corr}[1]{{\color{red} #1}}
\newcommand{\corre}[1]{{\color{blue} #1}}

\newcommand{\e}[0]{\mathrm{e}}
\newcommand{\I}[0]{\mathrm{i}}
\newcommand{\Dee}[0]{D}
\newcommand{\ft}[1]{\widehat{#1}}
\newcommand{\FT}[1]{\reallywidehat{#1}}

\newcounter{subfigcounter}
\setcounter{subfigcounter}{1}
\DeclareRobustCommand{\lecID}{\refstepcounter{subfigcounter}}
\DeclareRobustCommand{\subfigtopleft}[1]{\lecID\def\stackalignment{l}\topinset{\text{\footnotesize(\alph{subfigcounter})}}{#1}{0.1in}{-.05in}}

\maketitle

{
\let\latexthesection\thesection %
    \renewcommand{\thesection}{\arabic{section}} %(*)
    \renewcommand{\thefigure}{\arabic{figure}} %(*)
    \setcounter{section}{0}
\subimport{./}{article.tex}

\let\stdthebibliography\thebibliography
\renewcommand{\thebibliography}{%
\let\chapter\section
\stdthebibliography}
  {
  \small
  \bibliography{./biblio}
  }
}

\end{document}

%% file: article.tex
\begin{abstract}
The subgrid-scale modelling of a low Mach number strongly anisothermal turbulent
flow is investigated using direct numerical simulations.  The study is based on the filtering of the low
Mach number equations, suited to low Mach number flows with highly variable
fluid properties.
The results are relevant to formulations of the filtered low Mach number
equations established with the classical filter or the Favre filter.
The two most significant subgrid terms of the filtered low Mach number
equations are considered.  They are associated with the momentum convection and
the density-velocity correlation.
We focus on eddy-viscosity and eddy-diffusivity models. Subgrid-scale models
from the literature are analysed and two new models are proposed.
The subgrid-scale models are compared to the exact subgrid term using the
instantaneous flow field of the direct numerical simulation of a strongly
anisothermal fully developed turbulent channel flow.
There is no significant differences between the use of the classical
and Favre filter regarding the performance of the models.
We suggest that the models should take into account the asymptotic near-wall
behaviour of the filter length.
Eddy-viscosity and eddy-diffusivity models are able to represent the energetic
contribution of the subgrid term but not its effect in the flow governing
equations.
The AMD and scalar AMD models are found to be in better agreement with the exact
subgrid terms than the other investigated models in the a priori tests.
\end{abstract}

\section{Introduction}

This paper addresses the large-eddy simulation subgrid-scale modelling of low Mach
number strongly anisothermal turbulent flows.  Flows subjected to a strong
temperature gradient are prevalent in many industrial processes, such as heat
exchangers, propulsion systems or solar power towers \citep{serra2012}.  They
are characterised by strong coupling between turbulence and temperature, along
with high variations of the fluid properties (density, viscosity and thermal
conductivity) with temperature \citep{toutant2013, aulery2016,
dupuy2018turbulence}.  In many cases, the direct numerical simulation (DNS) of
strongly anisothermal turbulent flows is unpracticable because too many scales
of temperature and velocity are produced and not enough resolution is
available to resolve all the relevant scales. In order to predict the
large-scale behaviour of low Mach number strongly anisothermal turbulent flows,
thermal large-eddy simulation (LES) is an effective alternative. Large-eddy
simulation is based on the explicit resolution of the large scales of
turbulence and the use of subgrid-scale models to account for the effect of
the smaller scales on the large scales. The scale separation may be
represented by the application of a low-pass spatial filter on the flow
governing equations.

The filtering of the low Mach number equations gives rise to specific subgrid
terms. Using a priori tests, \citet{dupuy2018study} assessed the amplitude of
all subgrid terms in several formulations. The expression of the filtered low
Mach equations with the unweighted classical filter and the density-weighted
Favre filter \citep{favre1965equationsb} leads to two different set of
equations involving the same non-negligible subgrid terms \citep{dupuy2016,
dupuy2017sft, dupuy2018study}. The two most significant subgrid terms are
associated with the momentum convection and the density-velocity correlation.
The adequate modelling of these subgrid terms is required for the large-eddy
simulation of low Mach number strongly anisothermal turbulent flows.

Various modelling strategies have been devised to represent the subgrid terms.
Two main types of model are found: structural models, established with no
prior knowledge of the nature of the effect of the subgrid term, and functional
models, which assume that the effect of the subgrid term is similar to
molecular diffusion and therefore acts as a dissipative action
\citep{sagaut98}.
The subgrid-scale models should be consistent with important mathematical and
physical properties of the Navier--Stokes equations and the turbulent stresses
\citep{silvis2017physical}.
With regard to the subgrid term associated with momentum convection, the
functional eddy-viscosity models are by far the most used because they are
simple, inexpensive and robust. A review of eddy-viscosity models may be
found in \citep{sagaut98, trias2015building, silvis2017physical}.
The eddy-viscosity assumption can be extended to the density-velocity
correlation subgrid term using the constant subgrid-scale Prandtl or Schmidt number
assumption.
This is referred to as eddy-diffusivity models.

In this paper, we assess the subgrid-scale models a priori using the flow field
from the direct numerical simulation of a strongly anisothermal turbulent
channel flow.
In the literature, a priori studies of the subgrid-scale models have been
carried out
in incompressible flows \citep{clark1979evaluation, abba2003analysis, lu2007priori, piomelli1988model, li2013priori},
passive and active scalar decaying homogeneous turbulence \citep{chumakov2005dynamic, ghaisas2014priori}
and in flows with purely compressible effects,
in a temporal shear layer \citep{vreman1995priori, vreman1995subgrid, vreman1995direct},
a multi-species mixing layer \citep{borghesi2015priori},
and in freely decaying homogeneous isotropic turbulence \citep{martin2000subgrid}.
Besides, there are several works in the literature dealing with the large-eddy simulation of multiphase flows \citep{breuer2017influence, chen2017large, marchioli2017large, park2017simple, rosa2017impact, weiner2017advanced, frohlich2018validation, zhou2018structural}.
A priori tests have been carried out for two-phase divergence-free flows \citep{toutant2008dns, toutant2009jump1, toutant2009jump2, ketterl2018priori}.
The analysis is here extended to low Mach number strongly anisothermal
turbulent flows.
We focus on eddy-viscosity and
eddy-diffusivity models.
Structural models, such as the scale-similarity \citep{bardina1980improved}
and gradient model \citep{leonard74}, are known to display high degrees of
correlation with the exact subgrid term in a priori tests despite easily
leading to instabilities when used in an actual large-eddy simulation
\citep{bardina1980improved, sagaut98, sarghini1999scale, berselli2005mathematics, ketterl2018priori}.
Eddy-viscosity models,
which assume that the subgrid term is aligned with the rate of deformation
tensor or the scalar gradient, are purely dissipative and have desirable
property for numerical stability.
Besides, by restricting the study to a single family of models, we may hope that
the a priori tests have
a more easy-to-interpret relevance for a posteriori results.
The subgrid-scale models investigated are
the Smagorinsky model \citep{smagorinsky1963general},
the WALE model \citep{nicoud99b},
the Vreman model \citep{vreman2004eddy},
the Sigma model \citep{nicoud2011using},
the AMD model \citep{rozema2015minimum},
the scalar AMD model \citep{abkar2016minimum},
the VSS model \citep{ryu2014subgrid} and
the Kobayashi model \citep{kobayashi2005subgrid}.
In addition, two new eddy-viscosity and eddy-diffusivity
models are proposed and investigated,
the Anisotropic Smagorinsky model,
which attempts to improve anisotropy of the Smagorinsky model
by involving three filter length scales instead of one,
and
the MMG model,
which may be viewed as multiplicative mixed model.

The filtering of the low Mach number equations is described in section 2.
The subgrid-scale models are presented in section 3. 
The channel flow configuration and the numerical method are given in section 4.
The section 5 discusses the asymptotic near-wall behaviour of the models.
The results are analysed in section 6.

\section{Filtering of the low Mach number equations}

The low Mach number equations are an approximation of the Navier--Stokes
equations suited to turbulent flows with a low Mach number ($Ma < 0.3$) but
subjected to large variations of the fluid properties.
Using Paolucci's method \citep{paolucci1982a}, each variable of the
Navier--Stokes equations is written as a power series of the squared Mach
number. Neglecting all but the smaller-order terms, the pressure is split in
two parts:
The thermodynamical pressure $P$ (constant in space), which represents the
mean pressure in the domain,
and the mechanical pressure $P_0$, associated with the momentum variations.
The resulting equations are free from acoustic waves.

Considering in addition an ideal gas and neglecting gravity, the low Mach
number equations are given by:
\begin{itemize}
\item Mass conservation equation
\begin{equation}
\der{\rho}{t} + \der{\rho U_{j}}{x_j} = 0,
\end{equation}
\item Momentum conservation equation
\begin{equation}
\der{\rho U_i}{t} = - \der{\rho U_j U_i}{x_j} - \der{P}{x_i} + \der{\varSigma_{ij}(\vv{U}, T)}{x_j},
\end{equation}
\item Energy conservation equation
\begin{equation}\label{energyequationdiv}
\der{U_j}{x_j} = - \frac{1}{\gamma P_{0}}\left[ (\gamma - 1)\der{Q_j(T)}{x_j} + \der{P_{0}}{t} \right],
\end{equation}
\item Ideal gas law
\begin{equation}\label{idealgaslawn}
T = \frac{P_{0}}{\rho r},
\end{equation}
\end{itemize}
with $\rho$ the density, $T$ the temperature, $\varSigma_{ij}(\vv{U}, T)$ the shear-stress
tensor, $Q_j(T)$ the conductive heat flux, $\gamma$ the heat capacity ratio,
$r$ the ideal gas specific constant, $t$ the time, $P$ the mechanical pressure,
$P_0$ the thermodynamical pressure, $U_i$ the $i$-th component of velocity and
$x_i$ the Cartesian coordinate in $i$-th direction. Einstein summation
convention is used.
The low Mach number equations impose the local energy conservation by a
constraint (\ref{energyequationdiv}) on the divergence of the velocity \citep{nicoud2000conservative}.

The filtering of the low Mach number equations may lead to different
formulations of the filtered low Mach number equations depending on
the variables we express the equations with and
the manner the equations are arranged upon filtering.
Two formulations of the filtered low Mach number equations are selected, the
Velocity formulation and the Favre formulation.
In the Velocity formulation, a spatial filter ($\f{\,\,\,\cdot\,\,}$, classical filter) is
applied on the low Mach number equations with the momentum conservation
equation rewritten as the velocity transport equation.
The equations are then expressed in terms of classical-filtered variables.
The Favre formulation is based on the use of a density-weighted filter ($\fa{\,\,\,\cdot\,\,}$, Favre filter),
defined for any~$\phi$, as $\fa{\phi} = \f{\rho \phi} / \f{\rho}$.
In the Favre formulation, the low Mach number equations are filtered with the
classical filter and expressed in terms of Favre-filtered variables. 

Retaining only the most significant subgrid terms \citep{dupuy2016, dupuy2017sft, dupuy2018study},
the filtered low Mach number equations are given in the Velocity formulation by:
\begin{gather}
\der{\f{\rho}}{t} + \der{}{x_j}\left(\f{\rho} \f{U}_{j} + F_{\rho U_j}\right) = 0, \\
\der{\f{U}_i}{t} = - \der{}{x_j}\left(\f{U}_j\, \f{U}_i + F_{U_j U_i}\right) + \f{U}_i \der{\f{U}_j}{x_j} - \frac{1}{\f{\rho}}\der{\f{P}}{x_i} + \frac{1}{\f{\rho}} \der{\varSigma_{ij}(\vv{\f{U}}, \f{T})}{x_j}, \label{vte} \\
\der{\f{U}_j}{x_j} = - \frac{1}{\gamma P_{0}}\left[ (\gamma - 1)\der{Q_j(\f{T})}{x_j} + \der{P_{0}}{t} \right], \\
\f{T} = \frac{P_{0}}{r \f{\rho}},
\end{gather}
and in the Favre formulation by:
\begin{gather}
\der{\f{\rho}}{t} + \der{\f{\rho} \fa{U_{j}}}{x_j} = 0, \\
\der{\f{\rho} \fa{U}_i}{t} = - \der{}{x_j}\left(\f{\rho} \fa{U}_j \fa{U}_i + \f{\rho} G_{U_j U_i}\right) - \der{\f{P}}{x_i} + \der{\varSigma_{ij}(\vv{\fa{U}},\fa{T})}{x_j}, \label{mce} \\
\der{}{x_j}\left(\fa{U}_j + \f{\rho} G_{U_j/\rho}\right) = - \frac{1}{\gamma P_{0}}\left[ (\gamma - 1)\der{Q_j(\fa{T})}{x_j} + \der{P_{0}}{t} \right], \\
\fa{T} = \frac{P_{0}}{\f{\rho} r},
\end{gather}
with the subgrid terms:
\begin{align}
F_{U_j U_i} ={}& \f{U_j U_i} - \f{U}_j\, \f{U}_i
\\
G_{U_j U_i} ={}& \fa{U_j U_i} - \fa{U}_j \fa{U}_i
\\
F_{\rho U_j} ={}& \f{\rho U_j} - \f{\rho} \f{U}_j
\\
G_{U_j/\rho} ={}& \fa{U_j/\rho} - \fa{U}_j/\f{\rho}
\end{align}%

The Velocity and Favre formulations both involve a subgrid term
associated with the momentum convection, $F_{U_j U_i}$ or $G_{U_j U_i}$,
and a subgrid term associated with the density-velocity correlation,
$F_{\rho U_j}$ or $G_{U_j/\rho}$, such that
\begin{equation}
\frac{F_{\rho U_j}}{\f{\rho}} = - \f{\rho} G_{U_j/\rho}.
\end{equation}
The use of the Favre filter removes the need for the modelling of the density-velocity
correlation from the mass conservation equation but requires the
modelling of an additional subgrid term in the energy conservation equation \citep{dupuy2016, dupuy2017sft, dupuy2018study}.

The fluid (air) is assumed to be Newtonian to compute the shear-stress tensor,
\begin{equation}
\varSigma_{ij}(\vv{U}, T) = \mu(T) \left(\der{U_i}{x_j} + \der{U_j}{x_i}\right) - \frac{2}{3} \mu(T) \der{U_k}{x_k} \delta_{ij},
\end{equation}
with $\mu(T)$ the dynamic viscosity and $\delta_{ij}$ the Kronecker delta. The heat flux is given by
\begin{equation}
Q_j(T) = - \lambda(T) \der{T}{x_j},
\end{equation}
with $\lambda(T)$ the thermal conductivity.
The variations of viscosity with temperature are accounted for by
Sutherland's law~\citep{sutherland1893lii},
\begin{equation}
\mu(T) = \mu_0 \left(\frac{T}{T_0}\right)^{\frac{3}{2}} \frac{T_0 + S}{T + S},
\end{equation}
with $\mu_0 = 1.716\cdot 10^{-5}$~Pa~s, $S=110.4$~K and $T_0 = 273.15$~K.
The conductivity is deduced from the Prandlt number $Pr$ and
the heat capacity at constant pressure $C_p$, both assumed constant with
$Pr$ = 0.76 and~$C_p=1005$ J kg$^{-1}$~K$^{-1}$. The ideal gas
specific constant is~$r=287$~J~kg$^{-1}$~K$^{-1}$.

\section{Subgrid-scale models}

The subgrid terms of the Velocity and Favre formulations are formally
similar. Accordingly, the same modelling procedure is used in both cases.
To formalise this, we may express the subgrid-scale models as a function
of the filter length scales and of the filtered velocity and density
in the two formulations:
\begin{align}
F_{U_j U_i} \approx{}& \tau_{ij}^{\mathrm{mod}}(\vv{\f{U}}, \vv{\f{\Delta}}), \\
G_{U_j U_i} \approx{}& \tau_{ij}^{\mathrm{mod}}(\vv{\fa{U}}, \vv{\f{\Delta}}), \\
F_{\rho U_j} \approx{}& \pi_{j}^{\mathrm{mod}}(\vv{\f{U}}, \f{\rho}, \vv{\f{\Delta}}), \\
G_{U_j/\rho} \approx{}& \pi_{j}^{\mathrm{mod}}(\vv{\fa{U}}, 1/\f{\rho}, \vv{\f{\Delta}}),
\end{align}%
where the functions $\tau_{ij}^{\mathrm{mod}}(\vv{U},\vv{\f{\Delta}})$ and
$\pi_j^{\mathrm{mod}}(\vv{U},\phi,\vv{\f{\Delta}})$ are model-dependent but do not depend on the formulation.

Eddy-viscosity models for the subgrid term associated with momentum convection
may be written in the form
\begin{align}
\tau_{ij}^{\mathrm{mod}}(\vv{U}, \vv{\f{\Delta}}) ={}& - 2 \nu_e^{\mathrm{mod}}(\vv{g}, \vv{\f{\Delta}}) S_{ij}, \label{hela}
\end{align}%
with
$S_{ij} = \tfrac{1}{2}\left( g_{ij} + g_{ji} \right)$ the rate of deformation tensor
and
$\vv{g}$
the velocity gradient, defined by $g_{ij} = \partial_j U_i$.
Notice that $\tau_{ij}^{\mathrm{mod}}(\vv{U}, \vv{\f{\Delta}})$ may be
considered traceless without loss of generality, even in the incompressible case,
since the trace can be included as part of the filtered pressure $\f{P}$.
The eddy-viscosity $\nu_e^{\mathrm{mod}}(\vv{g}, \vv{\f{\Delta}})$
is given by the model used.
The following models from the literature are investigated in this paper using a priori tests:
\begin{flalign}
&\textrm{Smagorinsky model \citep{smagorinsky1963general}:} & \nu_e^{\mathrm{Smag.}}  (\vv{g}, \vv{\f{\Delta}}) ={}& \left( C^{\mathrm{Smag.}} \f{\Delta} \right)^2 \left|\vt{S}\right|, &&& \label{sma} \\
&\textrm{WALE model \citep{nicoud99b}:}                     & \nu_e^{\mathrm{WALE}}   (\vv{g}, \vv{\f{\Delta}}) ={}& \left( C^{\mathrm{WALE}} \f{\Delta} \right)^2 \frac{\left(\mathcal{S}^d_{ij} \mathcal{S}^d_{ij}\right)^{\tfrac{3}{2}}}{\left(S_{mn} S_{mn}\right)^{\tfrac{5}{2}} + \left(\mathcal{S}^d_{mn} \mathcal{S}^d_{mn}\right)^{\tfrac{5}{4}}}, &&& \\
&\textrm{Vreman model \citep{vreman2004eddy}:}              & \nu_e^{\mathrm{Vreman}} (\vv{g}, \vv{\f{\Delta}}) ={}& C^{\mathrm{Vreman}} \sqrt{\frac{\mathrm{II}_G}{g_{mn}g_{mn}}}, &&& \\
&\textrm{Sigma model \citep{nicoud2011using}:}              & \nu_e^{\mathrm{Sigma}}  (\vv{g}, \vv{\f{\Delta}}) ={}& \left( C^{\mathrm{Sigma}} \f{\Delta} \right)^2 \frac{\sigma_3\left(\sigma_1 - \sigma_2\right)\left(\sigma_2 - \sigma_3\right)}{\sigma_1^2}, &&& \\
&\textrm{AMD model \citep{rozema2015minimum}:}              & \nu_e^{\mathrm{AMD}}    (\vv{g}, \vv{\f{\Delta}}) ={}& C^{\mathrm{AMD}} \frac{\max(0, - G_{ij} S_{ij})}{g_{mn}g_{mn}}, &&& \\
&\textrm{VSS model \citep{ryu2014subgrid}:}                 & \nu_e^{\mathrm{VSS}}    (\vv{g}, \vv{\f{\Delta}}) ={}& \left( C^{\mathrm{VSS}} \f{\Delta} \right)^2  \frac{\left(R_{ij} R_{ij}\right)^{\frac{3}{2}}}{\left(S_{mn}S_{mn}\right)^{\frac{5}{2}}}, &&& \\
&\textrm{Kobayashi model \citep{kobayashi2005subgrid}:}     & \nu_e^{\mathrm{Koba.}}  (\vv{g}, \vv{\f{\Delta}}) ={}& C^{\mathrm{Koba.}} \f{\Delta}^2 \left|F_g\right|^{\frac{3}{2}} (1-F_g) \left|\vt{S}\right|, &&& \label{koba}
\end{flalign}%
where 
$\left|\vt{S}\right|=\sqrt{2 S_{ij} S_{ij}}$ is a norm of $\vt{S}$,
$\mathcal{S}^d_{ij} = \tfrac{1}{2}\left( g_{ik}g_{kj} + g_{jk}g_{ki} \right) - \tfrac{1}{3}g_{kp}g_{pk} \delta_{ij}$ the traceless symmetric part of the squared velocity gradient tensor, 
$\sigma_1 \geq \sigma_2 \geq \sigma_3$ the three singular values of $\vv{g}$,
$G_{ij} = \f{\Delta}_k^2 g_{ik} g_{jk}$ the gradient model for the subgrid term associated with momentum convection \citep{leonard74},
$\mathrm{II}_G = \tfrac{1}{2}\left(\tr^2\left(G\right) - \tr\left(G^2\right)\right)$ its second invariant,
$R_{ij}=\beta_i g_{jj}$ the volumetric strain-stretching, with $\beta=\left(S_{23}, S_{13}, S_{12}\right)$,
and $F_g = \left(\varOmega_{ij}\varOmega_{ij} - S_{ij}S_{ij}\right)/\left(\varOmega_{mn}\varOmega_{mn} + S_{mn}S_{mn}\right)$
the coherent structure function, with $\varOmega_{ij} = \tfrac{1}{2}\left( g_{ij} - g_{ji} \right)$ the spin tensor or rate of rotation tensor.
Only constant coefficient versions of eddy-viscosity and eddy-diffusivity models are considered.
The typical value of the coefficients from the literature is
$C^{\mathrm{Smag.}} = 0.10$, $C^{\mathrm{WALE}} = 0.55$, $C^{\mathrm{Vreman}}=0.07$, $C^{\mathrm{Sigma}}=1.5$, $C^{\mathrm{AMD}}=0.3$, $C^{\mathrm{VSS}}=1.3$ and $C^{\mathrm{Koba.}}=0.045$.
The corresponding dynamic versions of these models are not considered
in order to assess the relevance of the models before any dynamic correction \citep{germano91, lilly1992proposed, park2006dynamic}.
The filter length scale is computed following \citet{deardorff1970numerical} as  $\f{\Delta}=(\f{\Delta}_x\f{\Delta}_y\f{\Delta}_z)^{1/3}$.
A review of alternative possible definitions may be found in \citet{trias2017new}.

Following the same rationale, eddy-diffusivity models for the density-velocity
correlation subgrid term may be written in the form
\begin{align}
\pi_{j}^{\mathrm{mod}}(\vv{U}, \phi, \vv{\f{\Delta}}) ={}& - 2 \kappa_e^{\mathrm{mod}}(\vv{g}, \vv{d}, \vv{\f{\Delta}}) d_j. \label{helb}
\end{align}%
with $\vv{d}$
the scalar gradient, defined by $d_{j} = \partial_j \phi$.
It is common to express the eddy-diffusivity $\kappa_e^{\mathrm{mod}}(\vv{g}, \vv{\f{\Delta}})$
using the constant subgrid-scale Prandtl or Schmidt number assumption,
\begin{align}
\kappa_e^{\mathrm{mod}}    (\vv{g}, \vv{d}, \vv{\f{\Delta}}) ={}& \frac{1}{Pr_t} \nu_e^{\mathrm{mod}}    (\vv{g}, \vv{\f{\Delta}}), \label{dfv}
\end{align}%
where $Pr_t$ is the subgrid-scale Prandtl or Schmidt number. 
This provide a corresponding eddy-diffusivity model for each eddy-viscosity of equations (\ref{sma}--\ref{koba}).
The dimensionless number $Pr_t$ corresponds to a subgrid-scale Schmidt number
in the Velocity formulation and a subgrid-scale Prandtl number in the Favre
formulation.
Given the formal similarity between the density-velocity correlation subgrid
term in the Velocity and Favre formulation and the ideal gas law
(\ref{idealgaslawn}) which relates density and temperature, it is presumed that
the same value may be used in the two formulations.
Alternatively, some specific eddy-diffusivity models have been suggested in
the literature \citep{ghaisas2014priori, abkar2016minimum}.
We investigate using a priori tests the eddy-diffusivity models associated with equations (\ref{sma}--\ref{koba}) and the following specific model:
\begin{flalign}
&\textrm{Scalar AMD model \citep{abkar2016minimum}:}        & \kappa_e^{\mathrm{SAMD}}   (\vv{g}, \vv{d}, \vv{\f{\Delta}}) ={}& C^{\mathrm{SAMD}} \frac{\max(0, - D_j d_j)}{d_m d_m}, &&&
\end{flalign}%
with $D_j = \f{\Delta}_k^2 g_{jk} d_k$ the gradient model for the density-velocity correlation subgrid term.

In addition, we devised two new eddy-viscosity and eddy-diffusivity models
for the purpose of this study.
First, the Anisotropic Smagorinsky model is a modified version of the Smagorinsky model,
associated with a single
filter length scale, devised to involve the three filter length scales.
This aims to improve the anisotropy of the model.
The model is obtained by substituting in equations
(\ref{hela}) and (\ref{helb}) the velocity gradient $\vv{g}$
 and respectively the scalar
gradient $\vv{d}$
by the
scaled velocity gradient $\vv{g^a}$, defined by $g^a_{ij} = (\f{\Delta}_j/\f{\Delta}) \partial_j U_i$,
and respectively
the scaled scalar gradient $\vv{d^a}$, defined by $d^a_j = (\f{\Delta}_j/\f{\Delta}) \partial_j \phi$.
Namely,
\begin{align}
\tau_{ij}^{\mathrm{An. Smag.}}(\vv{U}, \vv{\f{\Delta}}) ={}& - 2 \nu_e^{\mathrm{Smag.}}(\vv{g^a}, \vv{\f{\Delta}}) S^a_{ij}, \\
\pi_{j}^{\mathrm{An. Smag.}}(\vv{U}, \phi, \vv{\f{\Delta}}) ={}& - 2 \kappa_e^{\mathrm{Smag.}}(\vv{g^a}, \vv{d^a}, \vv{\f{\Delta}}) d^a_j,
\end{align}
with $S^a_{ij} = \tfrac{1}{2}\left( g^a_{ij} + g^a_{ji} \right)$ the scaled rate of deformation tensor.
The eddy-viscosity and eddy-diffusivity are computed using equations (\ref{sma}) and (\ref{dfv}).
A similar procedure could be applied to obtain an anisotropic version
of the
WALE,
Sigma,
VSS and
Kobayashi models.

Besides, we study the multiplicative mixed model based on the gradient model (MMG model), a functional model constructed such that
its magnitude is determined by the gradient model \citep{leonard74} and
its orientation is aligned with the rate of deformation tensor or the scalar
gradient depending on the subgrid term.
This procedure is reminiscent of the multiplicative mixed model
of \citet{ghaisas2014priori, ghaisas2016dynamic} which had an opposite purpose.
The eddy-viscosity and eddy-diffusivity according to the MMG model are given
by,
\begin{flalign}
&\textrm{MMG model:}                     & \nu_e^{\mathrm{MMG}}     (\vv{g}, \vv{\f{\Delta}})         ={}& - C^{\mathrm{MMG}} \frac{G_{kk}}{\left|\vt{S}\right|}, &&& \\
&\textrm{Scalar MMG model:}              & \kappa_e^{\mathrm{SMMG}} (\vv{g}, \vv{d}, \vv{\f{\Delta}}) ={}& - C^{\mathrm{SMMG}} \frac{\sqrt{D_i D_i}}{\sqrt{d_m d_m}}. &&&
\end{flalign}%
A similar procedure can be applied to other structural
models, such as the scale-similarity model \citep{bardina1980improved}.
We may also view the MMG model as a multiplicative mixed model. 
Using the the Smagorinsky model and the isotropic part modelling of
\citet{yoshizawa1986statistical},
\begin{align}
\tau_{mm}^{\mathrm{Yosh.}}(\vv{U}, \vv{\f{\Delta}}) ={}& 2 C^{\mathrm{Yosh.}} \f{\Delta}^2 \left|\vt{S}\right|^2,
\end{align}
the MMG model $\tau_{ij}^{\mathrm{MMG}}(\vv{U}, \vv{\f{\Delta}}) = - 2 \nu_e^{\mathrm{MMG}}(\vv{g}, \vv{\f{\Delta}}) S_{ij}$
can be reformulated as
\begin{align}
\tau_{ij}^{\mathrm{MMG}}(\vv{U}, \vv{\f{\Delta}}) ={}& G_{kk}\frac{ \tau_{ij}^{\mathrm{Smag.}}(\vv{U}, \vv{\f{\Delta}})}{\tau_{mm}^{\mathrm{Yosh.}}(\vv{U}, \vv{\f{\Delta}})}
\end{align}
emphasising that the MMG model combines the magnitude of the gradient model
and the structure of the Smagorinsky model.
This leads by identification $C^{\mathrm{MMG}} = (C^{\mathrm{Smag.}})^2/(2C^{\mathrm{Yosh.}})$.
Note that the Vreman, AMD and scalar AMD models also directly involve the gradient model \citep{leonard74}. 

\section{Numerical study configuration}

\subsection{Channel flow configuration}

\begin{figure}[t!]
\centerline{\includegraphics[scale=0.6]{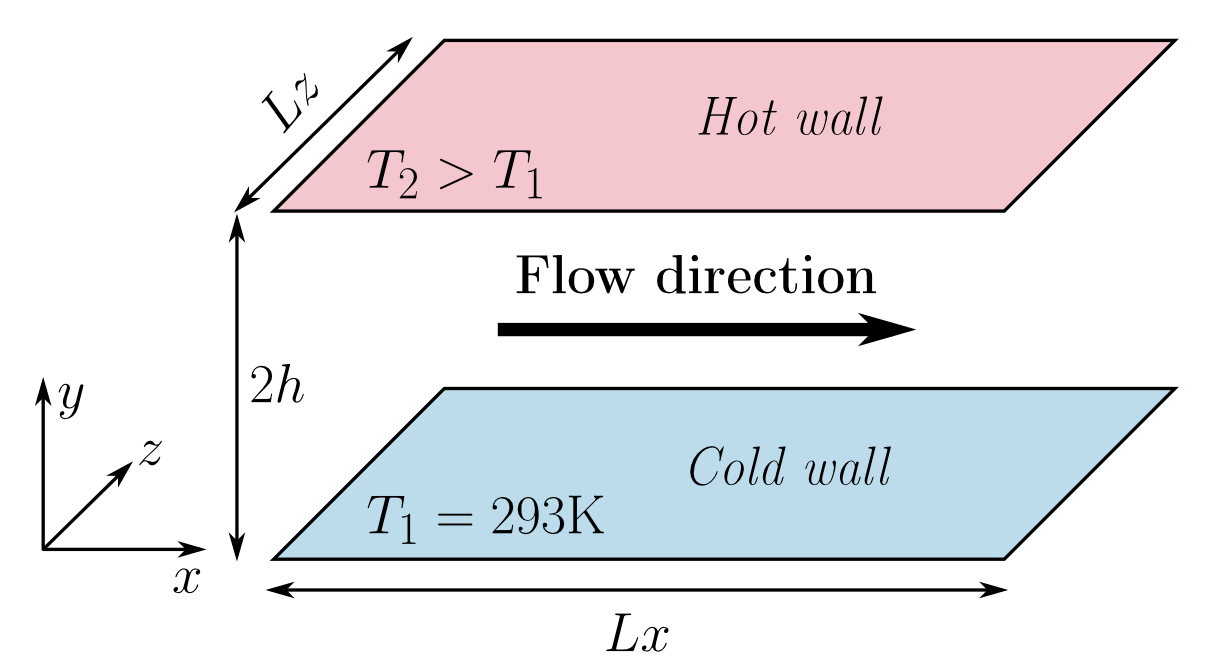}}
\caption{
Biperiodic anisothermal channel flow.
\label{lf1}}
\end{figure}

We consider a  fully developed three-dimensional anisothermal channel flow, as
represented in figure \ref{lf1}. This geometry is one of the simpler that reproduces
the distinctive features of low Mach number strongly anisothermal turbulent flows.
The channel is periodic in the streamwise ($x$) and spanwise
($z$) directions. The wall-normal direction is denoted ($y$).
The domain size is $4\pi h \times 2h \times 2\pi h$, with  $h$ = 15 mm.
The temperature at the channel walls is imposed at $T_1$ =
293 K at the cold wall ($y=0$) and $T_2$ = 586 K at the hot wall ($y=2h$).
This creates a temperature gradient in the wall-normal direction.
The mean friction
Reynolds number is $Re_{\tau}$ = 180,
where $Re_{\tau}$ is defined as the average
of the friction Reynolds numbers $Re_{\tau, \omega}$ calculated at the hot and cold wall,
\begin{equation}
Re_{\tau, \omega} = \frac{U_{\tau} h}{\nu_{\omega}},
\end{equation}
with $U_{\tau}=\nu_{\omega} (\partial_y \m{U}_x)_{\!\omega}^{0.5}$ the
friction velocity and $\nu_{\omega}$ the wall kinematic viscosity.

\subsection{Numerical settings}

The mesh contains $384 \times 266 \times 384$ grid points and
is regular in the homogeneous directions. It follows a hyperbolic
tangent law in the wall-normal coordinate direction. The wall-normal grid
coordinates are symmetrical with respect to the plane $y=h$. In the first
half of the channel, they are given by
\begin{equation}
y_k = h \left( 1 + \frac{1}{a} \tanh\left[ \left(\frac{k-1}{N_y-1} - 1\right)\tanh^{-1}(a)\right] \right), \label{eqmesh}
\end{equation}
with $a=0.97$ the mesh dilatation parameter and $N_y$ the number of grid points
in the wall-normal direction. The cell sizes in wall units are $\Delta_x^+$~=~8.5,
$\Delta_y^+$~=~0.13~at the wall and 4.2 at the centre of the channel and $\Delta_z^+$~=~4.2.
A finite volume method is used with a third-order Runge--Kutta time scheme
and a fourth-order centred momentum convection scheme. This is performed using
the TrioCFD software \citep{calvin2002object}.

\subsection{Filtering process}

The subgrid terms and the models are computed from the filtering of the
instantaneous DNS data at the resolution of a large-eddy simulation mesh.
The filter corresponds to a mesh with $48 \times 50 \times 48$ grid points
($\Delta_x^+= 68$; $\Delta_y^+= 0.5$ -- $25$; $\Delta_z^+= 34$)
constructed as the DNS mesh.
Due to the inhomogeneity of the mesh,
the filter width is variable in the wall-normal direction.

A top-hat filter is used. In one dimension, it is given in the physical space by
\begin{equation}
\f{\psi} (x) = \frac{1}{\f{\Delta}(x)} \int_{x-\tfrac{1}{2}\f{\Delta}(x)}^{x+\tfrac{1}{2}\f{\Delta}(x)} \psi(\xi) d\xi. \label{eqfilter}
\end{equation}
Multidimensional filtering is carried out by sequentially applying the
one-dimen\-sional filter in the three spatial directions. 
In order to carry out the filtering with an arbitrary filter length, the
DNS data are first interpolated using a cubic spline.
The top-hat filter is then computed from the interpolated value without mesh
restrictions.

The discretisation of the differential operator of the models is carried out
on the DNS grid, thus using data not available in an a posteriori large-eddy
simulation \citep{liu1994properties}. This assesses the relevance of the models
without regard to numerical errors.
The data from 100 uncorrelated timesteps were averaged in order to obtain a
satisfactory convergence of the results.

\section{Asymptotic near-wall behaviour of the models}

The WALE, Sigma, VSS and Kobayashi models have been designed to have an
eddy-viscosity with a proper asymptotic near-wall behaviour for the subgrid
term associated with momentum convection.
While the components of the subgrid terms have different wall orders,
the preferable asymptotic near-wall behaviour of the
eddy-viscosity is cubic with respect to the distance to the wall,
that is $\left.\nu_e^{\mathrm{mod}}(\vv{g}, \vv{\f{\Delta}})\right|_\omega = \mathcal{O}(y^{3})$.
A reason is that it is the order that the eddy-viscosity should have for
the near-wall behaviour of the subgrid kinetic energy dissipation to be consistent with the
exact subgrid kinetic energy dissipation.
The asymptotic near-wall behaviour of the models given in the literature 
\citep{nicoud99b, nicoud2011using, trias2015building, silvis2017physical}
considers the behaviour of the differential operator the models are based on,
assuming that the filter length does not tend to zero at the wall,
$\left.\f{\Delta}\right|_\omega = \mathcal{O}(y^{0})$.
The near-wall order of the models can be obtained from the Taylor
series expansion of the velocity and the scalar (density or inverse of density)
\citep{chapman1986limiting, sagaut98, chong2012topology, nicoud2011using}:
\begin{align}
\left.U_x\right|_\omega = \mathcal{O}(y^{1}),  \\
\left.U_y\right|_\omega = \mathcal{O}(y^{2}),  \\
\left.U_z\right|_\omega = \mathcal{O}(y^{1}),  \\
\left.\phi\right|_\omega = \mathcal{O}(y^{0}).
\end{align}
The quadratic behaviour of the wall-normal velocity follows from the mass
conservation equation, provided that the density is constant at the walls.
This assumption is valid in our case if the time variations of thermodynamical
pressure are neglected, since the wall temperatures are imposed.
The filter is considered to not alter the asymptotic near-wall behaviour of
the variables. This assumption is valid for a top-hat filter as defined in
equation (\ref{eqfilter}) with varying filter width.
The cubic asymptotic near-wall behaviour of
the subgrid term can be recovered, for the ``$xy$'' component,
from the linear near-wall order of the streamwise velocity and
the quadratic near-wall order of the wall-normal velocity \citep{sagaut98, silvis2017physical}.

We find that this procedure is not satisfactory for the density-velocity
correlation subgrid term.
Indeed, it is not able to take into account the fact that
$F_{\rho U_j} = \f{\rho U_j} - \f{\rho} \f{U}_j$
cannot have a near-wall order below $2$
because the filter used, given in equation (\ref{eqfilter}),
preserves constant and linear functions.
To determine the
asymptotic near-wall behaviour of the subgrid terms, we carry out
a Taylor series expansion of the filter, leading to the gradient model \citep{leonard74}.
Next, the near-wall order of the gradient model is expressed considering
a filter with a non-zero order at the wall.
For a continuous filter whose size in the wall-normal direction ($y$) tends to zero at the wall,
it is natural to consider
\begin{align}
\left.\f{\Delta}_x\right|_\omega ={}& \mathcal{O}(y^{0}), \label{eqof1}  \\
\left.\f{\Delta}_y\right|_\omega ={}& \mathcal{O}(y^{1}), \label{eqof2}  \\
\left.\f{\Delta}_z\right|_\omega ={}& \mathcal{O}(y^{0}). \label{eqof3}
\end{align}
It follows
\begin{align}
\left.\f{\Delta}\right|_\omega ={}& \mathcal{O}(y^{1/3}).
\end{align}
Note also that the near-wall order of the streamwise and spanwise derivatives of the scalar
is at least $\mathcal{O}(y^{1})$ under the hypothesis of constant density at the walls.
With these assumptions,
the expected asymptotic near-wall behaviour of 
the eddy-diffusivity for
models of the density-velocity correlation subgrid term is also
cubic with respect to the distance to the wall,
$\left.\kappa_e^{\mathrm{mod}}(\vv{g}, \vv{d}, \vv{\f{\Delta}})\right|_\omega = \mathcal{O}(y^{3})$.
This ensures that the order of the subgrid squared scalar dissipation
corresponds to that of the exact subgrid term.
For the subgrid term associated with momentum convection,
the results are consistent with the literature
since it leads to the same near-wall order for each component as 
the Taylor series expansion of the velocity tensor product.

The asymptotic near-wall behaviour of the investigated subgrid-scale models is given in table \ref{toder}
for a filter width of order
$\mathcal{O}(y^{0})$
at the wall and a filter which obeys equations (\ref{eqof1}--\ref{eqof3}).
With $\left.\f{\Delta}_y\right|_\omega = \mathcal{O}(y^{0})$,
the WALE, Sigma, VSS and Kobayashi models have the proper asymptotic near-wall
behaviour.
With $\left.\f{\Delta}_y\right|_\omega = \mathcal{O}(y^{1})$,
the AMD and scalar AMD models have the proper asymptotic near-wall
behaviour.

\begin{table}
\def~{\hphantom{0}}
\centerline{%
\small
\begin{tabular}{lcc}
Subgrid-scale model & With $\left.\f{\Delta}_y\right|_\omega = \mathcal{O}(y^{0})$ & With $\left.\f{\Delta}_y\right|_\omega = \mathcal{O}(y^{1})$\\[.5em]
\hline\\[-.5em]
Smagorinsky model \citep{smagorinsky1963general}  & $\mathcal{O}(y^{0})$ & $\mathcal{O}(y^{2})$  \\
WALE model \citep{nicoud99b}                      & $\mathcal{O}(y^{3})$ & $\mathcal{O}(y^{11/3})$ \\
Vreman model \citep{vreman2004eddy}               & $\mathcal{O}(y^1)$   & $\mathcal{O}(y^{3/2})$      \\
Sigma model \citep{nicoud2011using}               & $\mathcal{O}(y^{3})$ & $\mathcal{O}(y^{11/3})$ \\
AMD model \citep{rozema2015minimum}               & $\mathcal{O}(y^1)$   & $\mathcal{O}(y^3)$      \\
Scalar AMD model \citep{abkar2016minimum}         & $\mathcal{O}(y^0)$   & $\mathcal{O}(y^3)$      \\
VSS model \citep{ryu2014subgrid}                  & $\mathcal{O}(y^{3})$ & $\mathcal{O}(y^{11/3})$ \\
Kobayashi model \citep{kobayashi2005subgrid}      & $\mathcal{O}(y^{3})$ & $\mathcal{O}(y^{11/3})$ \\
Anisotropic Smagorinsky model                     & $\mathcal{O}(y^0)$   & $\mathcal{O}(y^2)$      \\
MMG model                                         & $\mathcal{O}(y^0)$   & $\mathcal{O}(y^2)$      \\
Scalar MMG model                                  & $\mathcal{O}(y^0)$   & $\mathcal{O}(y^2)$
\end{tabular}
}
\caption[Asymptotic near-wall behaviour of the models.]{Asymptotic near-wall behaviour of the models, for a constant and
linear near-wall behaviour of the filter width. The expected order is
$\mathcal{O}(y^3)$ for the subgrid term associated with momentum convection and
the density-velocity correlation subgrid term.
\label{toder}}
\end{table}                

\section{Results and discussion}

The performance of the subgrid-scale models is assessed from the comparison
of the models and the subgrid terms computed from the DNS data.
It is customary \citep[see e.g.][]{clark1979evaluation, vreman1995priori, borue1998local, martin2000subgrid, pruett2000priori, abba2003analysis, lu2007priori, ghaisas2014priori, ketterl2018priori}
to compare the model to the exact subgrid terms using a linear regression
analysis.
The correlation coefficient is an index scaled to between $-1$ and $1$ which
measures the linear correlation between two variables, that is the closeness
of the relationship between the two variables with a linear relationship.
A value of
$-1$ indicates a perfect negative linear relationship,
$0$ no linear relationship and
$1$ a perfect positive linear relationship.
Let us note $b$ a model for the subgrid term of exact value $a$.
The correlation coefficient between $a$ and $b$ is defined by,
\begin{equation}
\mathrm{Corr}(a,b) = \frac{\left\langle ab \right\rangle - \left\langle a \right\rangle \left\langle b\right\rangle}{\sqrt{( \left\langle a^2 \right\rangle - \left\langle a \right\rangle^2 ) ( \left\langle b^2 \right\rangle - \left\langle b\right\rangle^2 )}},
\end{equation}
where the angle brackets $\left\langle\,\,\cdot\,\,\right\rangle$ denote an ensemble
averaging.
The regression coefficient gives the slope of the linear relationship,
\begin{equation}
\mathrm{Regr}(a,b) = \frac{\left\langle ab \right\rangle - \left\langle a \right\rangle \left\langle b\right\rangle}{\left\langle a^2 \right\rangle - \left\langle a \right\rangle^2}.
\end{equation}
The concordance correlation coefficient \citep{lin1989concordance} is a
correlation-like index scaled to between $-1$ and $1$ which measure the
agreement between two variables, that is the closeness of the relationship
between the two variables with identity,
\begin{equation}
\mathrm{Conc}(a,b) = \frac{\left\langle ab \right\rangle - \left\langle a \right\rangle \left\langle b\right\rangle}{ \left\langle a^2 \right\rangle - \left\langle a \right\rangle^2  +  \left\langle b^2 \right\rangle - \left\langle b\right\rangle^2  + ( \left\langle a \right\rangle - \left\langle b\right\rangle )^2}.
\end{equation}
The correlation coefficient between the model and the exact subgrid term may be
interpreted as the ability of the model to capture the correct flow structure
and the regression coefficient of the correct magnitude level. The concordance
correlation coefficient combines the two types of information.
The optimal value of the correlation coefficient, the regression coefficient
and the concordance correlation coefficient is $1$. However, only a
concordance correlation coefficient of $1$ implies that the model and the exact
subgrid term are identical.

Given the homogeneity of the flow in the streamwise and spanwise directions,
the analysis is carried out as a function of the wall-normal coordinate.
The ensemble averaging $\left\langle\,\,\cdot\,\,\right\rangle$ is computed as an average
over time and the two homogeneous directions and the linear relationship
assessed for each value of $y$.
Notice that the addition for any value of $y$ of a constant scaling factor to
the model
does not modify the correlation coefficient,
multiply the regression coefficient by the constant and
has a non-trivial effect on the concordance correlation coefficient.

We first present some general results regarding the performance of the models.
Then, the subgrid-scale models are assessed for the subgrid term associated
with momentum convection and the density-velocity correlation subgrid term.

\subsection{General results}

\begin{figure}[t!]
\centerline{
 \includegraphics[width=0.51\textwidth, trim={0 10 10 5}, clip]{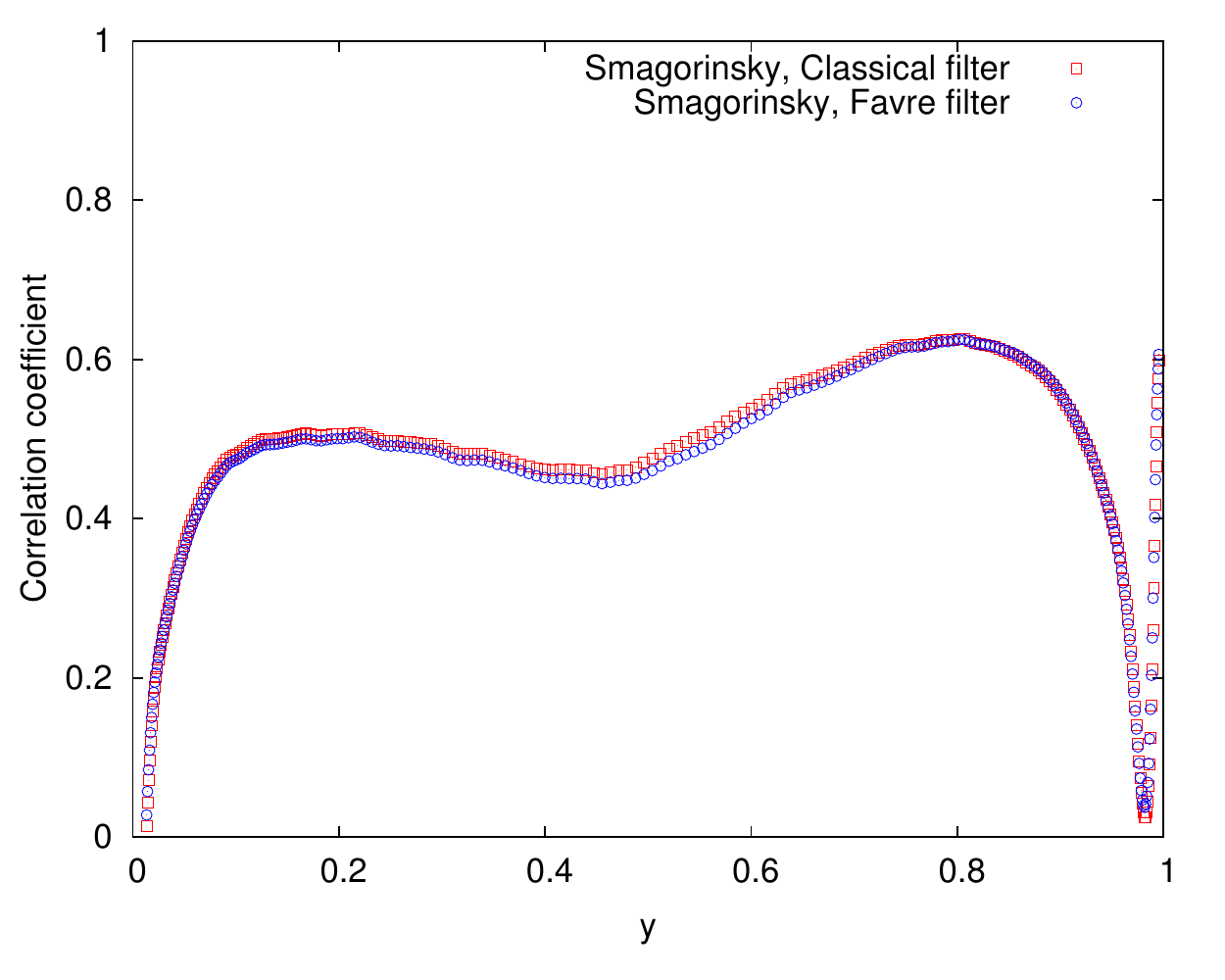}\includegraphics[width=0.51\textwidth, trim={0 10 10 5}, clip]{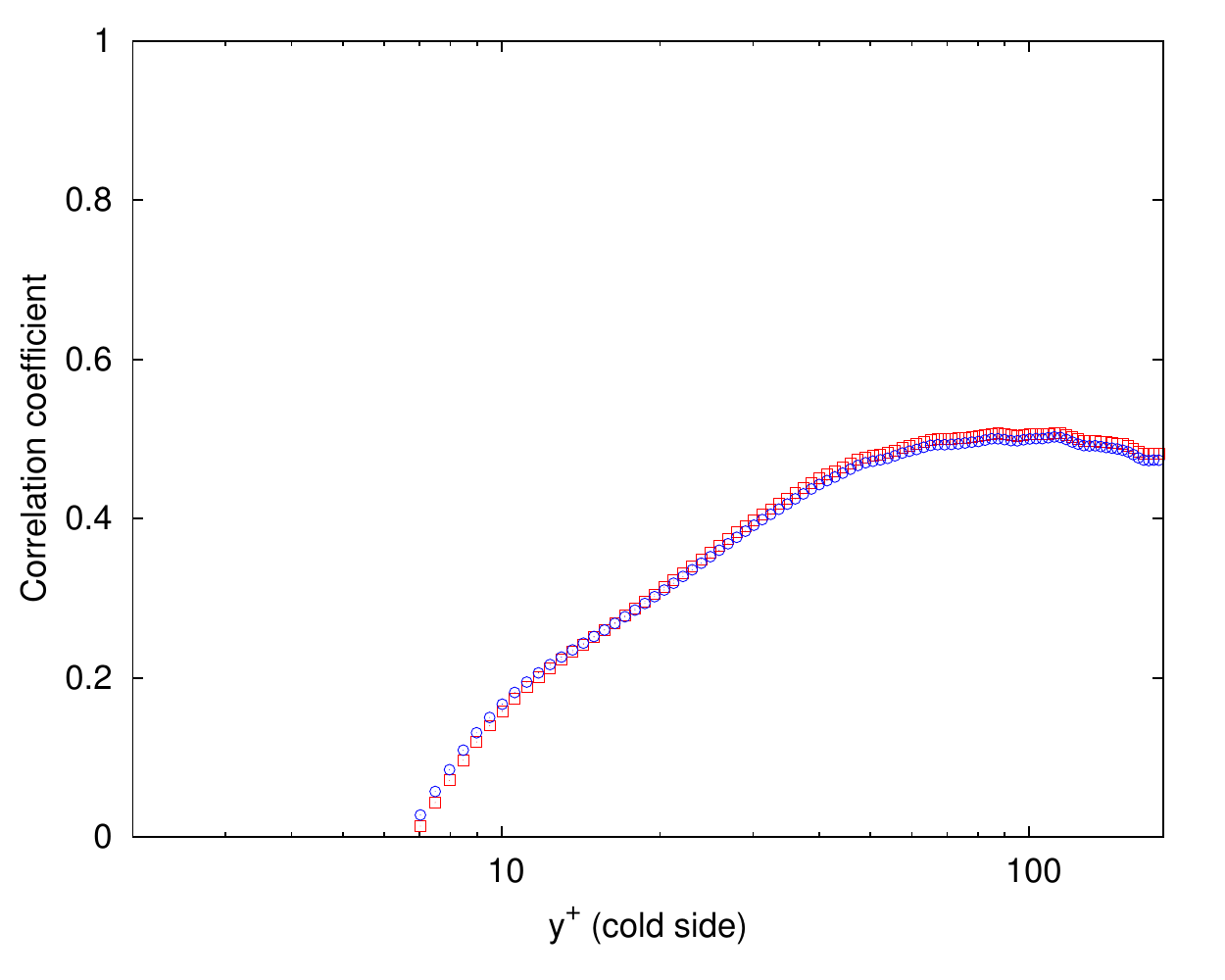}
 %}
}
\caption[Correlation coefficient between the exact momentum convection subgrid term and the Smagorinsky model for the term that appears in the streamwise momentum conservation equation.]{
Correlation coefficient between 
the exact momentum convection subgrid term and the Smagorinsky model
for the term that appears in the streamwise
velocity transport equation~(\ref{vte})
in the Velocity formulation, $\mathrm{Corr}(\partial_j F_{U_j U_x},\partial_j \tau_{xj}^{\mathrm{Smag.}}(\vv{\f{U}}, \vv{\f{\Delta}}))$,
and in the streamwise momentum conservation equation~(\ref{mce})
in the Favre formulation, $\mathrm{Corr}(\partial_j \f{\rho} G_{U_j U_x},\partial_j \f{\rho} \tau_{ij}^{\mathrm{Smag.}}(\vv{\fa{U}}, \vv{\f{\Delta}}))$.
\label{falun}}
\end{figure}

The subgrid-scale modelling in the Velocity and Favre formulations are compared
from the study of the subgrid terms and the models with the classical
filter and with the Favre filter.
The results show no differences between the classical and Favre filter with
regard to the performance of the models. 
For instance, the correlation coefficient between the Smagorinsky model and
the exact momentum convection subgrid term with the classical filter and with
the Favre filter are very similar (figure \ref{falun}).
The a priori study of the subgrid-scale models thus does not let us select
between the Velocity and Favre formulations of the filtered low Mach number
equations.
Thereafter, the subgrid-scale models are assessed in the Velocity formulation,
using the classical filter, but the results also apply to the Favre
formulation. 

\begin{figure}[t!]
\centerline{
\includegraphics[width=0.51\textwidth, trim={0 10 10 5}, clip]{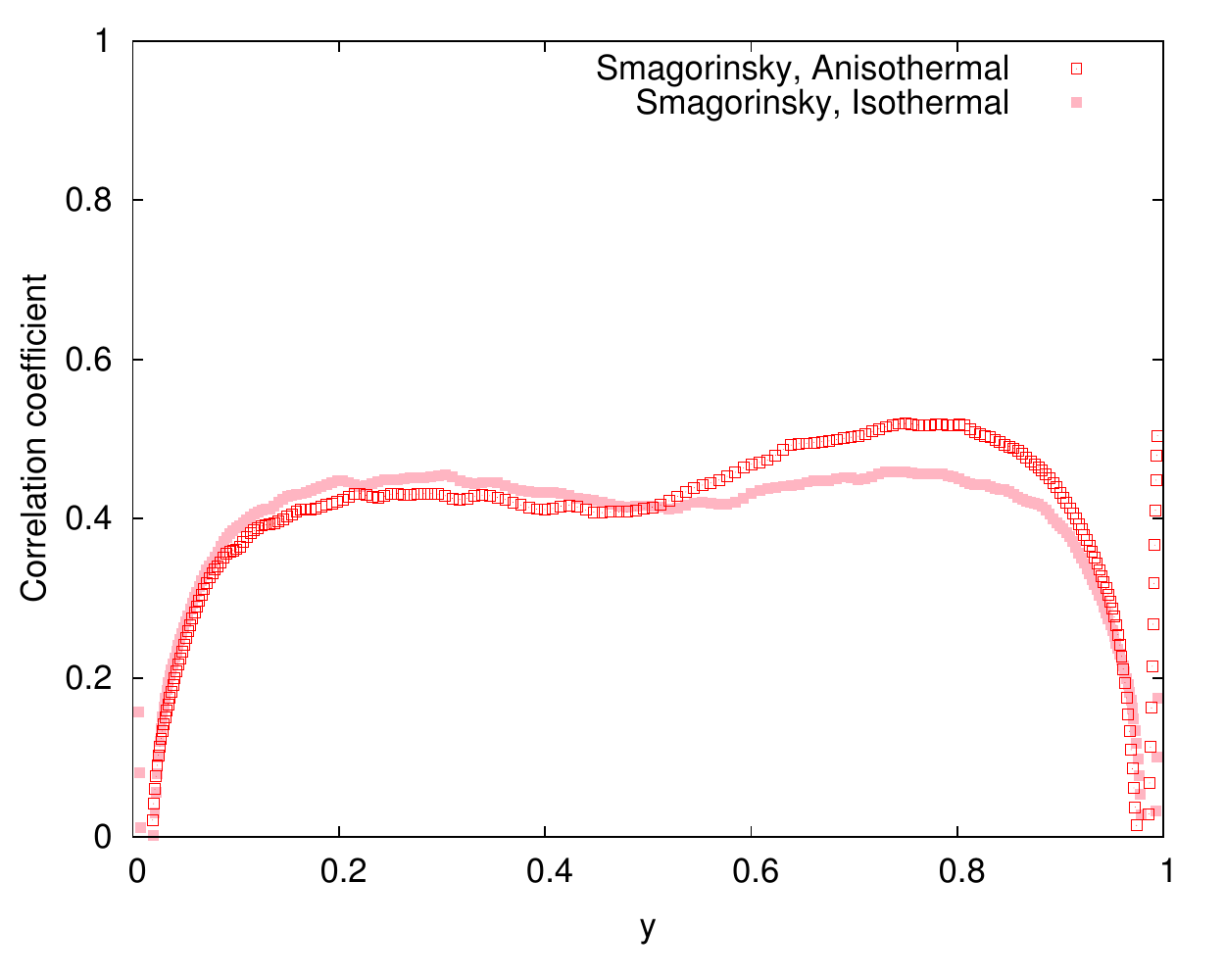}\includegraphics[width=0.51\textwidth, trim={0 10 10 5}, clip]{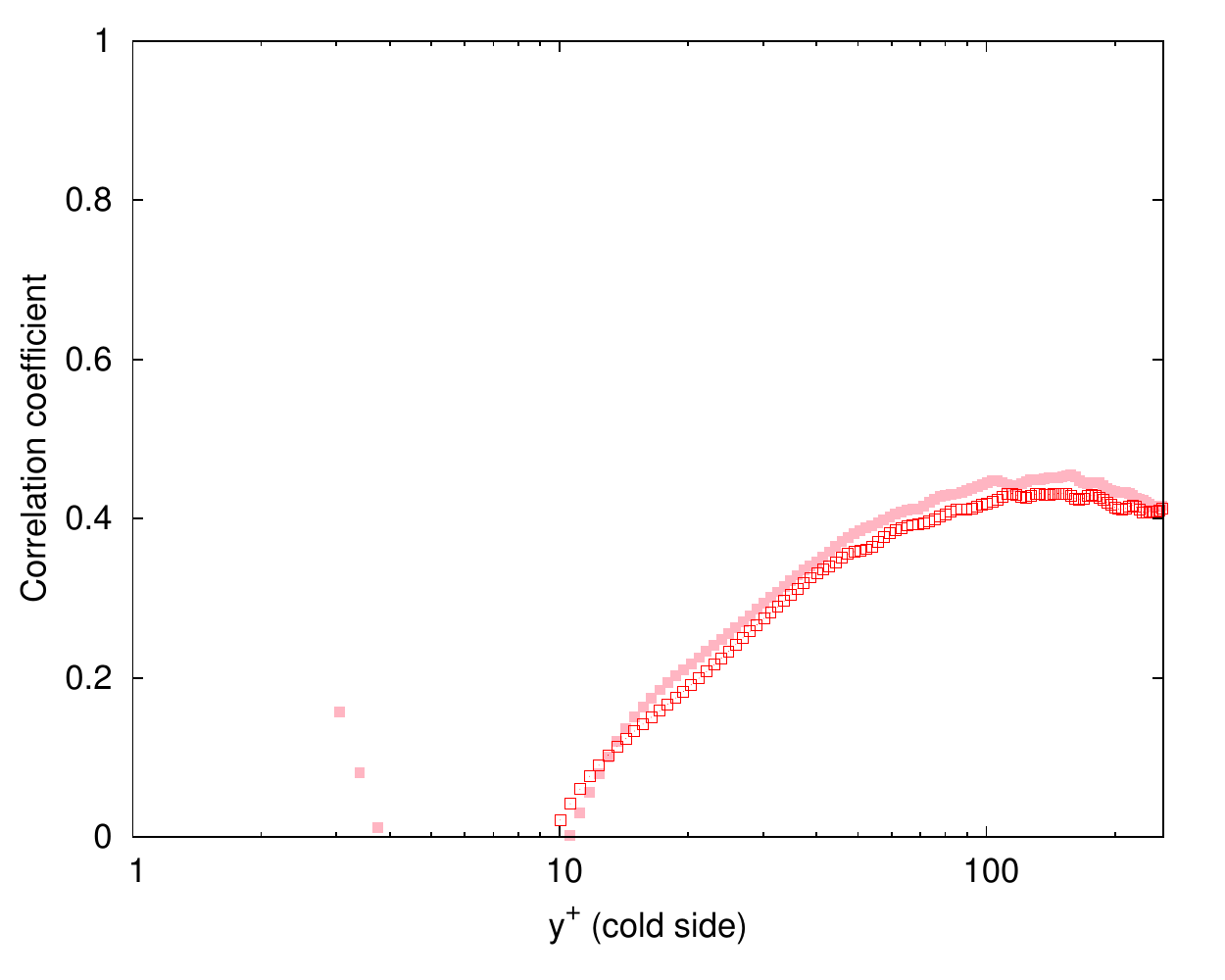}
}
\caption[Correlation coefficient between the divergence of the streamwise-related part of the exact momentum convection subgrid term and the Smagorinsky model in the isothermal and anisothermal configurations.]{
Correlation coefficient between the divergence of the streamwise-related part
of the exact momentum convection subgrid term and the Smagorinsky model,
$\mathrm{Corr}(\partial_j F_{U_j U_x},\partial_j \tau_{xj}^{\mathrm{Smag.}}(\vv{\f{U}}, \vv{\f{\Delta}}))$,
in the isothermal and anisothermal configurations.
\label{sonin}}
\end{figure}

The temperature gradient generates an asymmetry between the hot and cold sides
with regard to the performance of the models.
This is highlighted in figure \ref{sonin} by comparing in the case of the
Smagorinsky model the results with an isothermal simulation performed with the
same mesh, numerical settings, friction Reynolds number and filtering.
The correlation coefficient is larger at the hot side than in the isothermal
configuration, and lower at the cold side.
The asymmetry may be attributed to an asymmetry of filtering resolution
compared to the turbulence intensity.
Indeed, due to the variations of the fluid properties with temperature, the
local friction Reynolds number varies across the channel, from 105 at the hot
wall to 261 at the cold wall, leading to a lower turbulence intensity level at
the hot side than in the isothermal configuration,
and higher at the cold side.

\subsection{Subgrid term associated with momentum convection}

\begin{figure}
\centerline{
\subfigure[Correlation coefficient. \label{divia}]{\includegraphics[width=0.51\textwidth, trim={0 10 10 5}, clip]{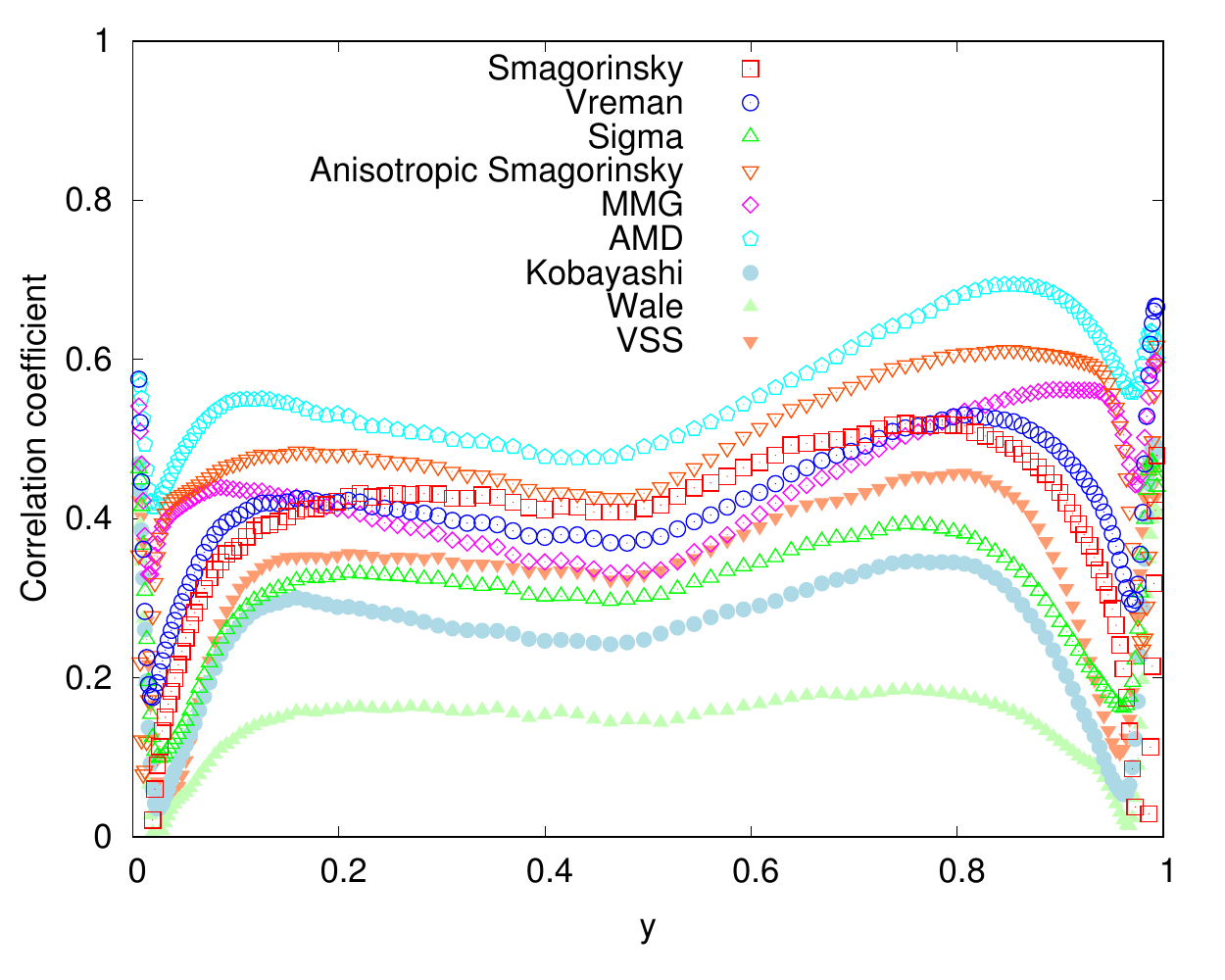}\includegraphics[width=0.51\textwidth, trim={0 10 10 5}, clip]{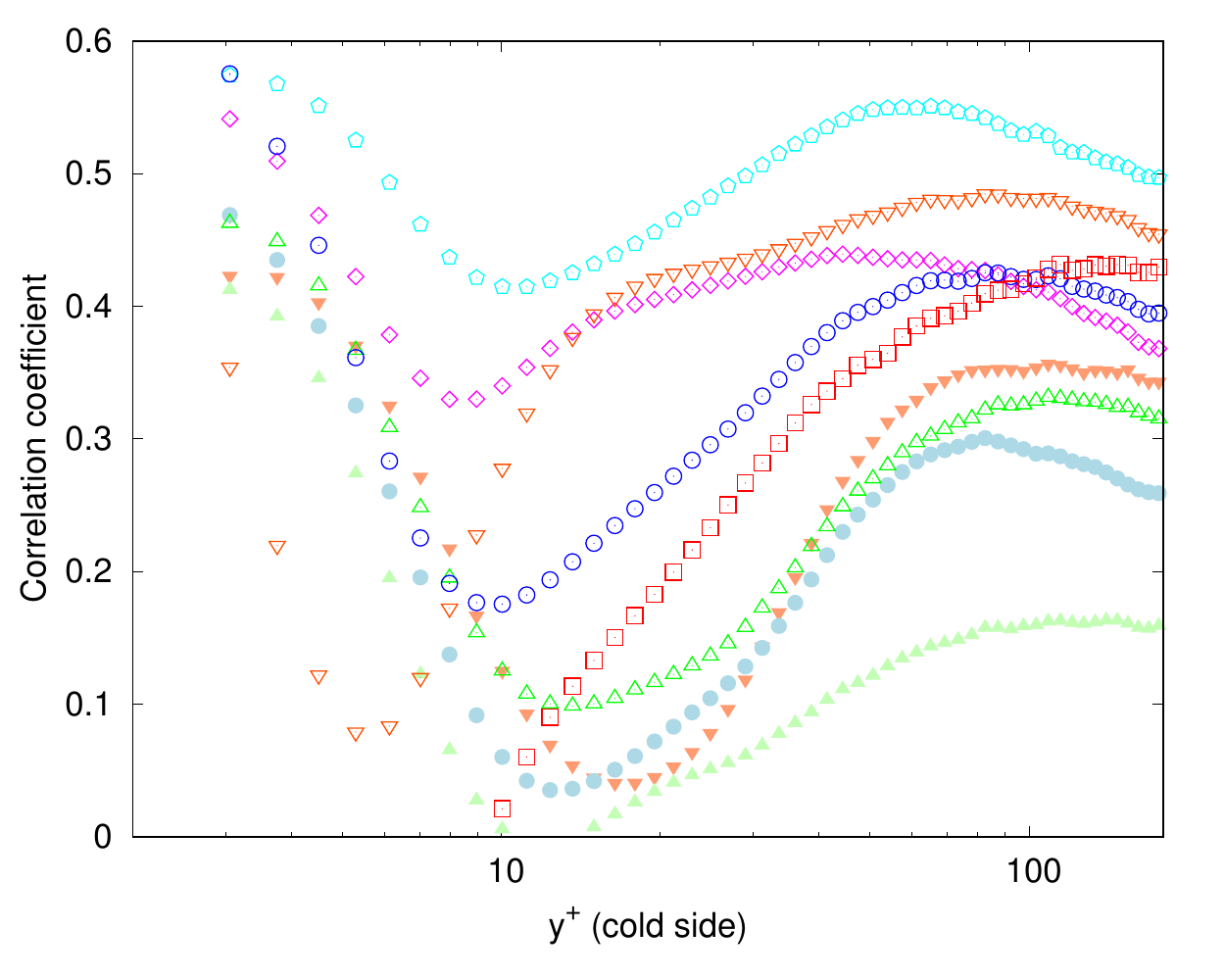}}
}\centerline{
\subfigure[Regression coefficient. \label{divib}]{\includegraphics[width=0.51\textwidth, trim={0 10 10 5}, clip]{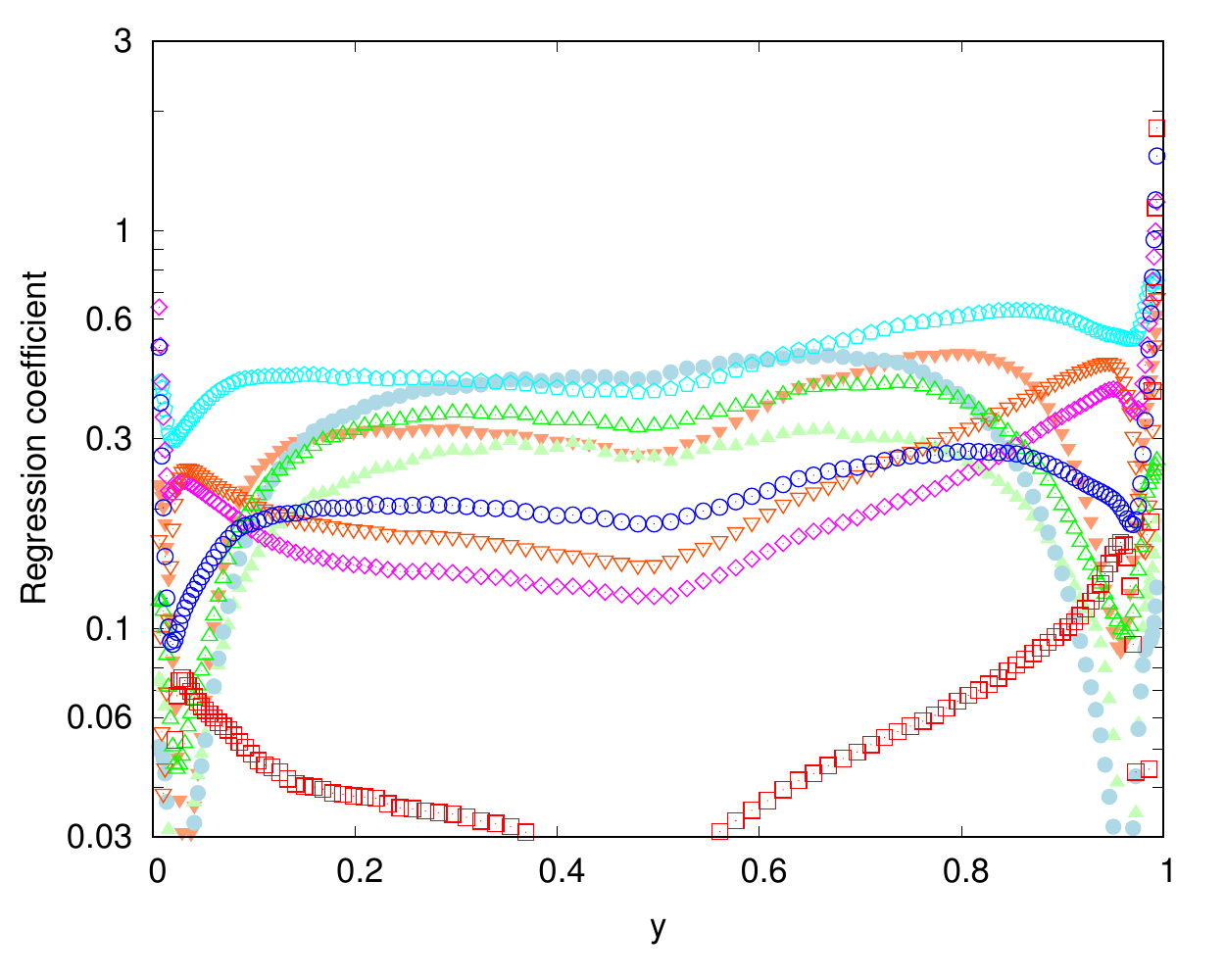}\includegraphics[width=0.51\textwidth, trim={0 10 10 5}, clip]{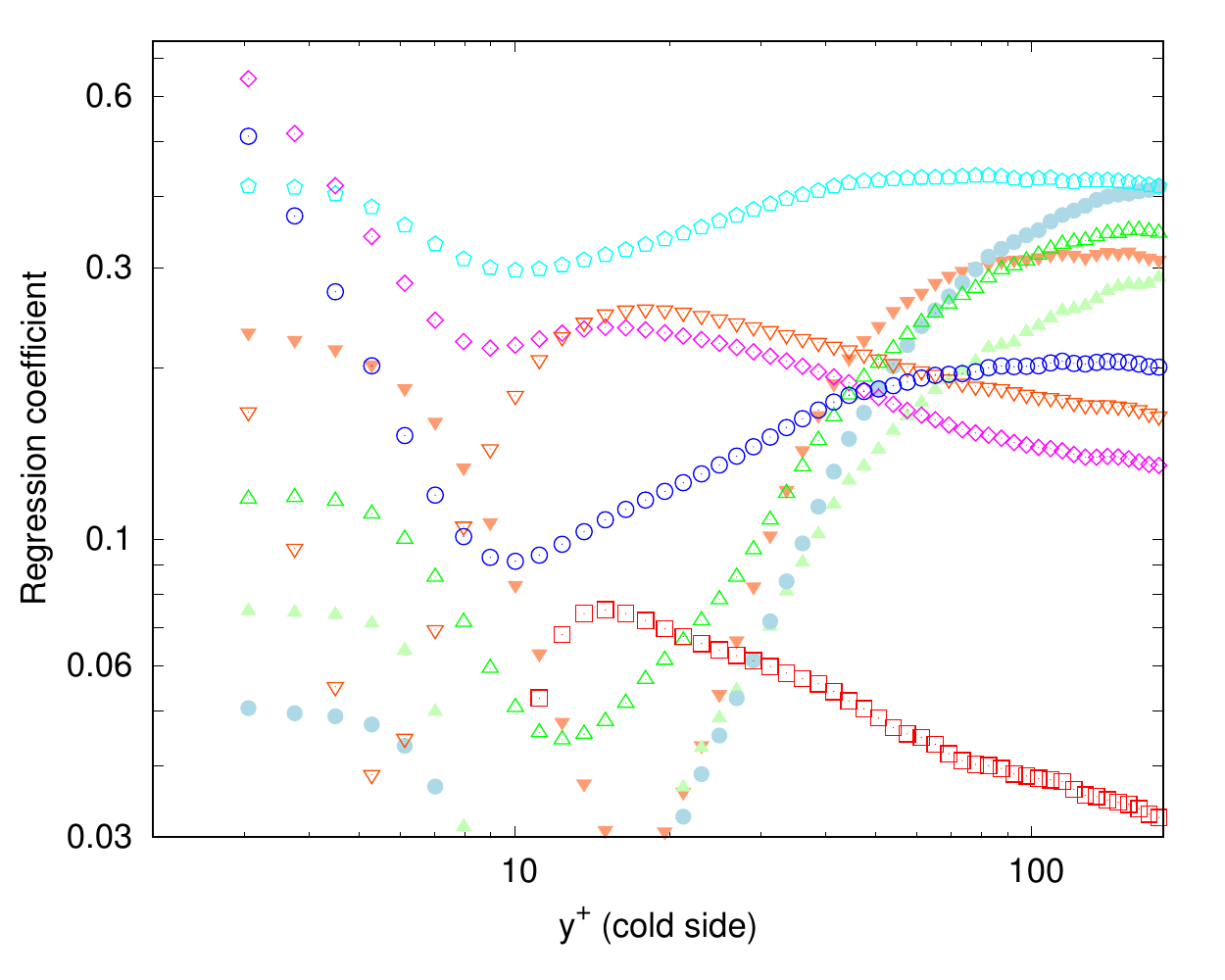}}
}\centerline{
\subfigure[Concordance correlation coefficient. \label{divic}]{\includegraphics[width=0.51\textwidth, trim={0 10 10 5}, clip]{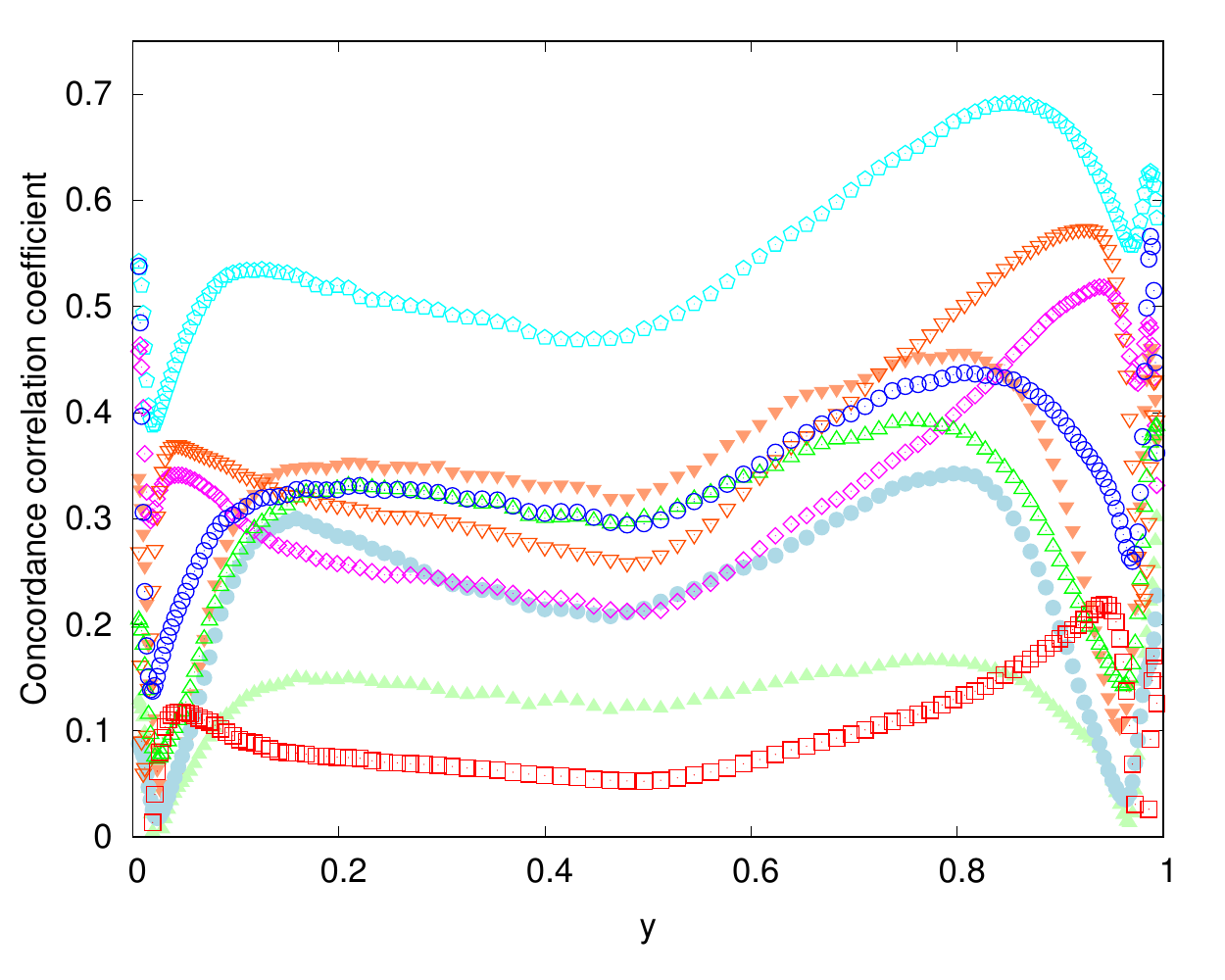}\includegraphics[width=0.51\textwidth, trim={0 10 10 5}, clip]{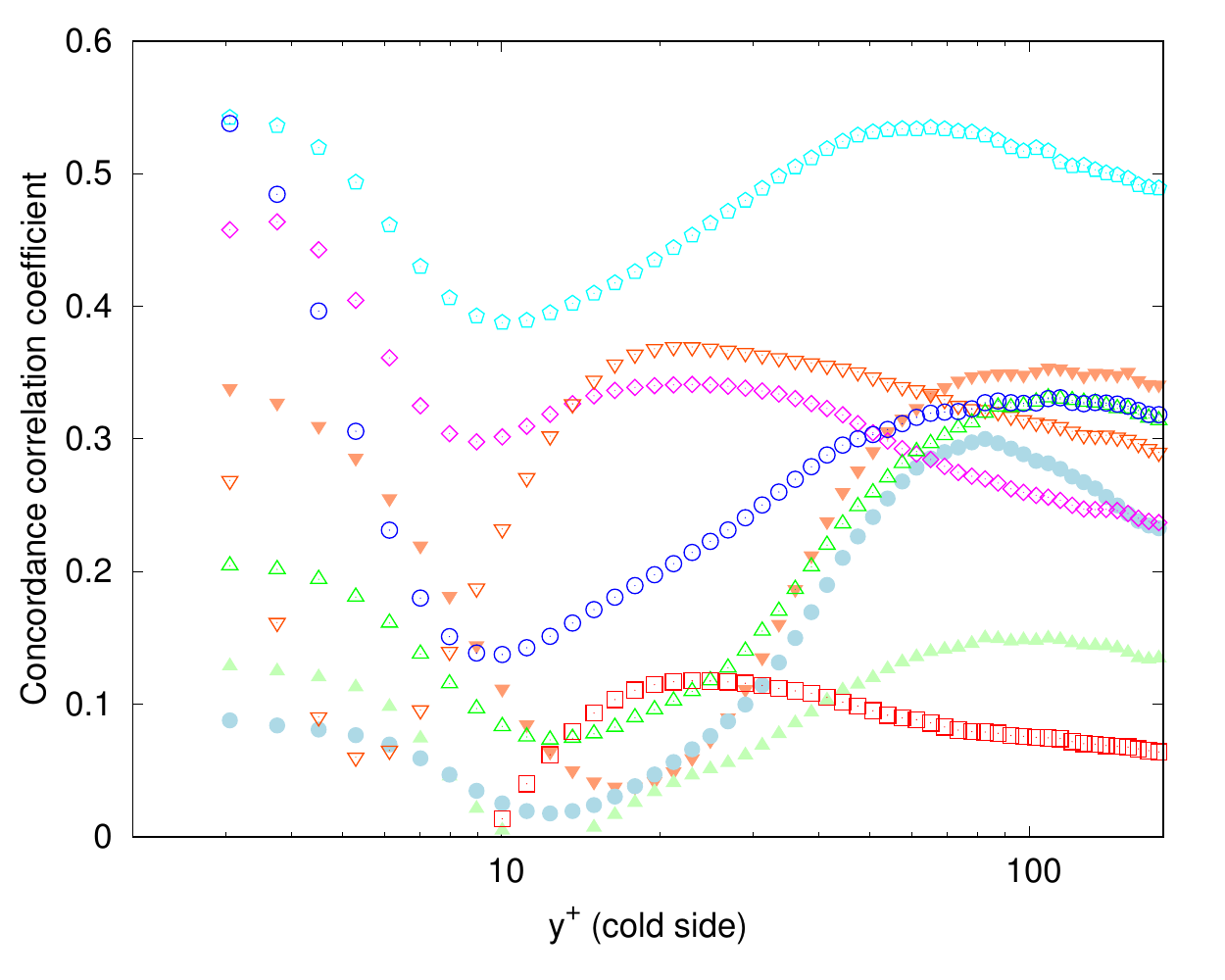}}
}
\caption[Correlation coefficient, regression coefficient, and concordance correlation coefficient between the divergence of the streamwise-related part of the exact momentum convection subgrid term and eddy-viscosity models.]{
Correlation coefficient,
regression coefficient,
and concordance correlation coefficient
between the divergence of the streamwise-related part
of the exact momentum convection subgrid term $\partial_j F_{U_j U_x}$
and eddy-viscosity models $\partial_j \tau_{xj}^{\mathrm{mod}}(\vv{\f{U}}, \vv{\f{\Delta}})$.
\label{divi}}
\end{figure}

\begin{figure}
\centerline{
\subfigure[Correlation coefficient. \label{divja}]{\includegraphics[width=0.51\textwidth, trim={0 10 10 5}, clip]{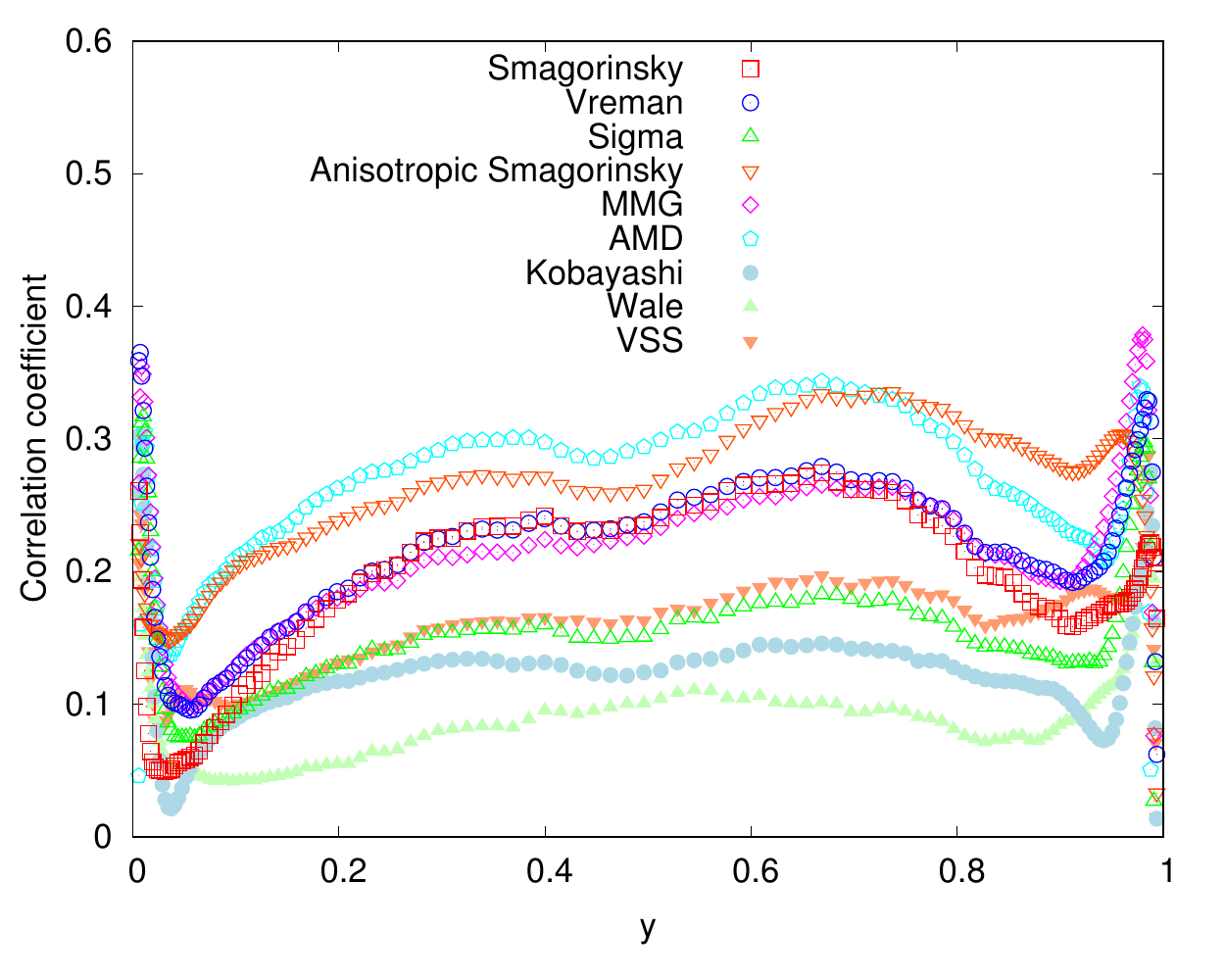}\includegraphics[width=0.51\textwidth, trim={0 10 10 5}, clip]{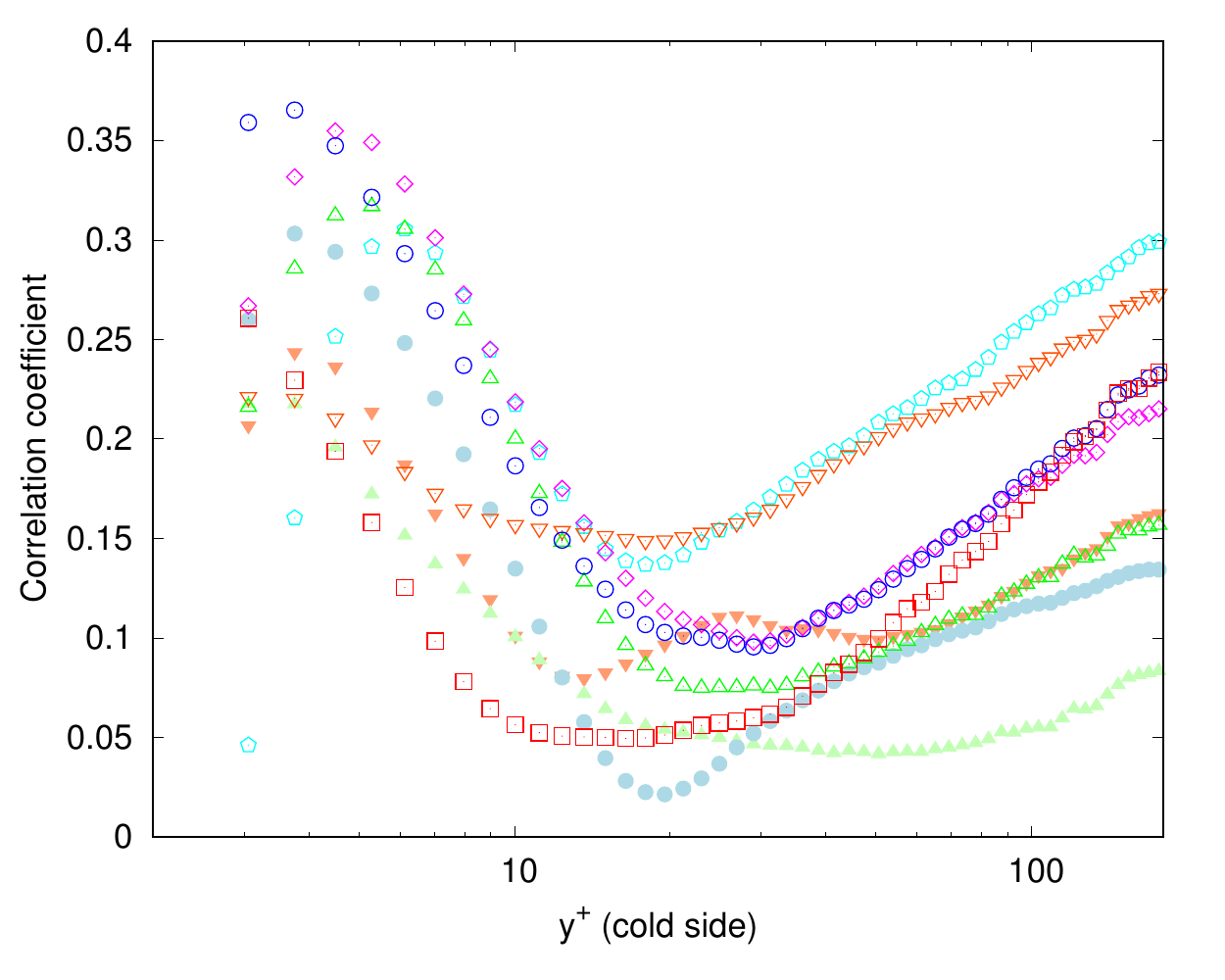}}
}\centerline{
\subfigure[Regression coefficient. \label{divjb}]{\includegraphics[width=0.51\textwidth, trim={0 10 10 5}, clip]{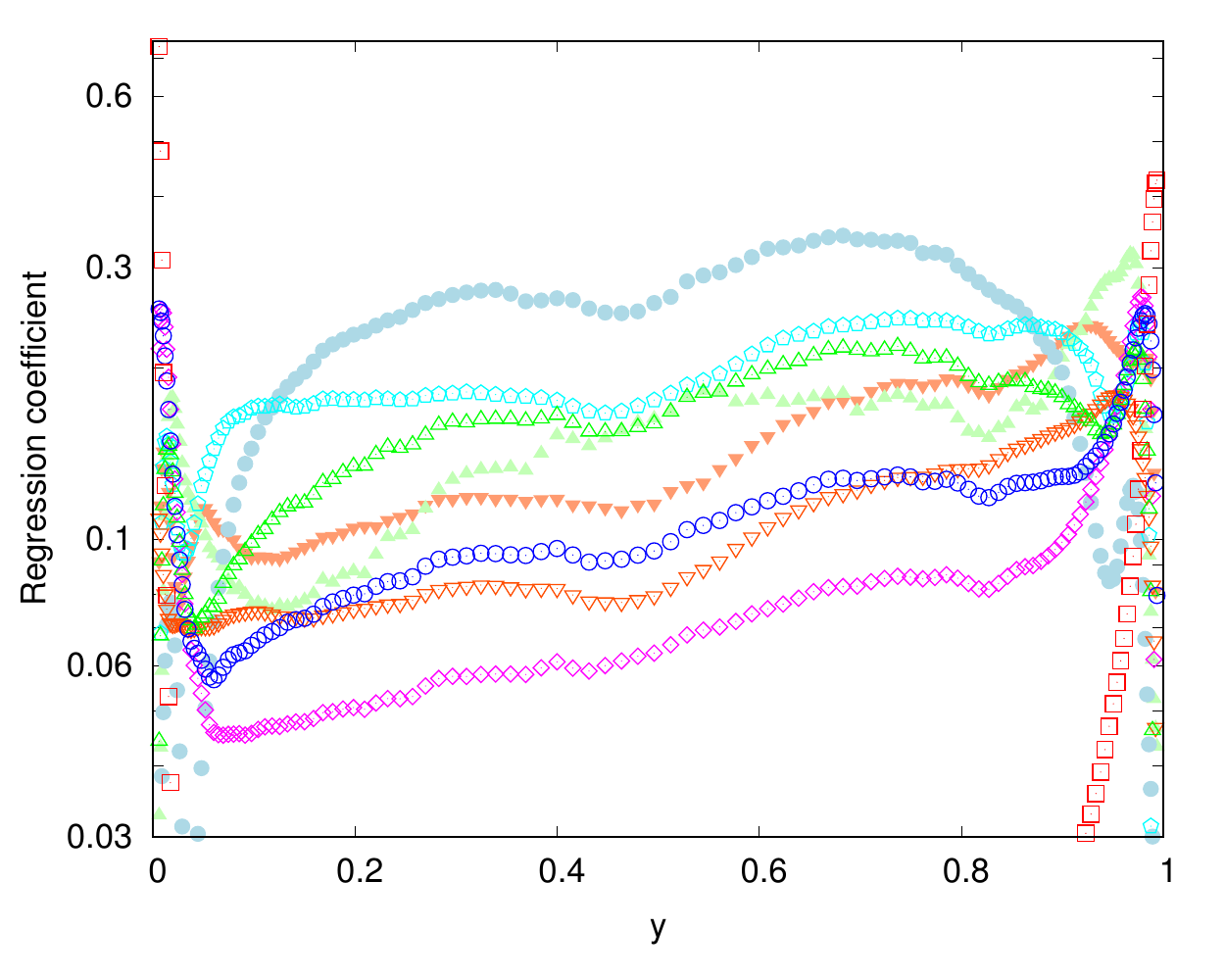}\includegraphics[width=0.51\textwidth, trim={0 10 10 5}, clip]{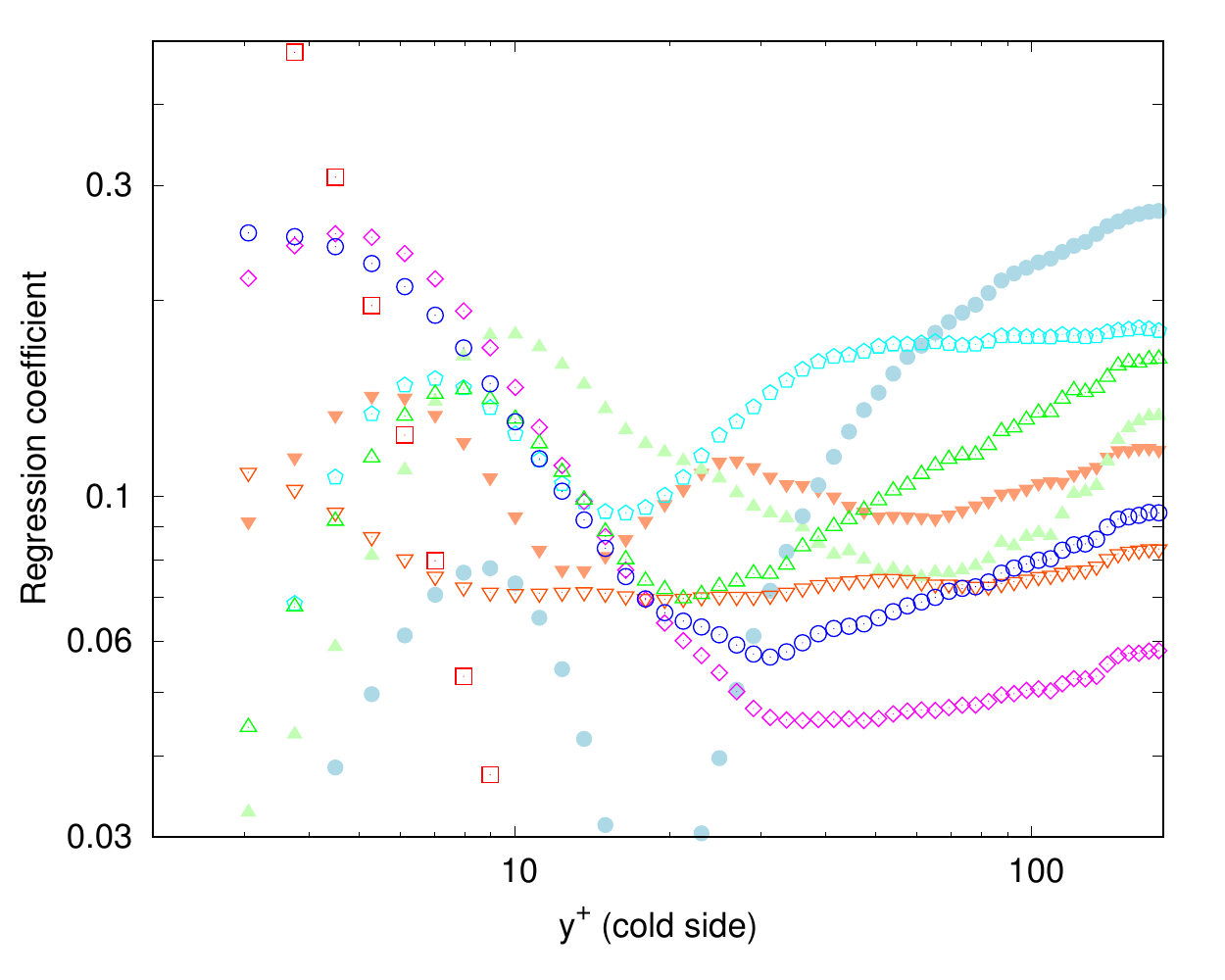}}
}\centerline{
\subfigure[Concordance correlation coefficient. \label{divjc}]{\includegraphics[width=0.51\textwidth, trim={0 10 10 5}, clip]{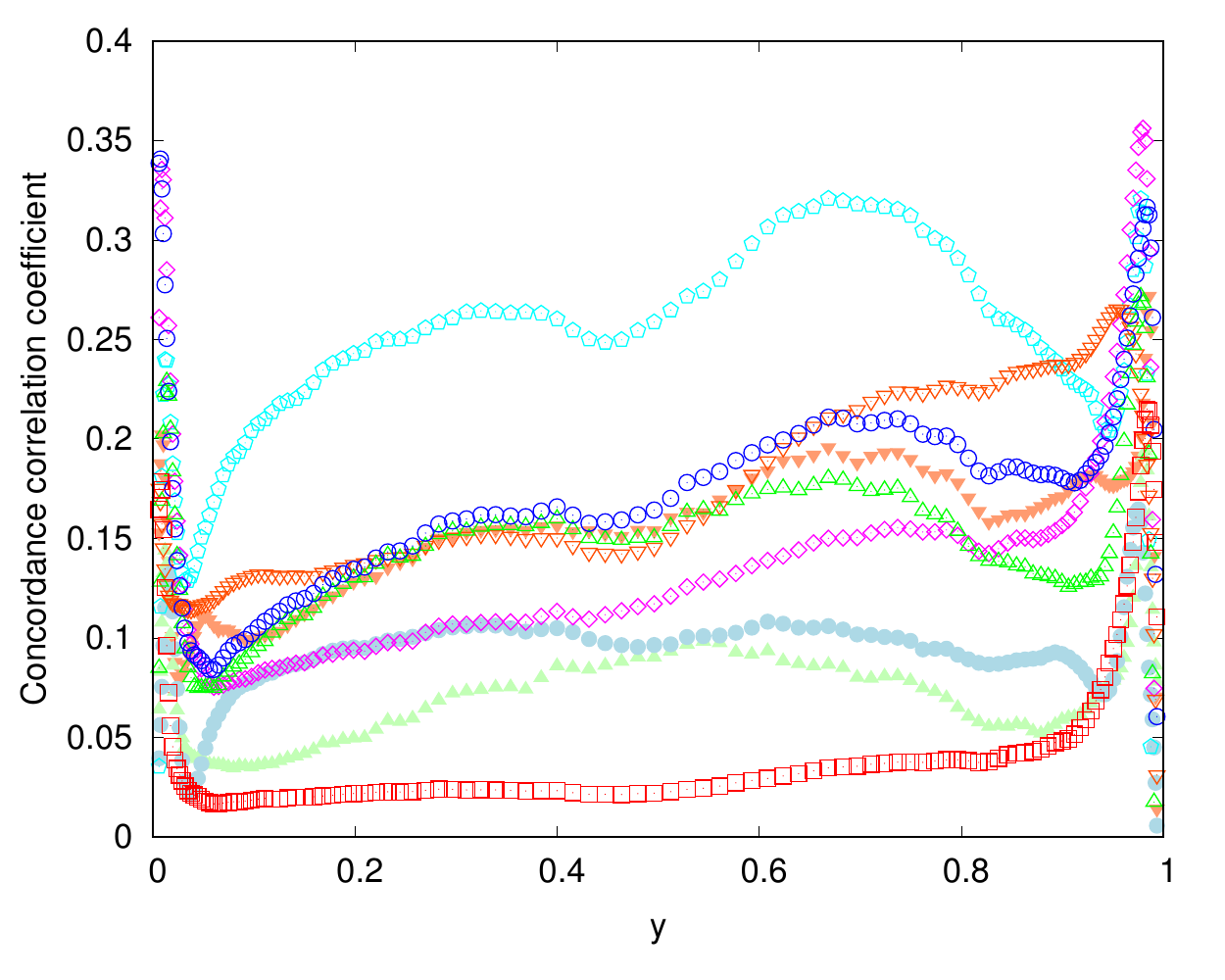}\includegraphics[width=0.51\textwidth, trim={0 10 10 5}, clip]{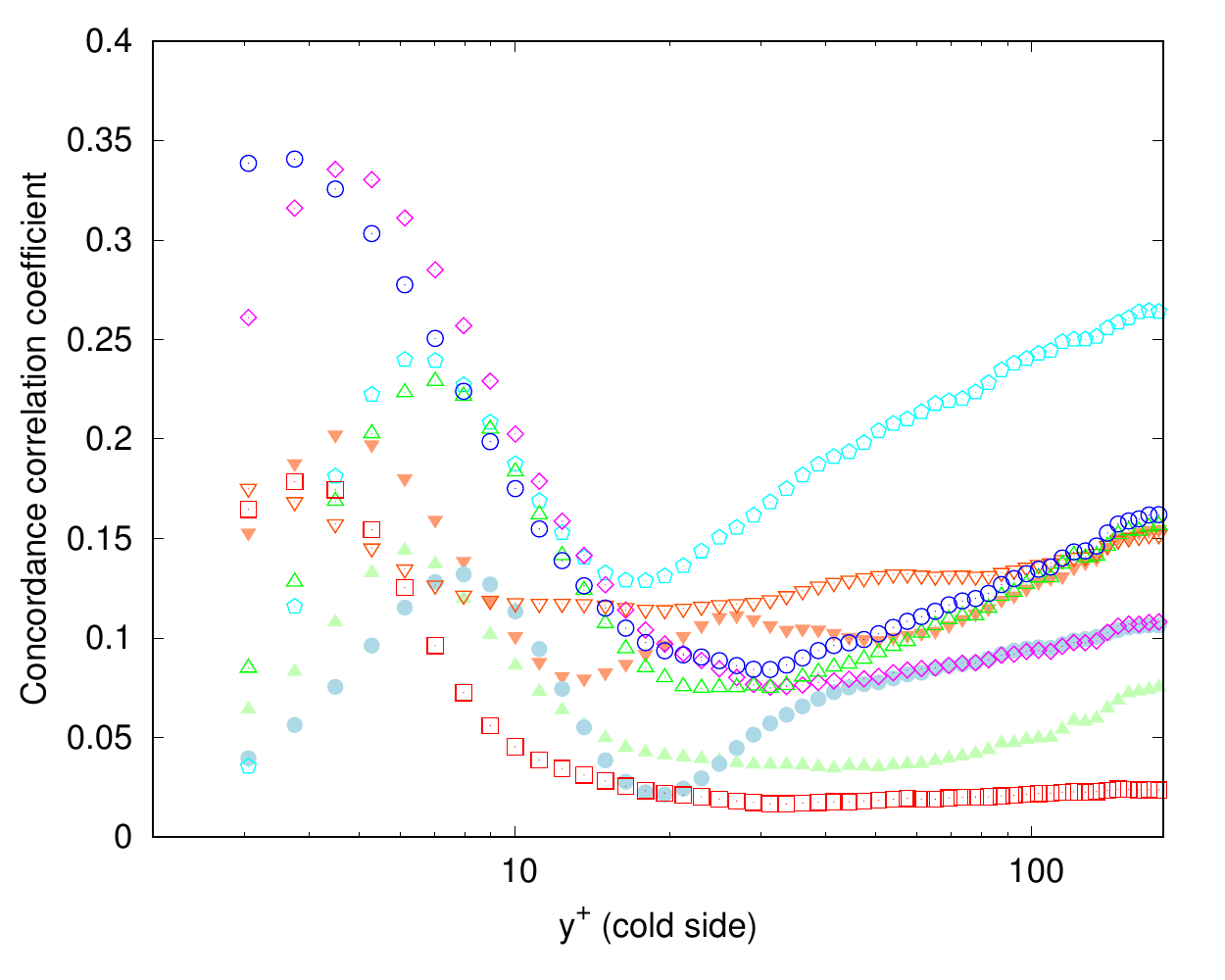}}
}
\caption[Correlation coefficient, regression coefficient, and concordance correlation coefficient between the divergence of the spanwise-related part of the exact momentum convection subgrid term and eddy-viscosity models.]{
Correlation coefficient,
regression coefficient,
and concordance correlation coefficient
between the divergence of the spanwise-related part
of the exact momentum convection subgrid term $\partial_j F_{U_j U_z}$
and eddy-viscosity models $\partial_j \tau_{zj}^{\mathrm{mod}}(\vv{\f{U}}, \vv{\f{\Delta}})$.
\label{divj}}
\end{figure}

\begin{figure}
\centerline{
\subfigure[Correlation coefficient. \label{divka}]{\includegraphics[width=0.51\textwidth, trim={0 10 10 5}, clip]{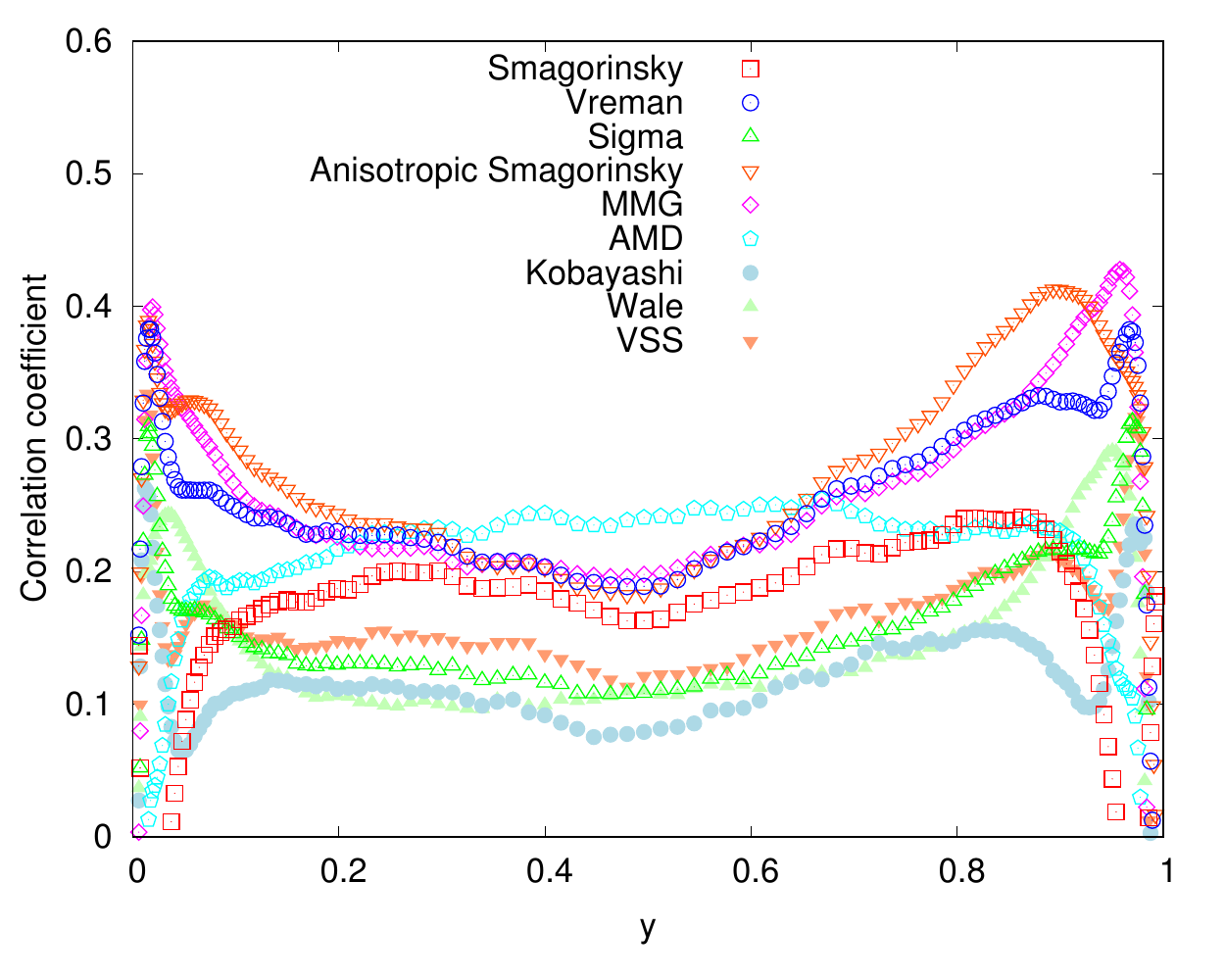}\includegraphics[width=0.51\textwidth, trim={0 10 10 5}, clip]{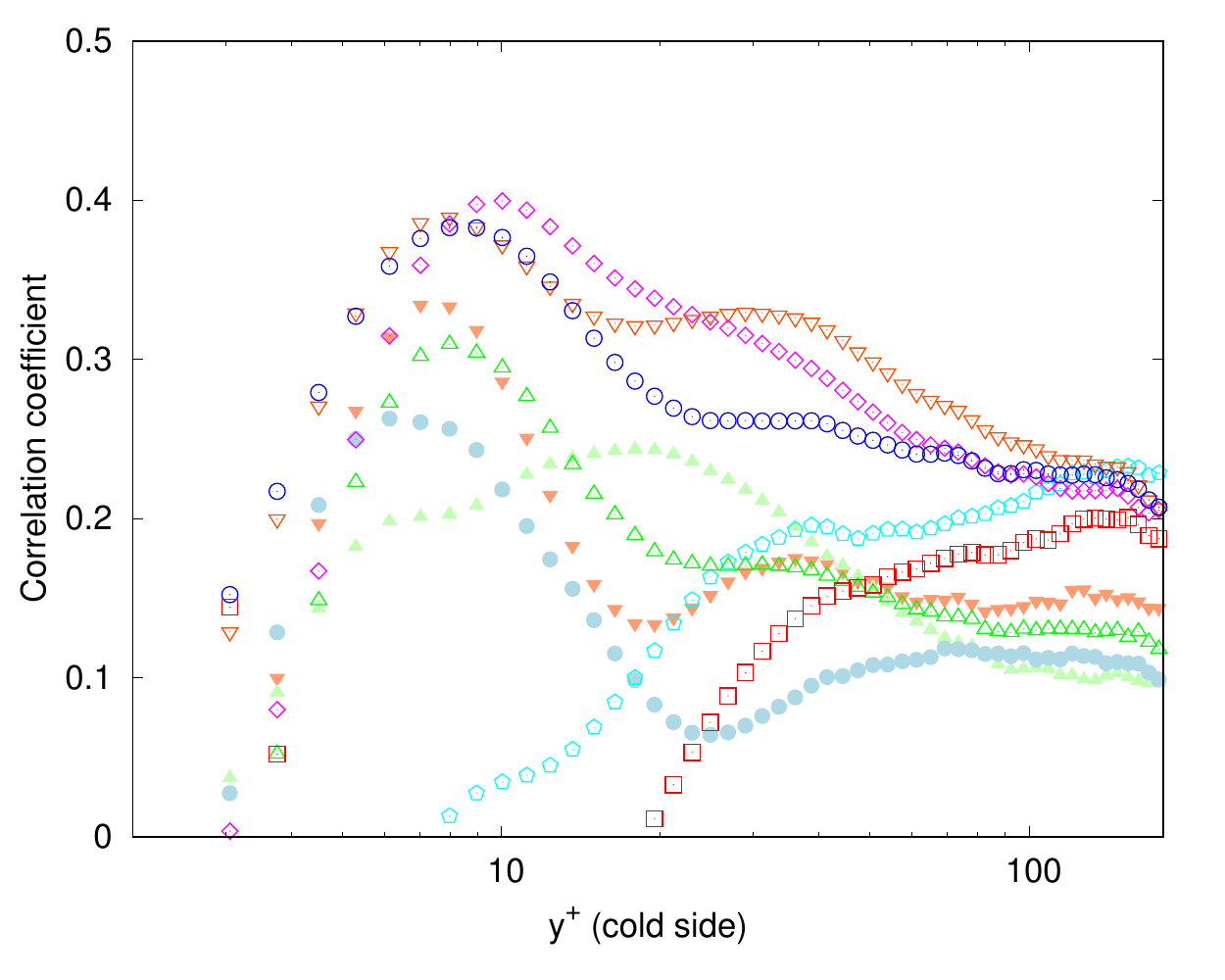}}
}\centerline{
\subfigure[Regression coefficient. \label{divkb}]{\includegraphics[width=0.51\textwidth, trim={0 10 10 5}, clip]{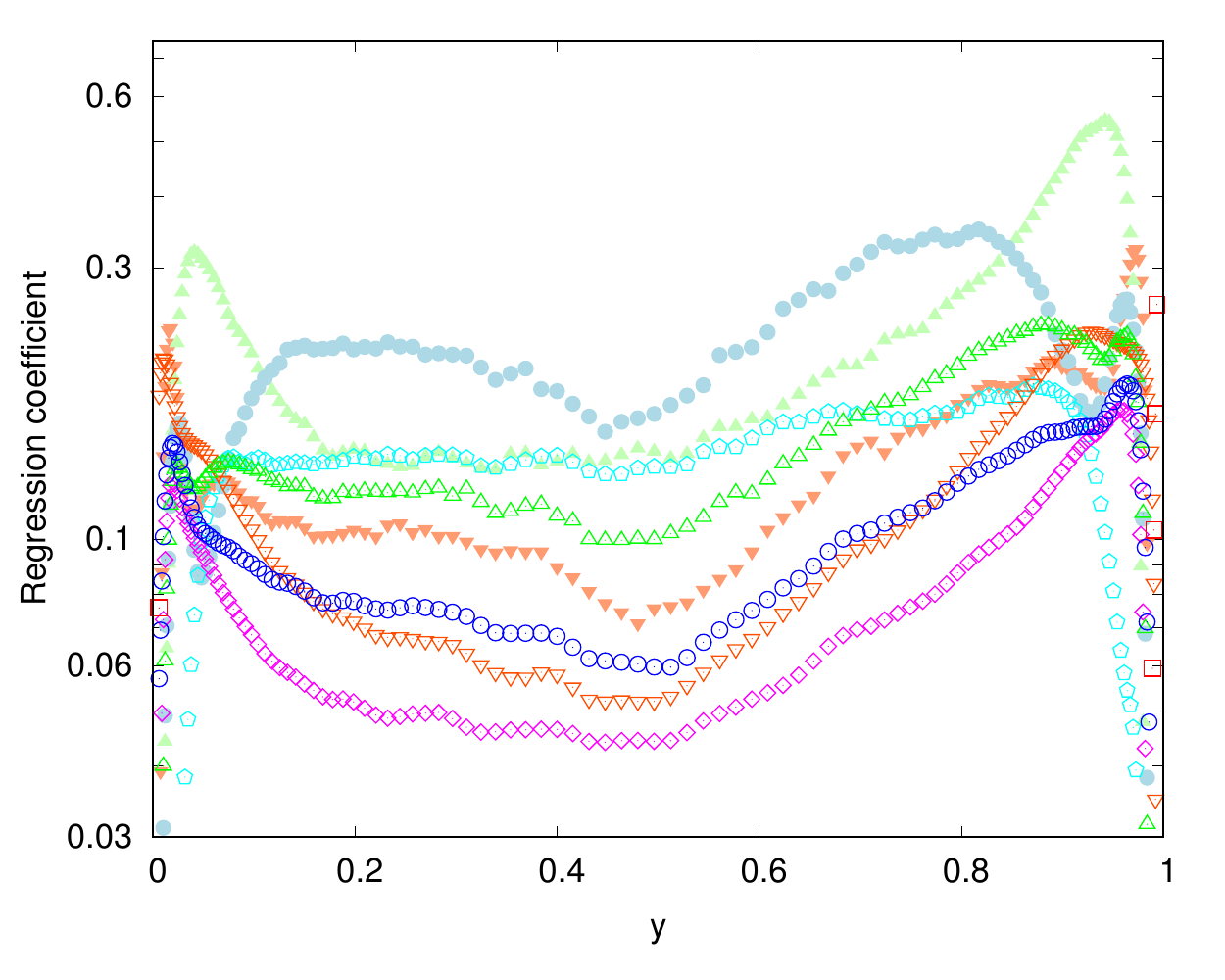}\includegraphics[width=0.51\textwidth, trim={0 10 10 5}, clip]{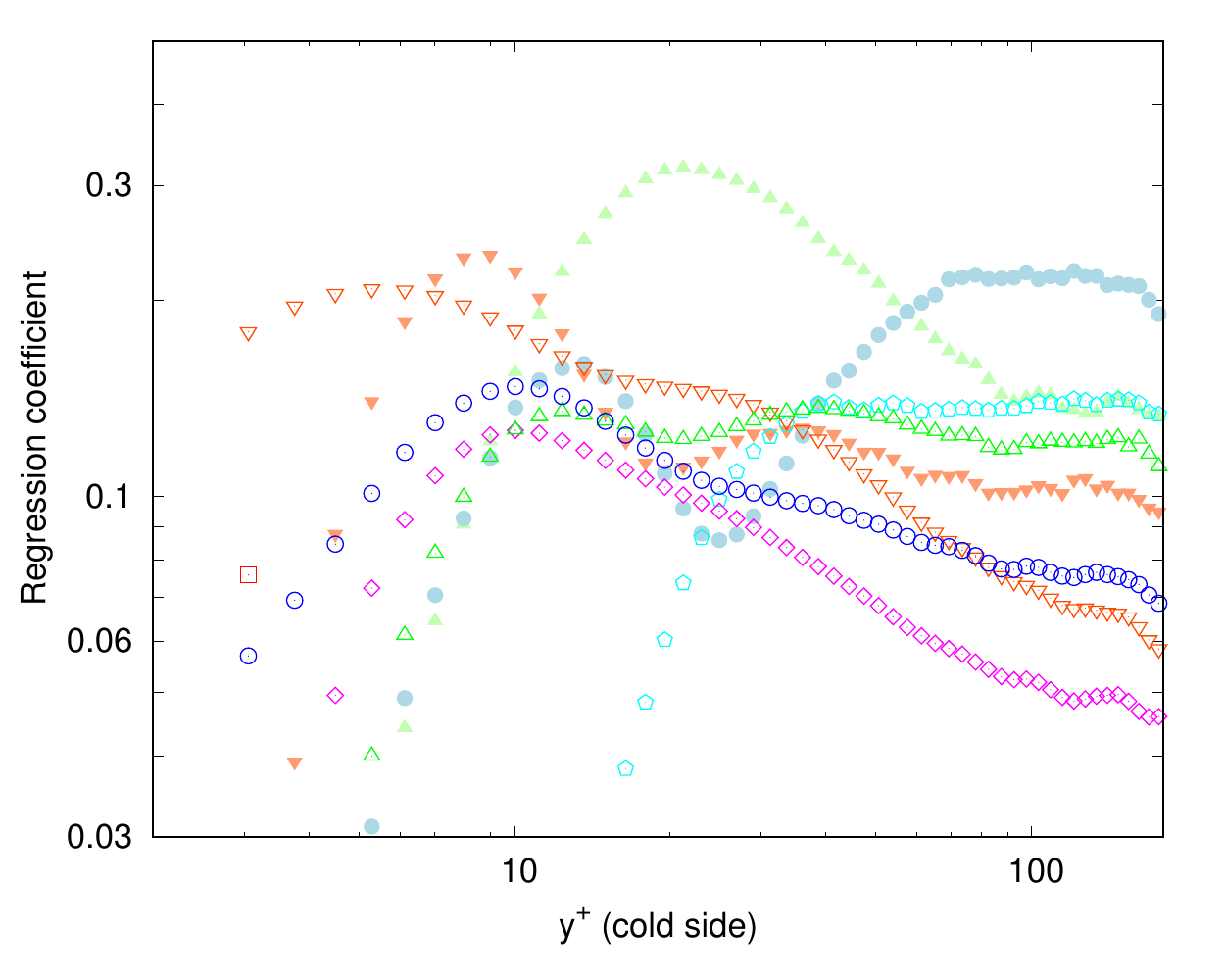}}
}\centerline{
\subfigure[Concordance correlation coefficient. \label{divkc}]{\includegraphics[width=0.51\textwidth, trim={0 10 10 5}, clip]{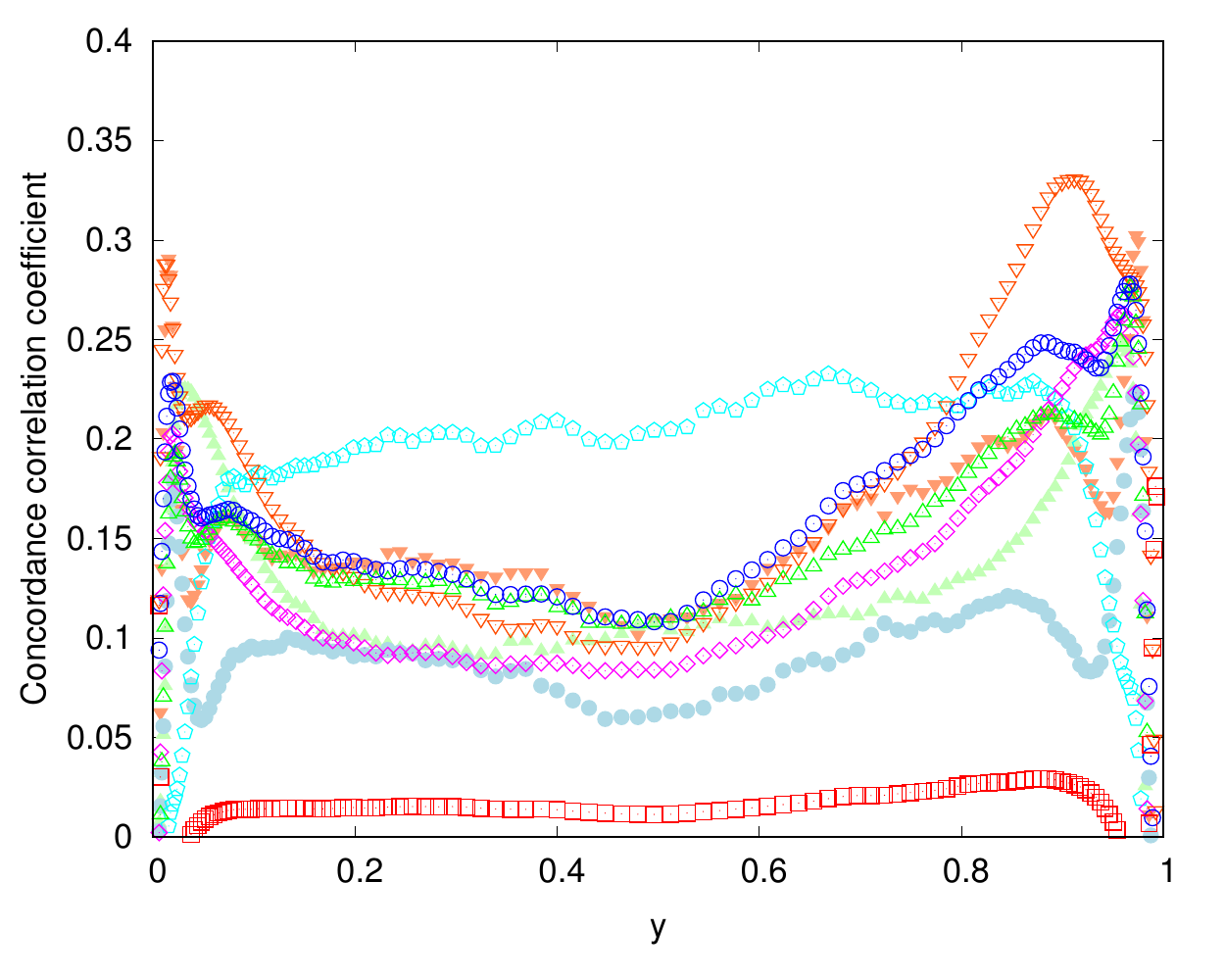}\includegraphics[width=0.51\textwidth, trim={0 10 10 5}, clip]{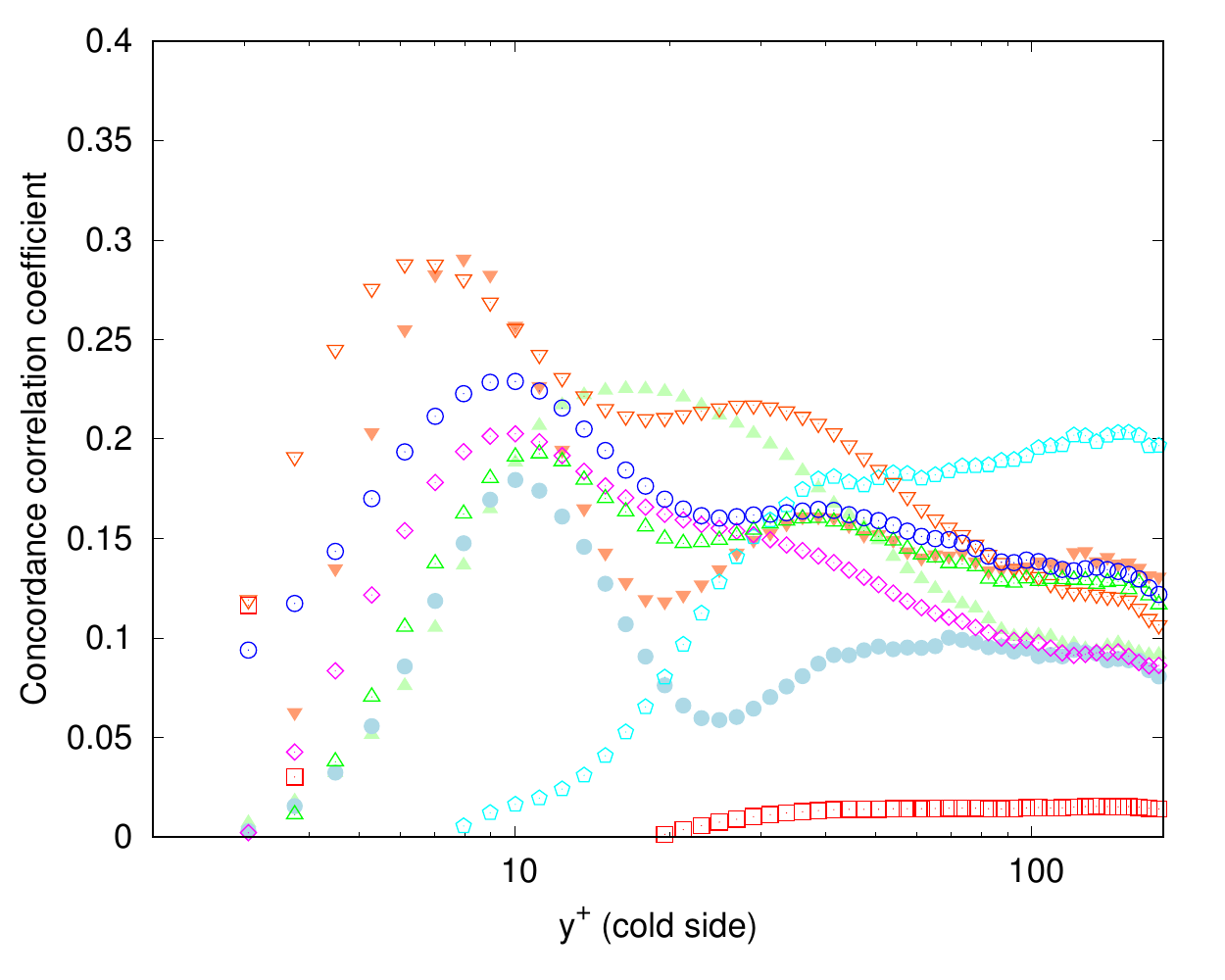}}
}
\caption[Correlation coefficient, regression coefficient, and concordance correlation coefficient between the divergence of the wall-normal-related part of the exact momentum convection subgrid term and eddy-viscosity models.]{
Correlation coefficient,
regression coefficient,
and concordance correlation coefficient
between the divergence of the wall-normal-related part
of the exact momentum convection subgrid term $\partial_j F_{U_j U_y}$
and eddy-viscosity models $\partial_j \tau_{yj}^{\mathrm{mod}}(\vv{\f{U}}, \vv{\f{\Delta}})$.
\label{divk}}
\end{figure}

\begin{figure}
\centerline{
\subfigure[Correlation coefficient. \label{fduidxja}]{\includegraphics[width=0.51\textwidth, trim={0 10 10 5}, clip]{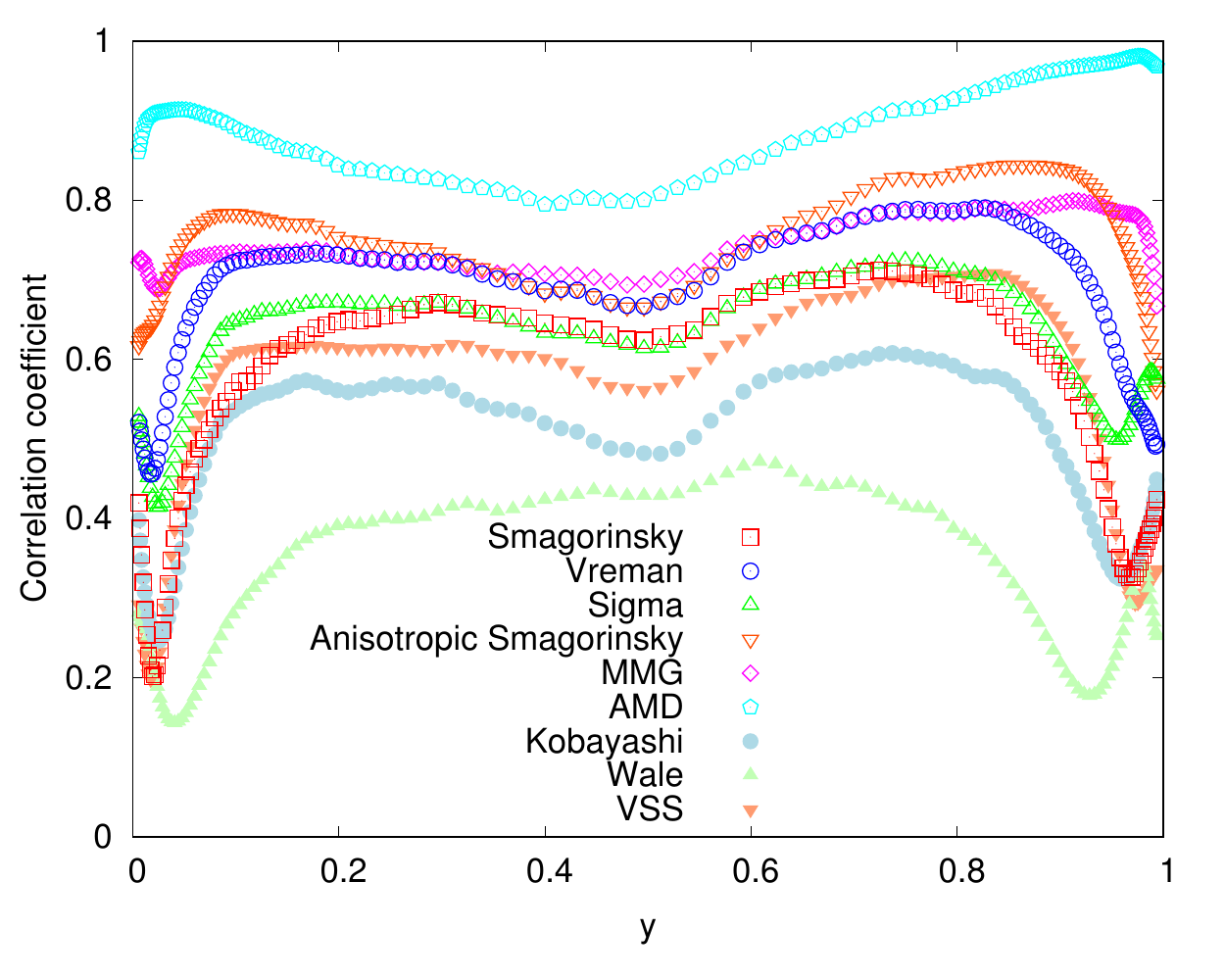}\includegraphics[width=0.51\textwidth, trim={0 10 10 5}, clip]{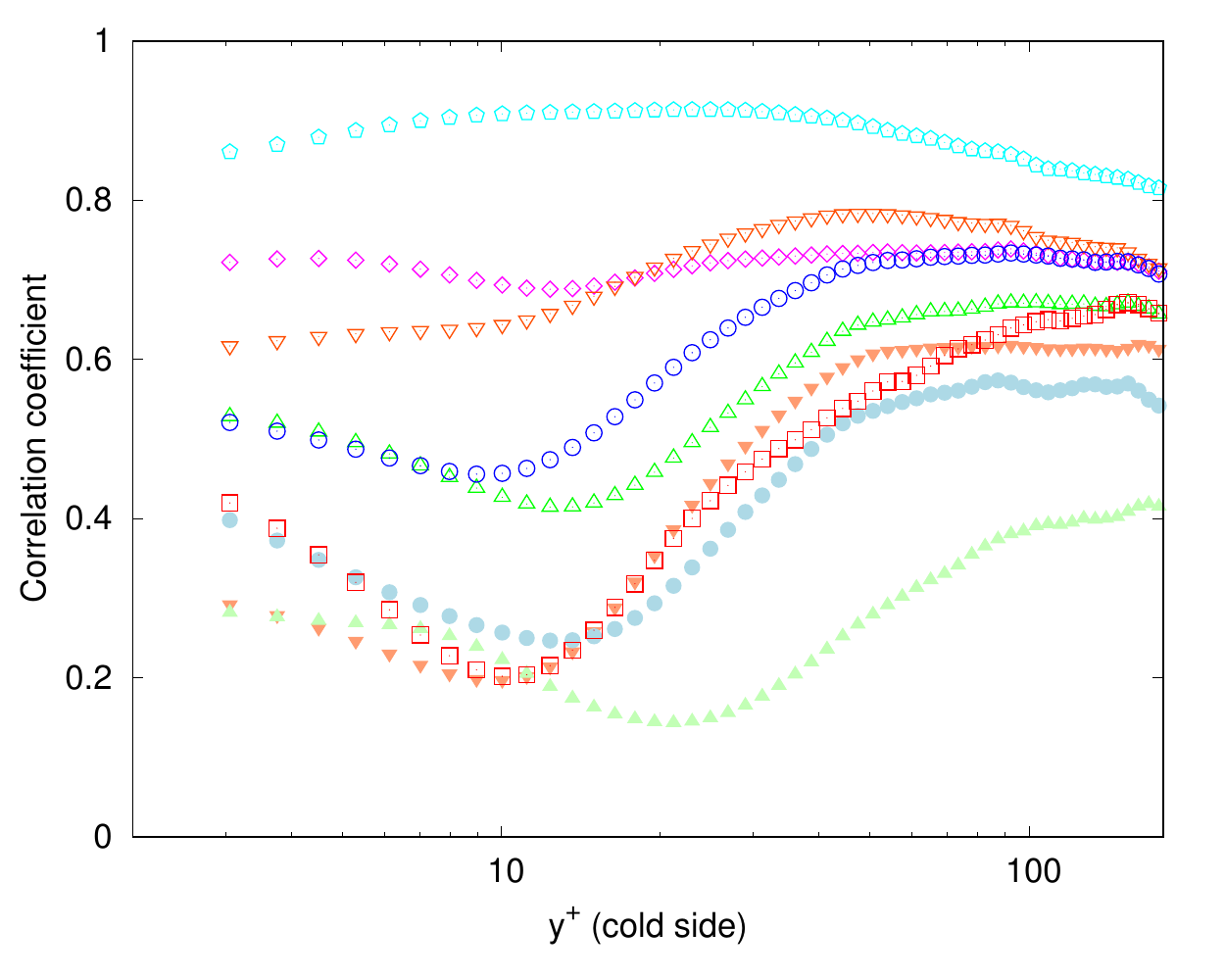}}
}\centerline{
\subfigure[Regression coefficient. \label{fduidxjb}]{\includegraphics[width=0.51\textwidth, trim={0 10 10 5}, clip]{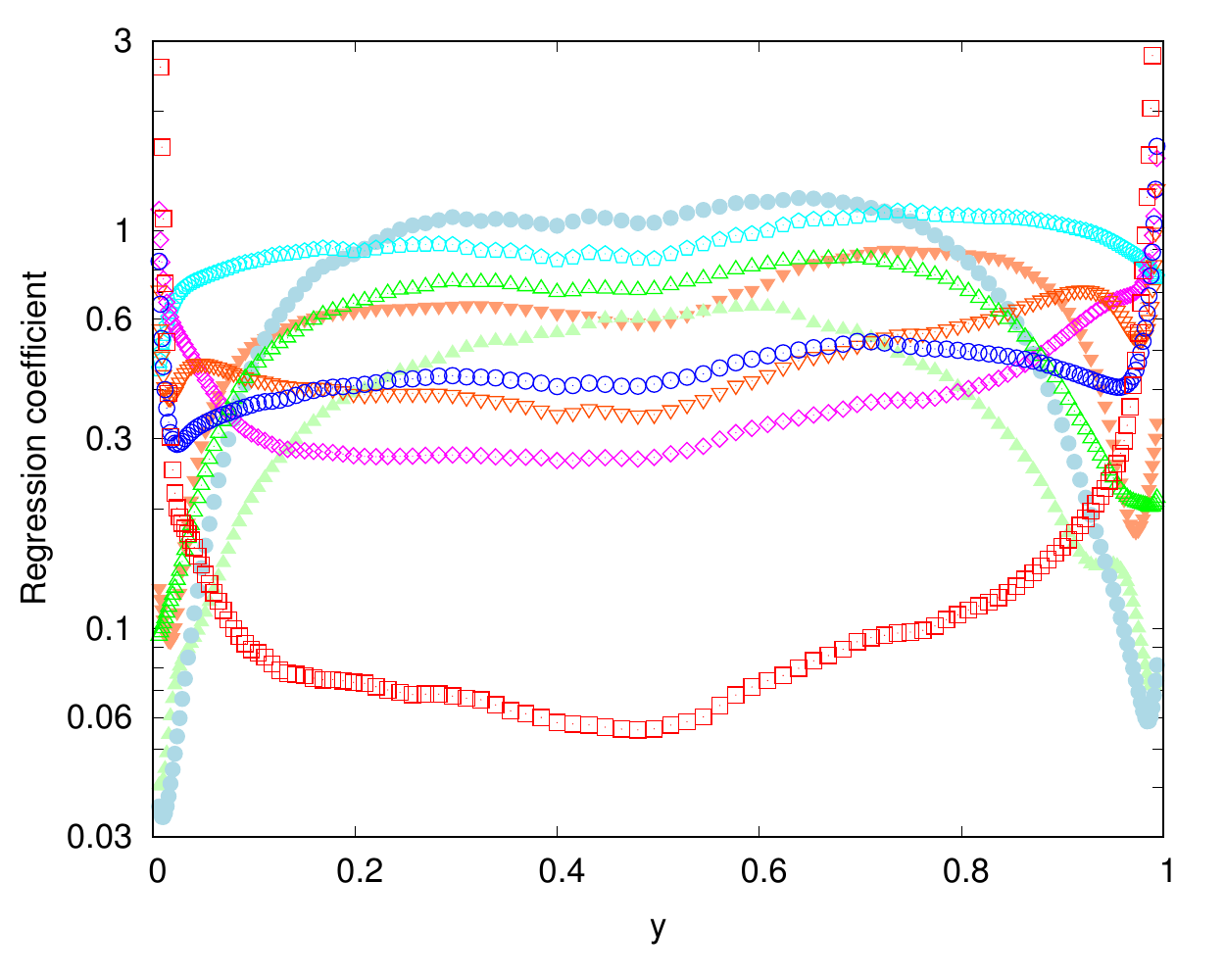}\includegraphics[width=0.51\textwidth, trim={0 10 10 5}, clip]{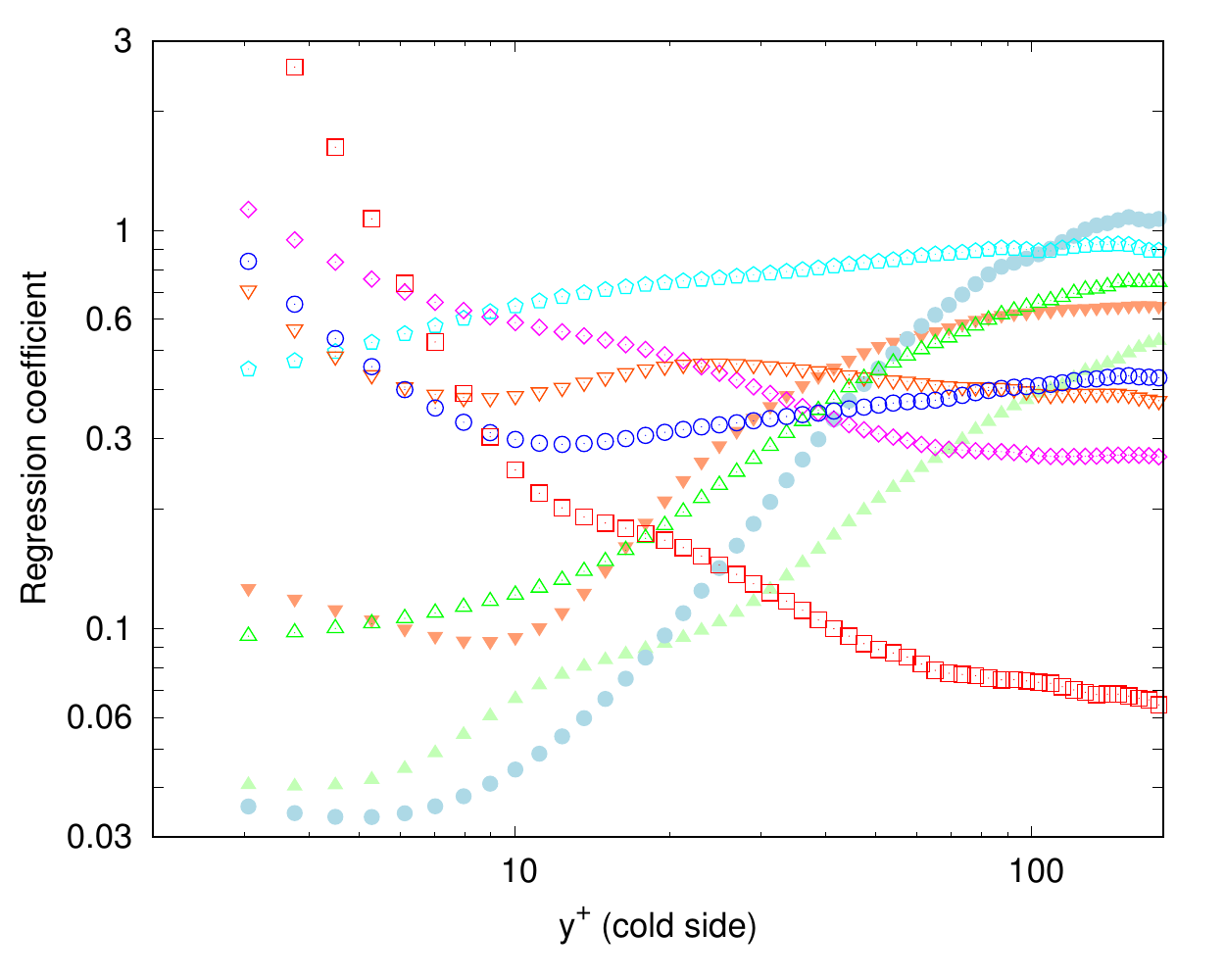}}
}\centerline{
\subfigure[Concordance correlation coefficient. \label{fduidxjc}]{\includegraphics[width=0.51\textwidth, trim={0 10 10 5}, clip]{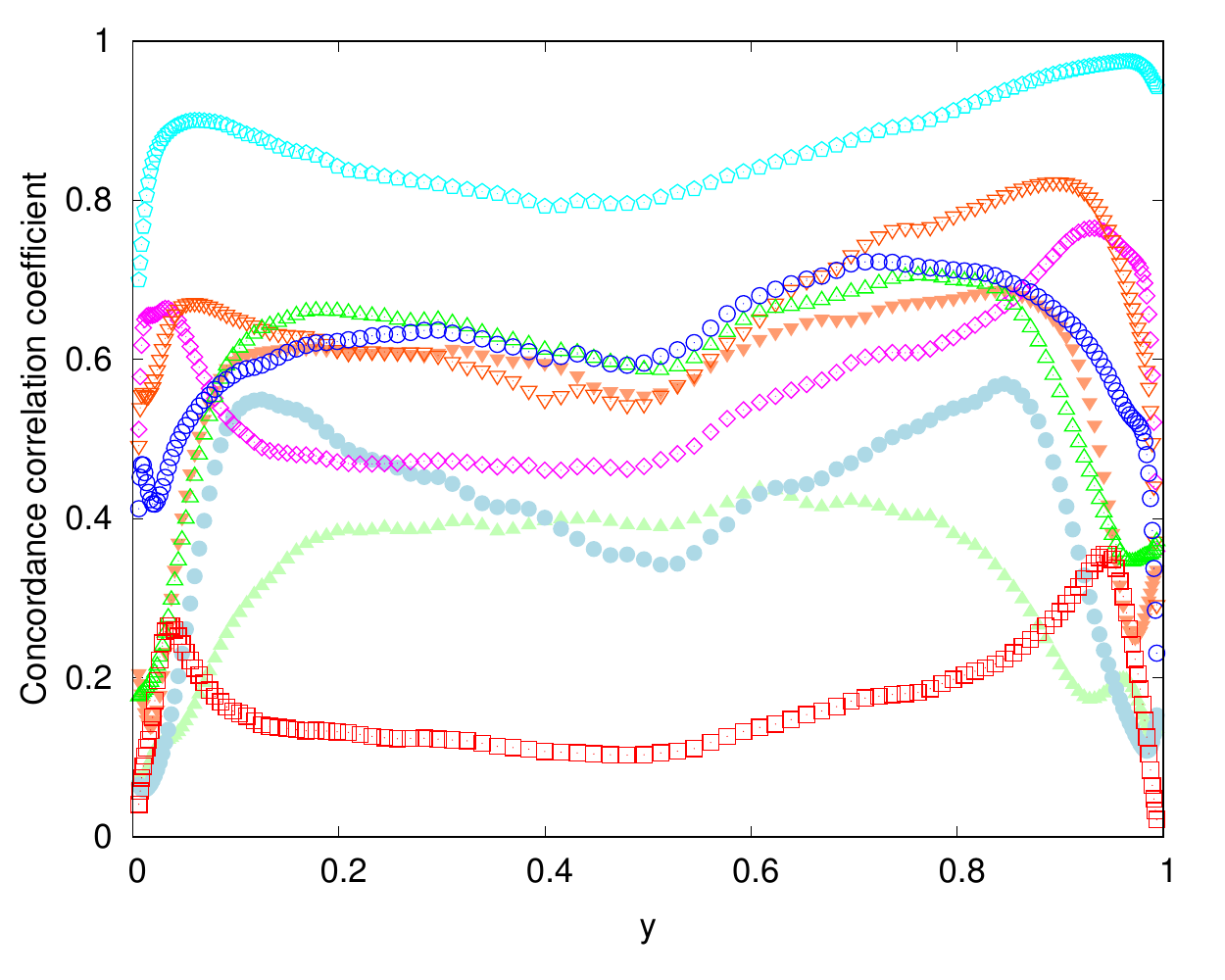}\includegraphics[width=0.51\textwidth, trim={0 10 10 5}, clip]{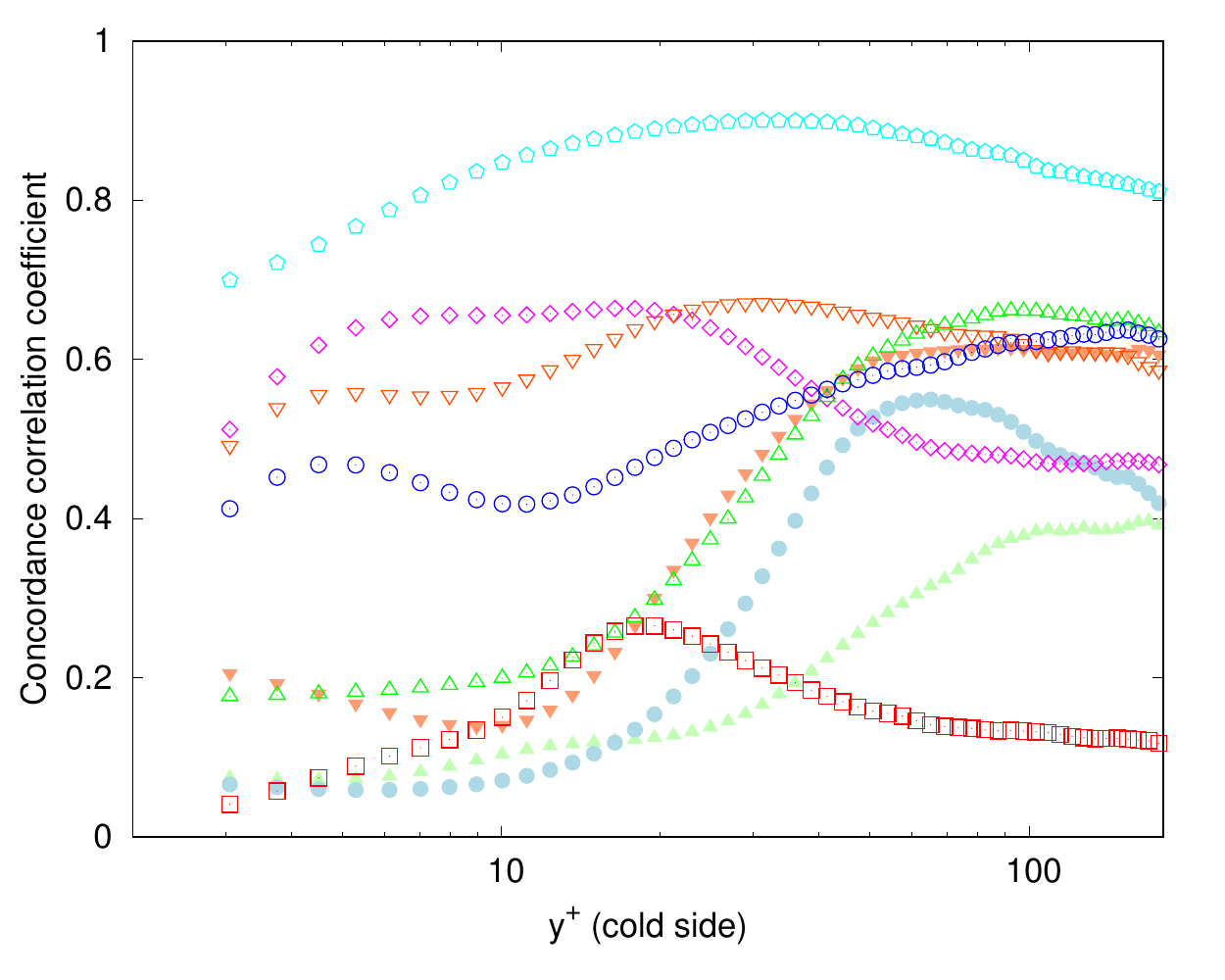}}
}
\caption[Correlation coefficient, regression coefficient, and concordance correlation coefficient between the subgrid kinetic energy dissipation of the exact momentum convection subgrid term and eddy-viscosity models.]{
Correlation coefficient,
regression coefficient,
and concordance correlation coefficient
between the subgrid kinetic energy dissipation
of the exact momentum convection subgrid term $\f{\rho} F_{U_j U_i} S_{ij}$
and eddy-viscosity models $\f{\rho} \tau_{ij}^{\mathrm{mod}}(\vv{\f{U}}, \vv{\f{\Delta}}) S_{ij}$.
\label{fduidxj}}
\end{figure}

The models for the subgrid term associated with momentum convection
are assessed
as it appears in the streamwise velocity transport equation in figure \ref{divi},
in the spanwise velocity transport equation in figure \ref{divj},
and in the wall-normal velocity transport equation in figure \ref{divk}.
The figure \ref{fduidxj} addresses the subgrid kinetic energy dissipation
$\f{\rho} F_{U_j U_i} S_{ij}$,
an important part of the contribution of the subgrid term to the kinetic energy
exchanges.
In each case, the profiles of
the correlation coefficient,
the regression coefficient and
the concordance correlation coefficient
are given
as a function of the wall-normal coordinate $y$,
scaled by the height of the channel,
and in the classical wall scaling
\begin{equation}
y^+ = Re_{\tau} \frac{y}{h} = \frac{y U_{\tau}}{\nu_{\omega}}.
\end{equation}
As a basis of comparison, each model is scaled in order to match 
the correct level of total subgrid kinetic energy dissipation in the volume.
This is equivalent to setting the constant of the models to
\begin{equation}
C^{\mathrm{mod}} = \frac{\int_T \int_V \f{\rho} F_{U_j U_i} S_{ij} \mathop{dx} \mathop{dy} \mathop{dz} \mathop{dt}}{\int_T \int_V \f{\rho} \tau_{ij}^{\mathrm{mod}}(\vv{U}, \vv{\f{\Delta}}) S_{ij} \mathop{dx} \mathop{dy} \mathop{dz} \mathop{dt}},
\end{equation}
where $V$ denotes the entire domain and $T$ the integration time.

All the investigated subgrid-scale models correlates rather poorly with
the exact subgrid term as it occurs in the velocity transport equation (figures \ref{divia}, \ref{divja}, \ref{divka}).
This is consistent with previous findings which showed that the exact subgrid term
correlates poorly with the rate of deformation tensor
\citep{lu2007priori, clark1979evaluation, liu1994properties},
and reflects the limits of the eddy-viscosity assumption.
The models are however better-correlated with
the exact subgrid term for the subgrid kinetic energy dissipation (figure \ref{fduidxja}),
with correlation coefficients
higher than $0.7$--$0.8$ throughout the channel for the best models.
Accordingly, the regression coefficient at the centre of the channel appears
too low for all models in the three components of the velocity transport equation (figures \ref{divib}, \ref{divjb}, \ref{divkb}),
but around an adequate level for the subgrid kinetic energy dissipation (figure \ref{fduidxjb}).
This discrepancy is related to the intrinsic nature of the models and may not be easily corrected
as increasing the magnitude level of the models to a sufficient amplitude in the
velocity transport equation would make the models overdissipative in the
kinetic energy transport equation.

The AMD model is significantly more well-correlated with the
exact subgrid term than the other investigated models (figures \ref{divia}, \ref{divja}, \ref{divka}, \ref{fduidxja}).
The Vreman, Anisotropic Smagorinsky and MMG models also have a high level of
correlation throughout the channel.
In the streamwise velocity transport equation (figure \ref{divia}), the WALE
model has a very low correlation coefficient ($<0.2$) in the bulk of the
channel but gives better results at the wall. In the kinetic energy transport
equation (figure \ref{fduidxja}), it is the opposite.
To a lesser extent, the Sigma, VSS and Kobayashi models obey to the same
pattern.

Near the wall, the correlation of the Smagorinsky model deteriorates and its
amplitude increases dramatically because the differential operator it is based
on does not vanish in near-wall regions, which conflicts with the near-wall
behaviour of the exact subgrid term.
The Anisotropic Smagorinsky model is able to improve greatly the
near-wall behaviour of the Smagorinsky model, the filter lengths in the
Anisotropic Smagorinsky model acting akin to a damping function.
The Vreman, Anisotropic Smagorinsky and MMG models vanish at the wall but
with a lower order than the exact subgrid term (table \ref{toder}).
Their magnitude compared to the exact subgrid term is increased near the wall.
Nevertheless,
their regression coefficient
is subject to less variations across throughout the channel than
the WALE, Sigma, VSS and Kobayashi models (figures \ref{divib}, \ref{divjb}, \ref{divkb}, \ref{fduidxjb}),
up to the first point of the LES mesh that the filter represents.

\afterpage{\clearpage}

\begin{figure}[t!]
\centerline{
\includegraphics[width=0.51\textwidth, trim={0 10 10 5}, clip]{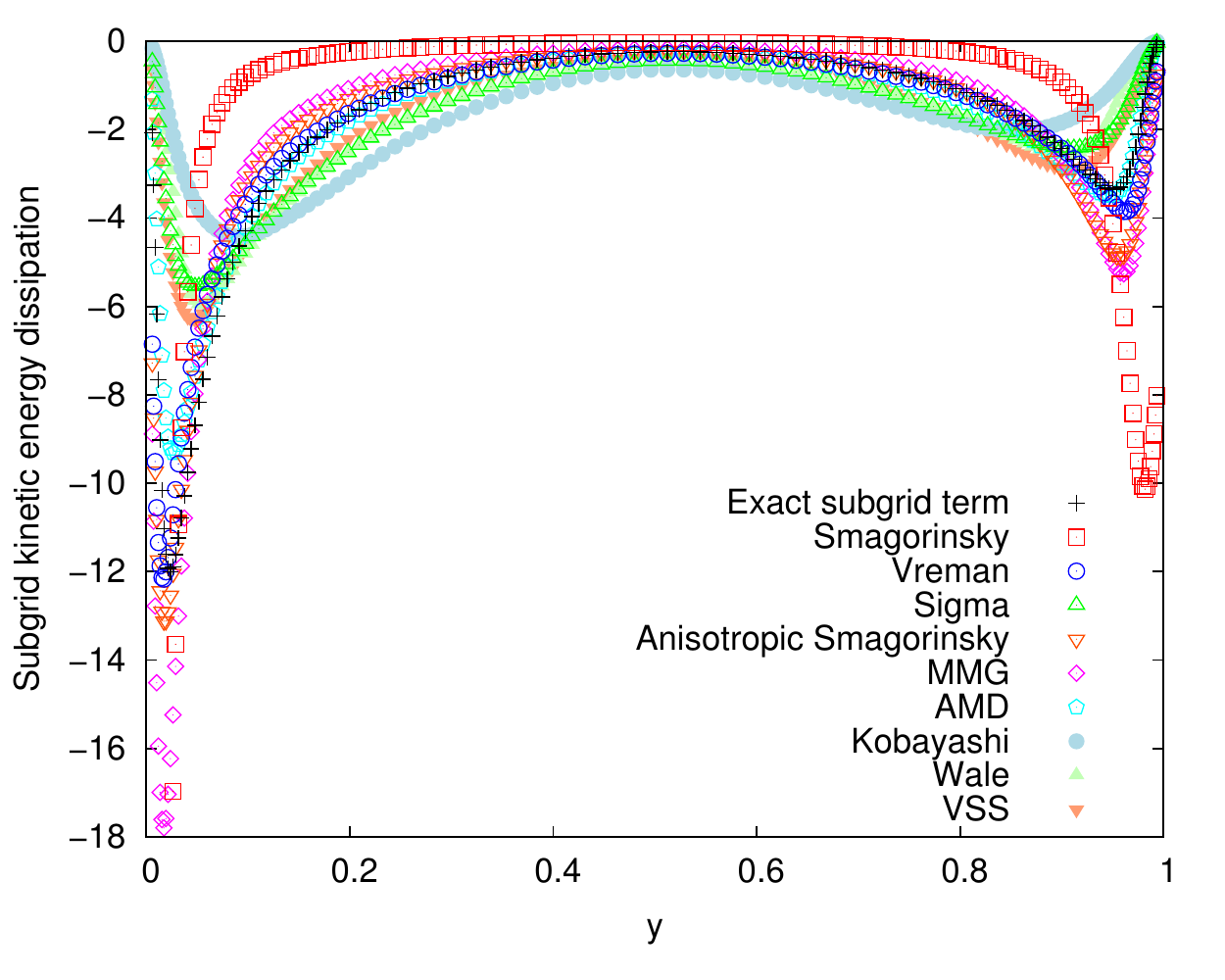}\includegraphics[width=0.51\textwidth, trim={0 10 10 5}, clip]{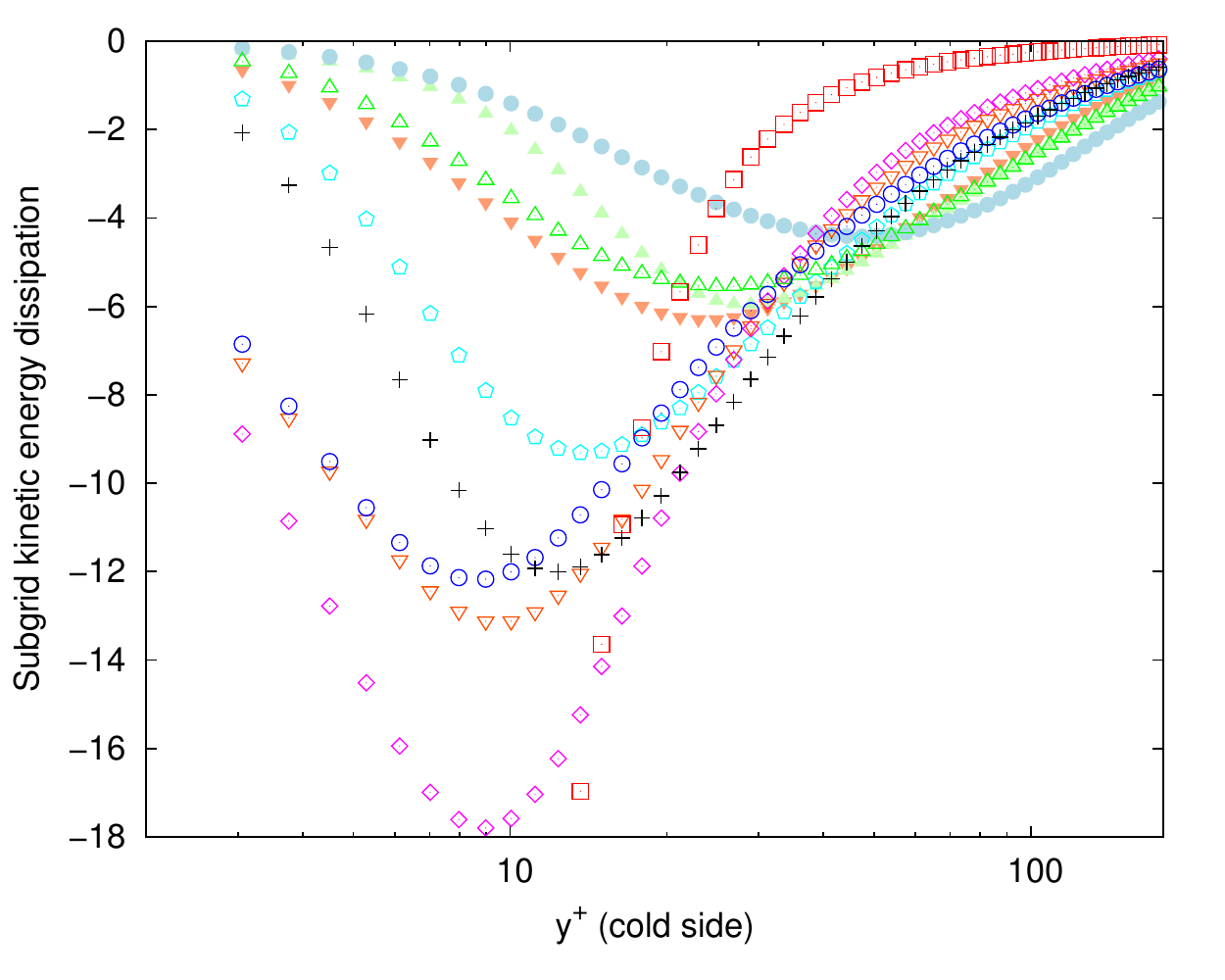}
}
\caption[Profile of the statistical average of the subgrid kinetic energy dissipation of the exact momentum convection subgrid term and eddy-viscosity models.]{
Profile of the statistical average
of the subgrid kinetic energy dissipation
of the exact momentum convection subgrid term $\f{\rho} F_{U_j U_i} S_{ij}$
and eddy-viscosity models $\f{\rho} \tau_{ij}^{\mathrm{mod}}(\vv{\f{U}}, \vv{\f{\Delta}}) S_{ij}$.
\label{fduidxjd}}
\end{figure}

The profile of the subgrid kinetic energy dissipation is given in figure \ref{fduidxjd}.
Compared to the exact subgrid term,
the Smagorinsky, Vreman, Anisotropic Smagorinsky and MMG models
are overdissipative in the near-wall
region and underdissipative at the centre of the channel,
while the WALE, Sigma, VSS and Kobayashi models
dissipates more at the centre of the channel and less near the wall.
This corresponds to the models theoretically predicted to
lead to, respectively, a lower and a higher near-wall order
than the exact subgrid term with
a filter such that $\left.\f{\Delta}_y\right|_\omega = \mathcal{O}(y^{1})$ (table \ref{toder}).

The maximum of subgrid kinetic energy dissipation is located at
$y^+=12$ at the cold side and $y^+=10$ at the hot side,
in the range of the turbulence kinetic energy production \citep{dupuy2018turbulence}.
Its location
is mispredicted towards the centre of the channel by
the WALE, Sigma, VSS and Kobayashi models
and towards the wall by
the Vreman, Anisotropic Smagorinsky and MMG models.
The AMD model predicts quite accurately the location of
the maximum of subgrid kinetic energy dissipation.
It is underdissipative at the cold side and slightly
overdissipative at the hot side, meaning that the asymmetry between the hot
and cold side is not fully captured by the model.

\begin{figure}[t!]
\centerline{
\includegraphics[width=0.67\textwidth, trim={0 0 0 0}, clip]{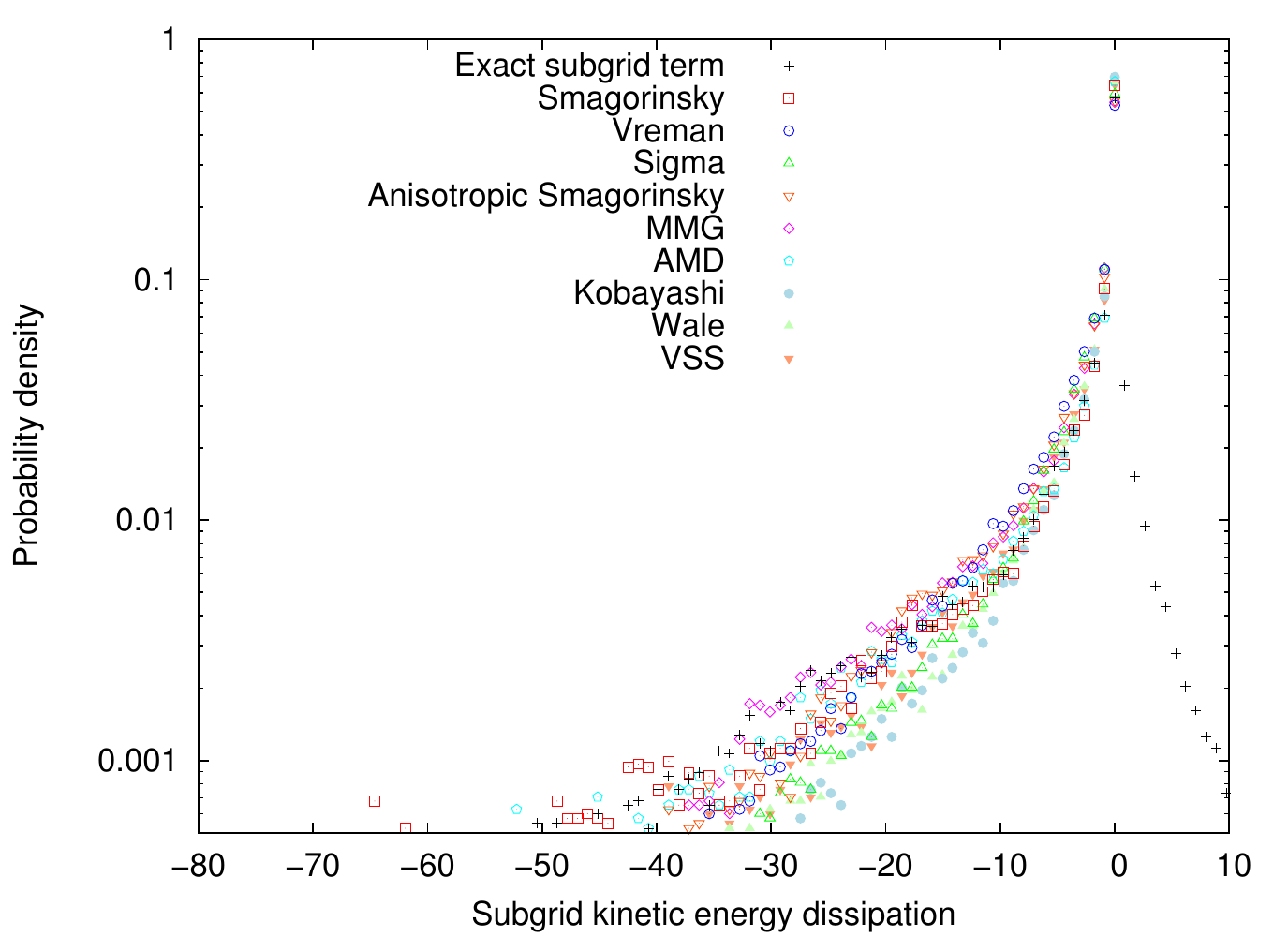}
}
\caption[Probability density function of the subgrid kinetic energy dissipation of the exact momentum convection subgrid term and eddy-viscosity models.]{
Probability density function of the subgrid kinetic energy dissipation
of the exact momentum convection subgrid term $\f{\rho} F_{U_j U_i} S_{ij}$
and eddy-viscosity models $\f{\rho} \tau_{ij}^{\mathrm{mod}}(\vv{\f{U}}, \vv{\f{\Delta}}) S_{ij}$.
\label{pedif}}
\end{figure}

Eddy-viscosity models are by construction purely dissipative.
They represent relatively well the exact subgrid term for the negative values
of the subgrid kinetic energy dissipation, which corresponds to a kinetic energy transfer from
the resolved to subgrid scales,
but cannot represent positive values of the subgrid kinetic energy dissipation.
This readily appears in
the probability density function of
the subgrid kinetic energy dissipation,
given in figure \ref{pedif}.
While this is a desirable characteristic for numerical stability, this is
inconsistent with the behaviour of the exact subgrid term which locally
transfer the energy from the subgrid to resolved scales.
The backscatter region amounts to $21$\% of the points in
the domain.

Overall, the models in better agreement with the exact subgrid term are
the AMD model, followed by the Vreman, Anisotropic Smagorinsky and MMG models (figures \ref{divic}, \ref{divjc}, \ref{divkc}, \ref{fduidxjc}).
Note that in the a priori tests, the performance of the AMD model is not
significantly undermined by the clipping of negative viscosity or diffusivity.

\subsection{Density velocity correlation subgrid term}

\begin{figure}
\centerline{
\subfigure[Correlation coefficient. \label{rdiva}]{\includegraphics[width=0.51\textwidth, trim={0 10 10 5}, clip]{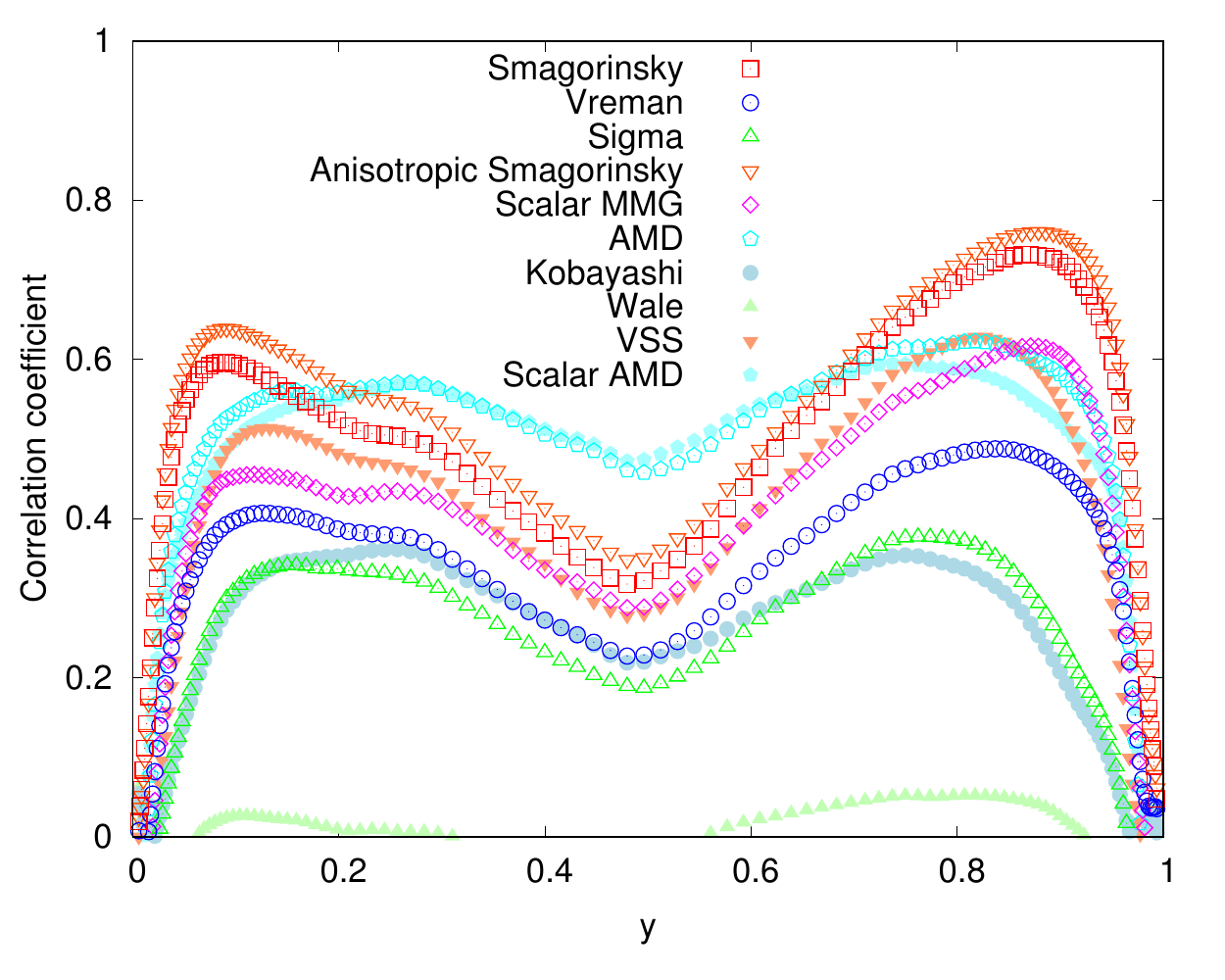}\includegraphics[width=0.51\textwidth, trim={0 10 10 5}, clip]{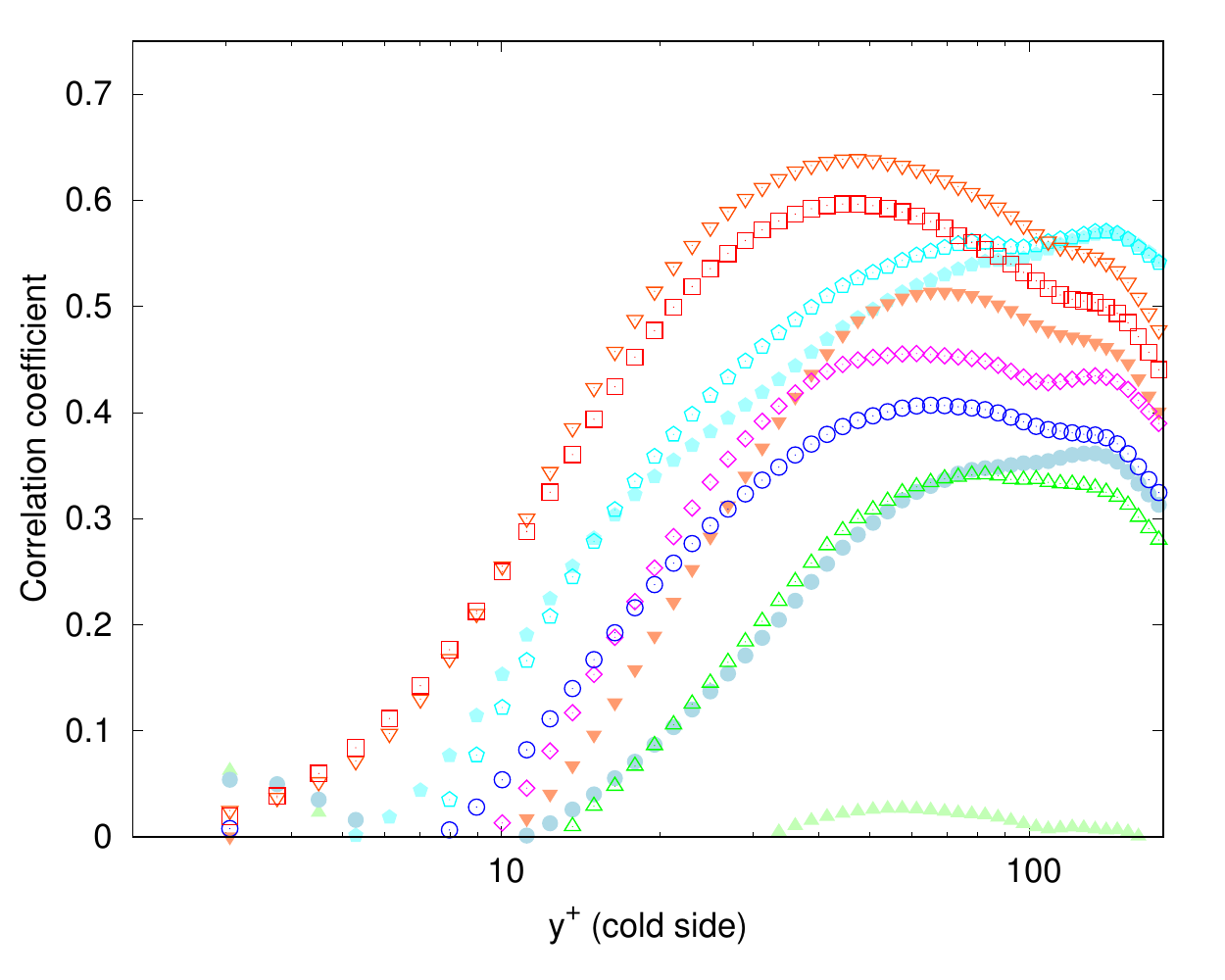}}
}\centerline{
\subfigure[Regression coefficient. \label{rdivb}]{\includegraphics[width=0.51\textwidth, trim={0 10 10 5}, clip]{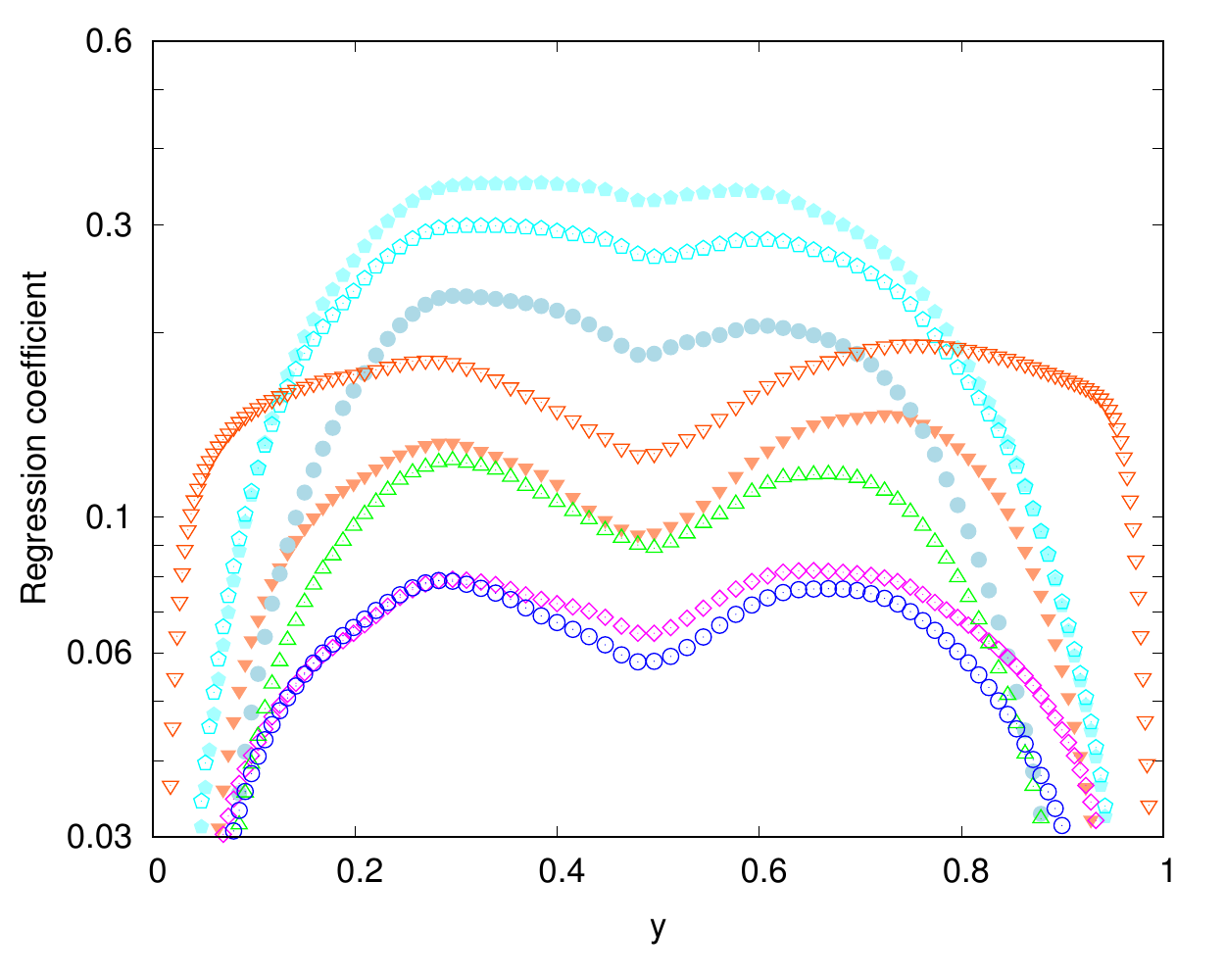}\includegraphics[width=0.51\textwidth, trim={0 10 10 5}, clip]{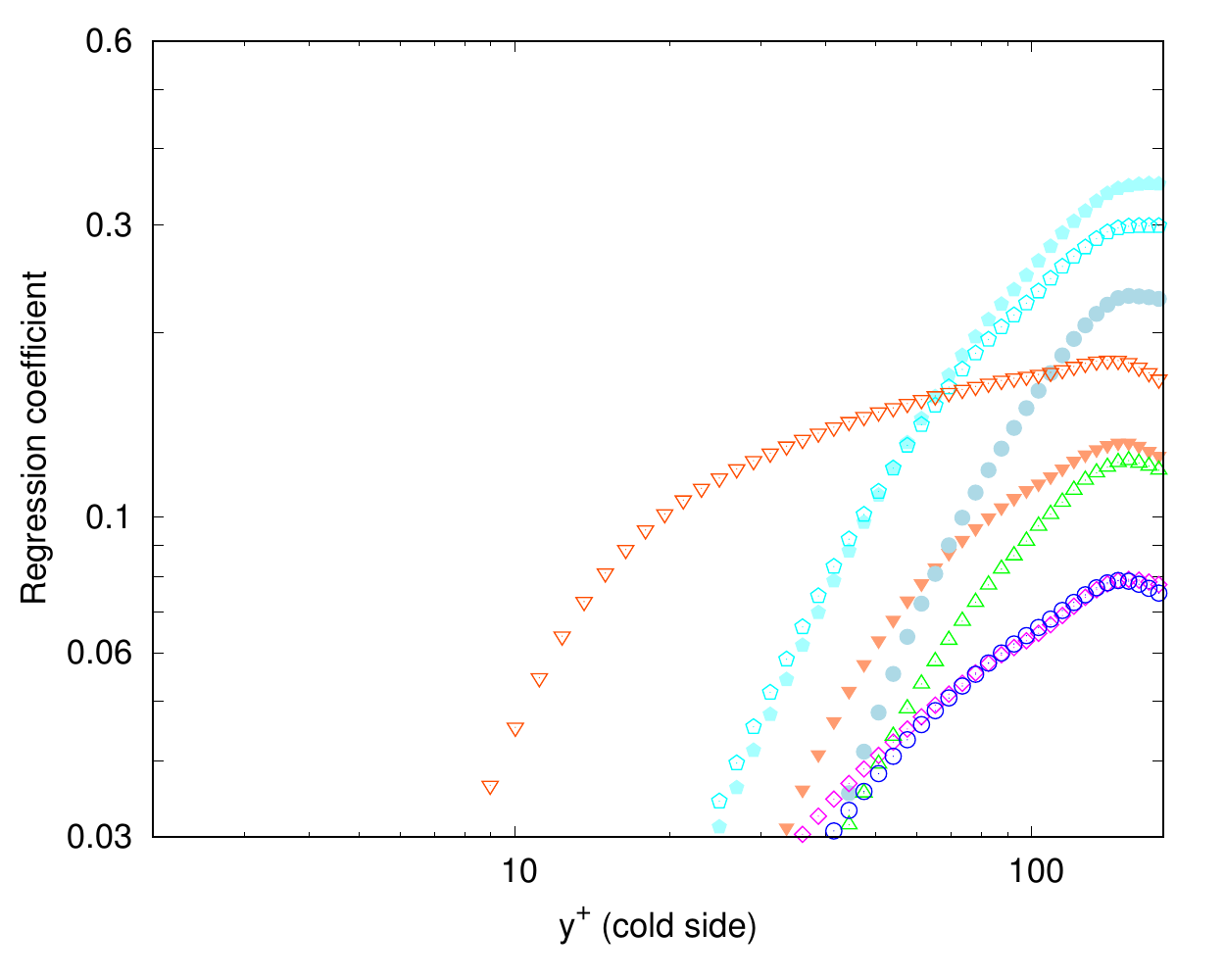}}
}\centerline{
\subfigure[Concordance correlation coefficient. \label{rdivc}]{\includegraphics[width=0.51\textwidth, trim={0 10 10 5}, clip]{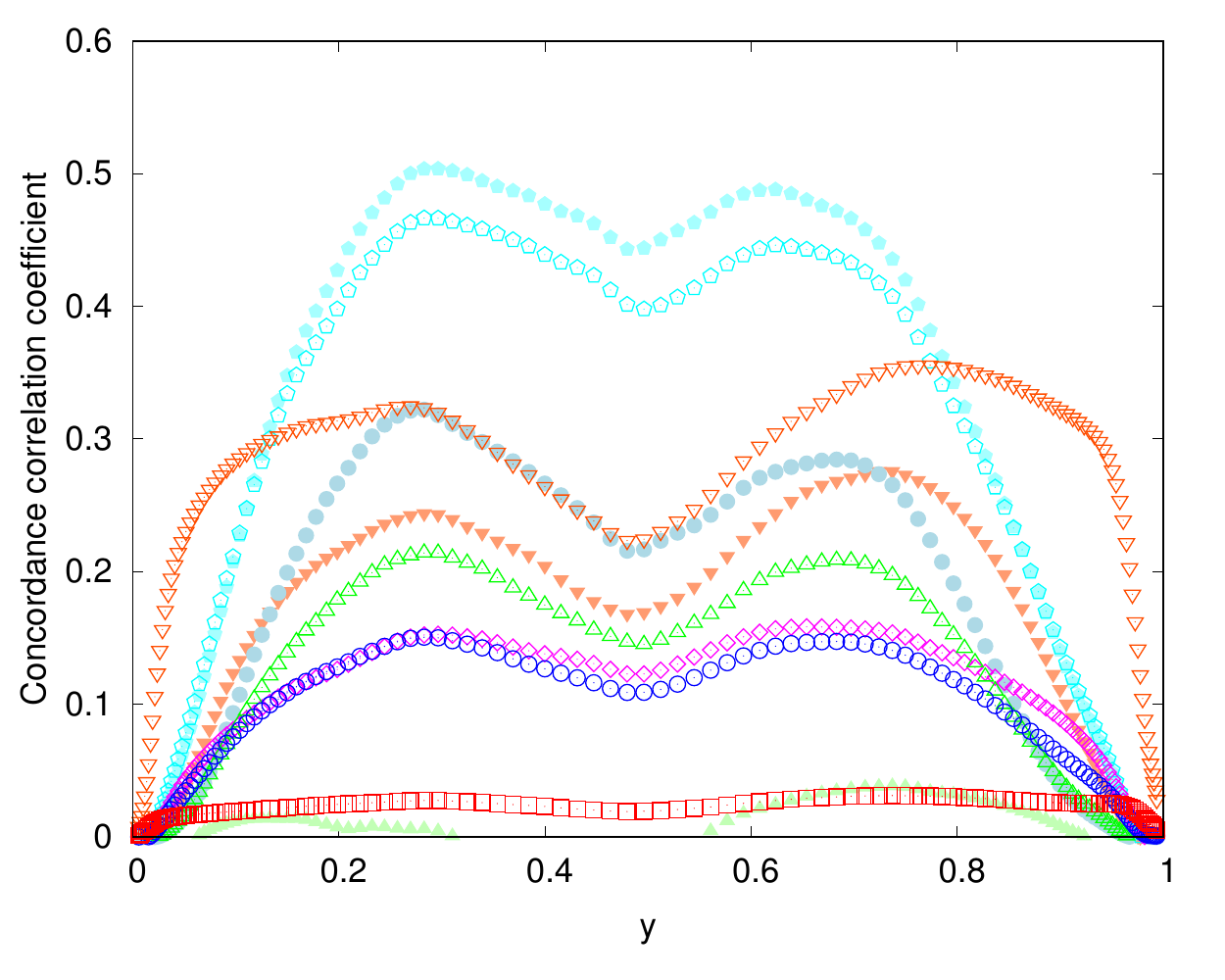}\includegraphics[width=0.51\textwidth, trim={0 10 10 5}, clip]{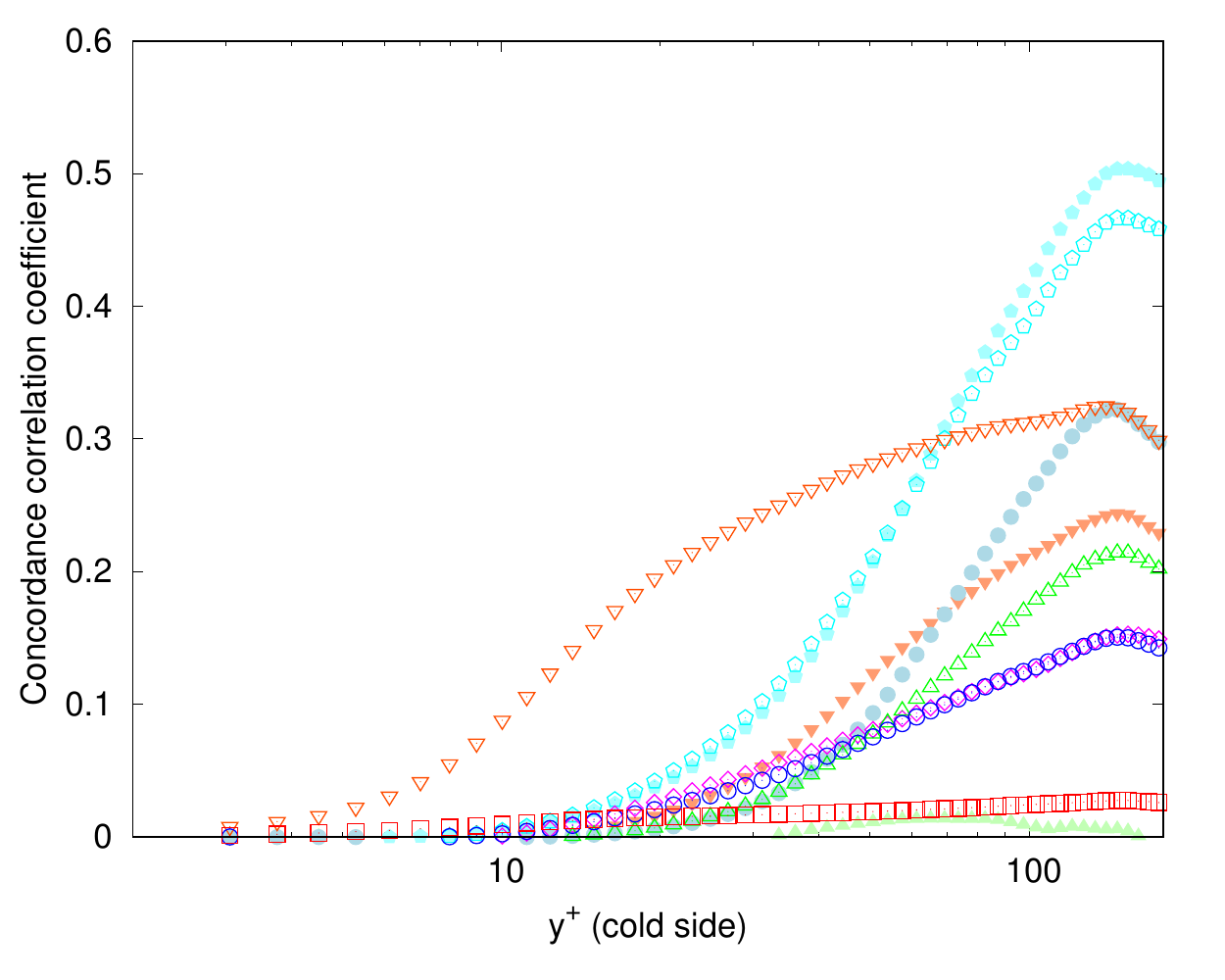}}
}
\caption[Correlation coefficient, regression coefficient, and concordance correlation coefficient between the divergence of the exact density-velocity correlation subgrid term and eddy-diffusivity models.]{
Correlation coefficient,
regression coefficient,
and concordance correlation coefficient
between the divergence
of the exact density-velocity correlation subgrid term $\partial_j F_{\rho U_j}$
and eddy-diffusivity models $\partial_j \pi_{j}^{\mathrm{mod}}(\vv{\f{U}}, \f{\rho}, \vv{\f{\Delta}})$.
\label{rdiv}}
\end{figure}

\begin{figure}
\centerline{
\subfigure[Correlation coefficient. \label{rfdrdxja}]{\includegraphics[width=0.51\textwidth, trim={0 10 10 5}, clip]{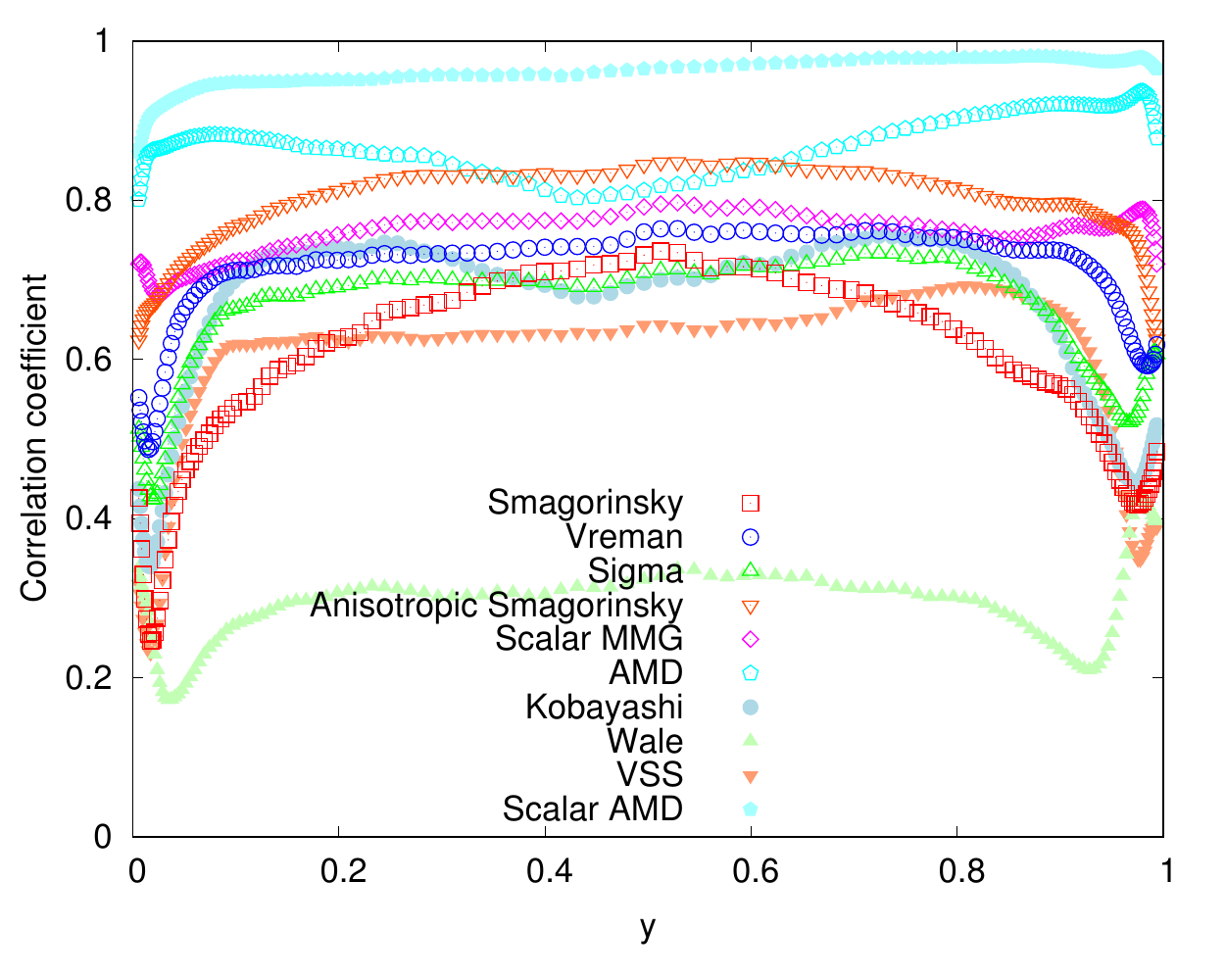}\includegraphics[width=0.51\textwidth, trim={0 10 10 5}, clip]{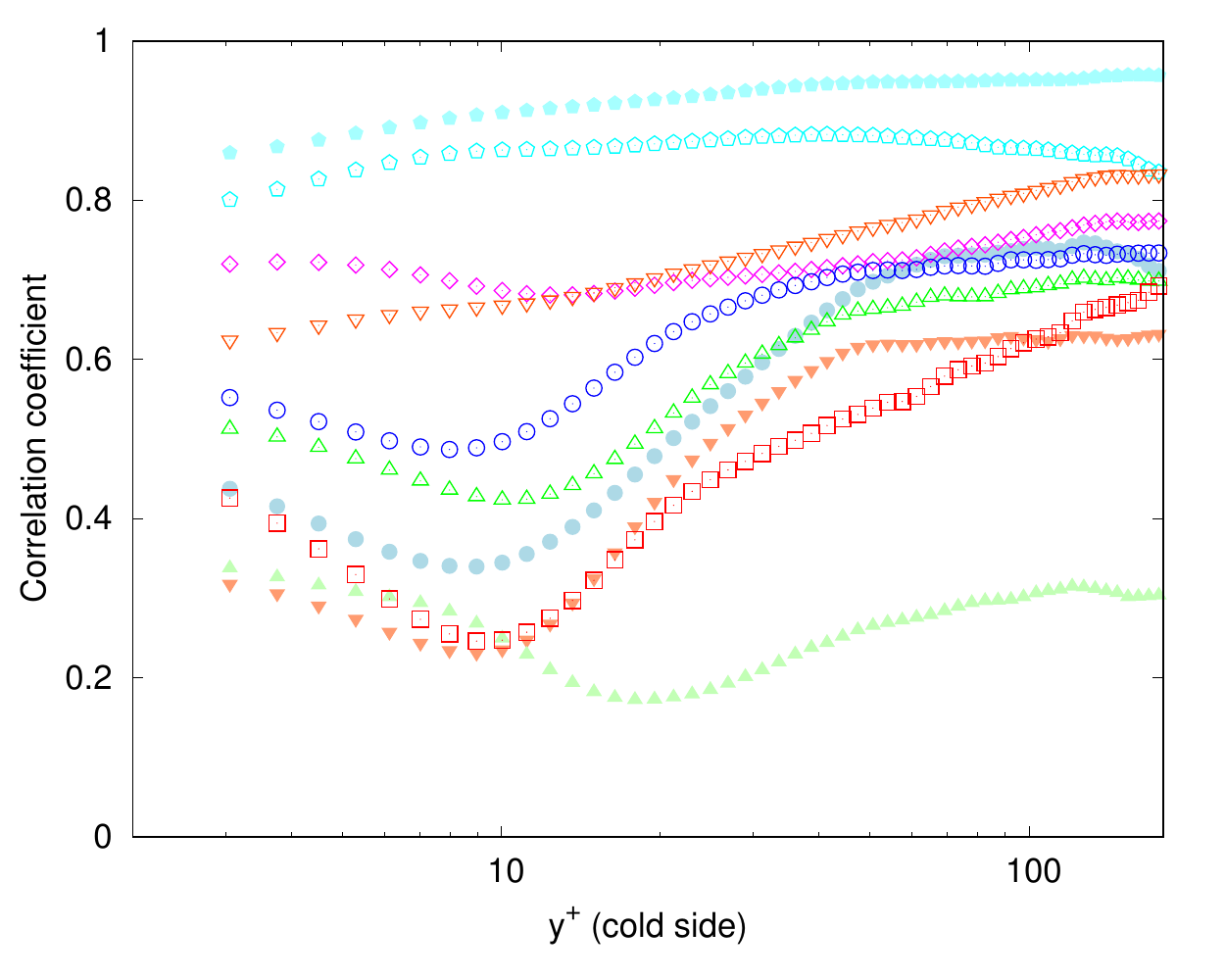}}
}\centerline{
\subfigure[Regression coefficient. \label{rfdrdxjb}]{\includegraphics[width=0.51\textwidth, trim={0 10 10 5}, clip]{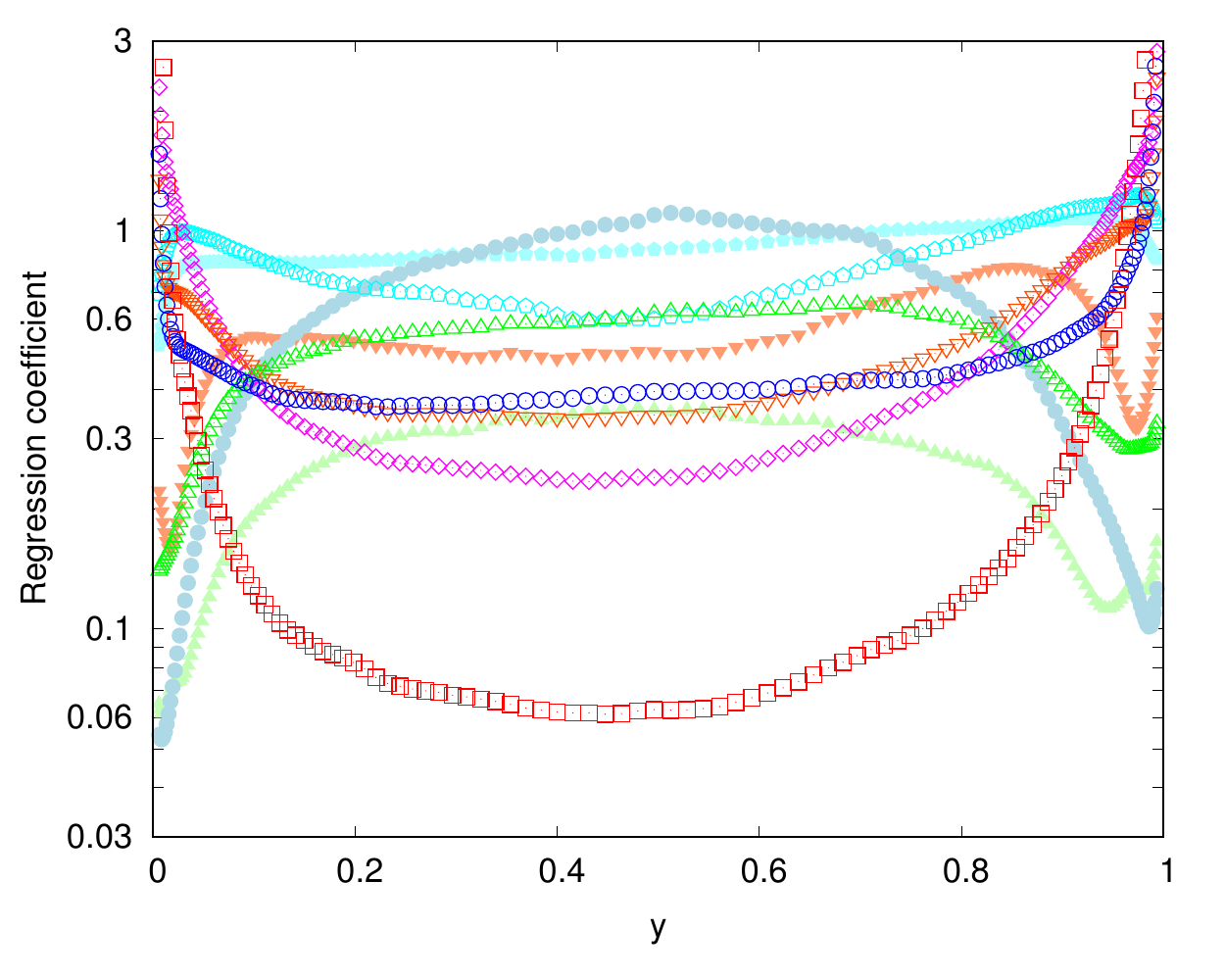}\includegraphics[width=0.51\textwidth, trim={0 10 10 5}, clip]{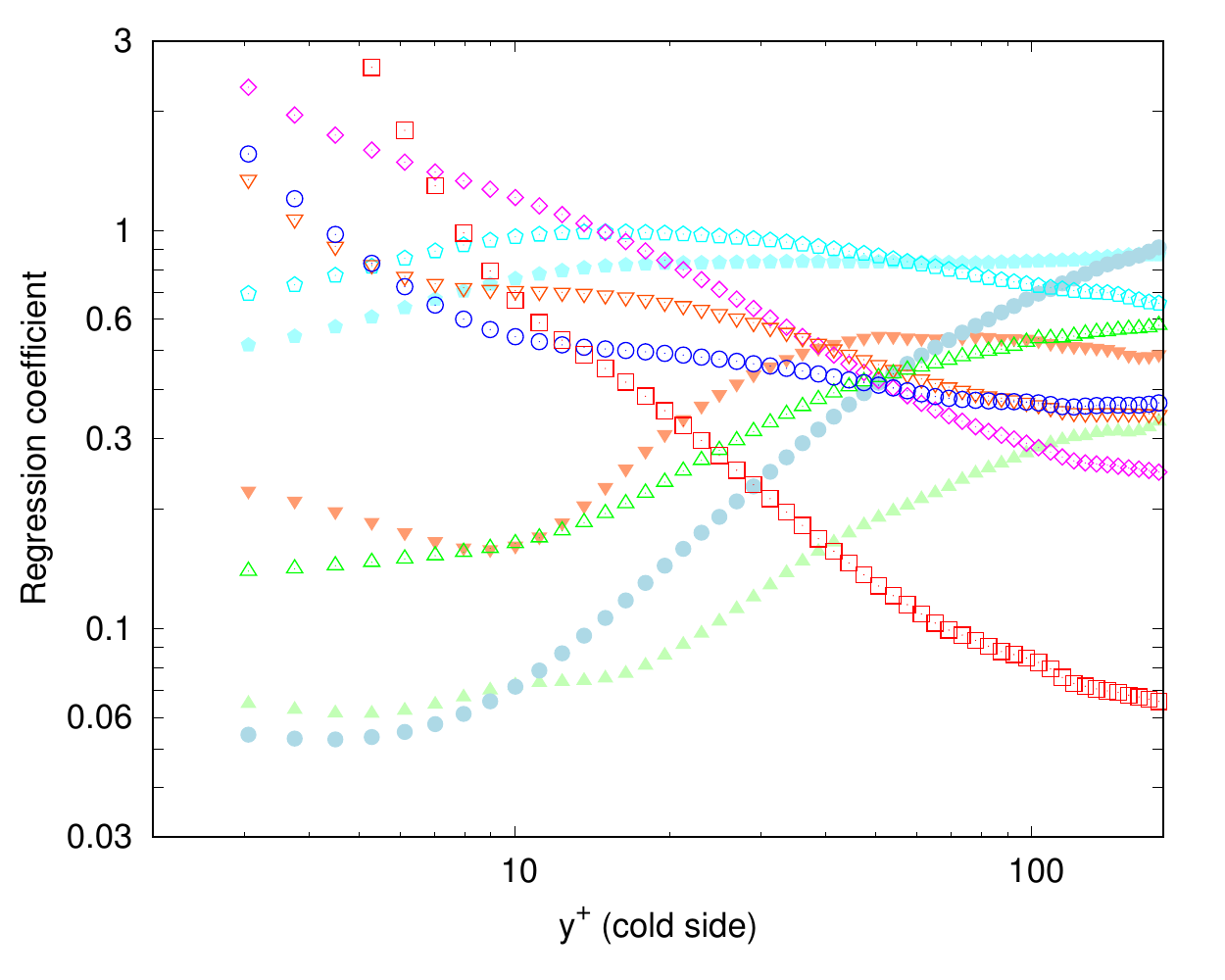}}
}\centerline{
\subfigure[Concordance correlation coefficient. \label{rfdrdxjc}]{\includegraphics[width=0.51\textwidth, trim={0 10 10 5}, clip]{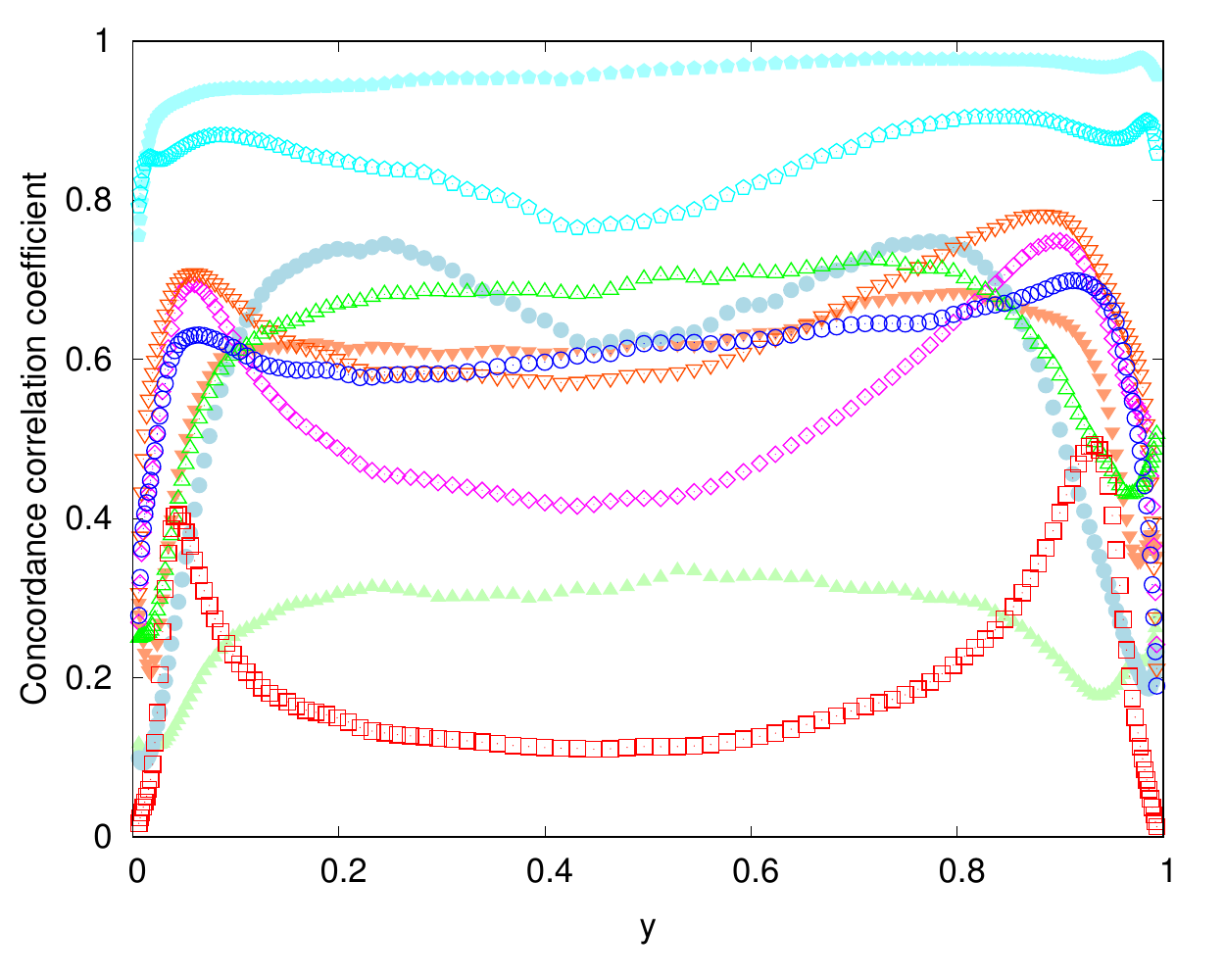}\includegraphics[width=0.51\textwidth, trim={0 10 10 5}, clip]{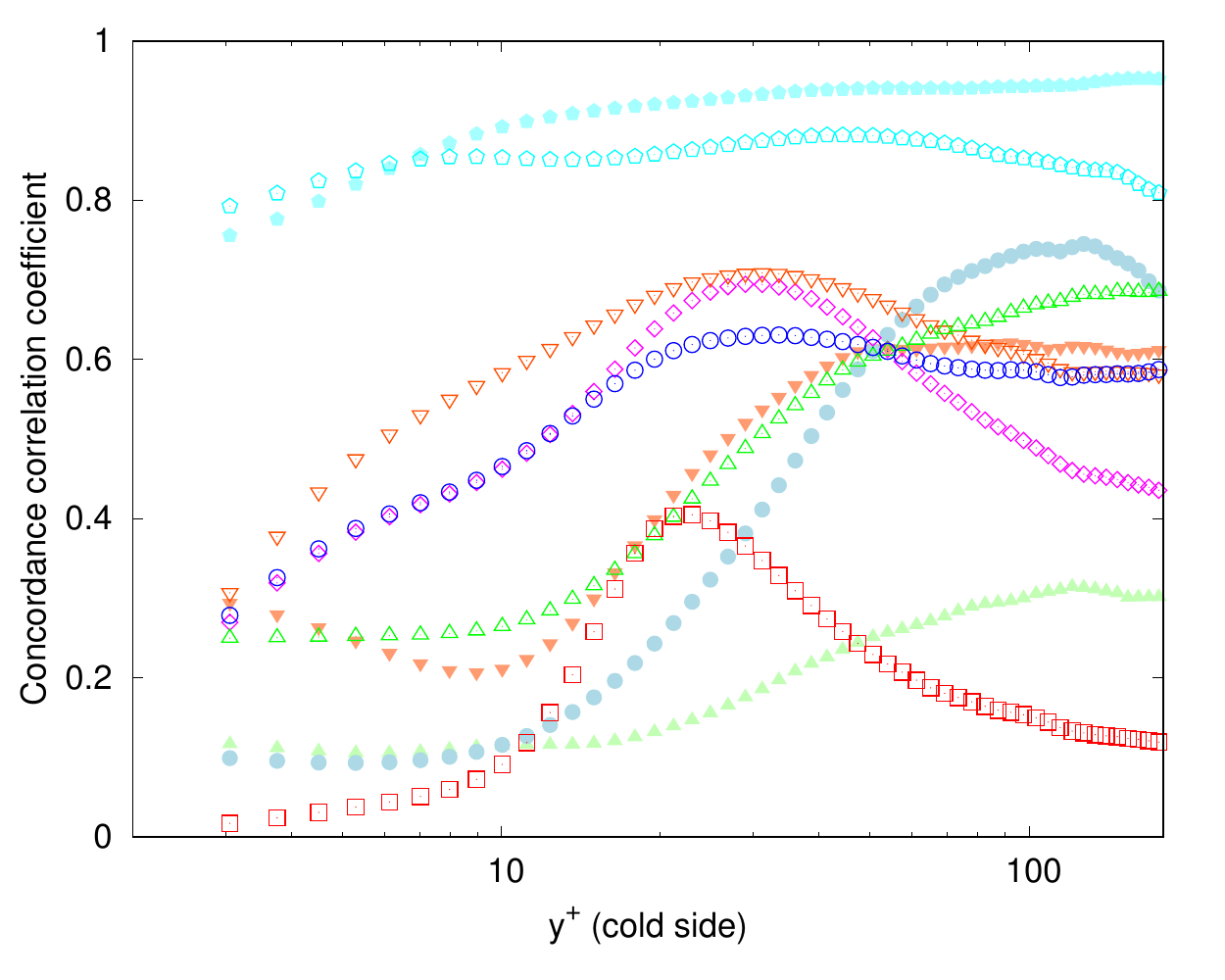}}
}
\caption[Correlation coefficient, regression coefficient, and concordance correlation coefficient between the subgrid squared scalar dissipation of the exact density-velocity correlation subgrid term and eddy-diffusivity models.]{
Correlation coefficient,
regression coefficient,
and concordance correlation coefficient
between the subgrid squared scalar dissipation
of the exact density-velocity correlation subgrid term $F_{\rho U_j} d_j$
and eddy-diffusivity models $\pi_{j}^{\mathrm{mod}}(\vv{\f{U}}, \f{\rho}, \vv{\f{\Delta}}) d_j$.
\label{rfdrdxj}}
\end{figure}

The models for the density-velocity correlation subgrid term
are assessed
as it appears in the mass conservation equation in figure \ref{rdiv}
and the subgrid squared scalar dissipation
$F_{\rho U_j} d_{j}$ is addressed in figure \ref{rfdrdxj}.
As a basis of comparison, each model is scaled in order to match 
the correct level of total subgrid squared scalar dissipation in the volume.
This is equivalent to a modification of the subgrid-scale Prandtl or Schmidt number,
or to setting the constant of the models to
\begin{equation}
C^{\mathrm{mod}} = \frac{\int_T \int_V F_{\rho U_j} d_{j} \mathop{dx} \mathop{dy} \mathop{dz} \mathop{dt}}{\int_T \int_V \pi_{j}^{\mathrm{mod}}(\vv{\f{U}}, \f{\rho}, \vv{\f{\Delta}}) d_{j} \mathop{dx} \mathop{dy} \mathop{dz \mathop{dt}}},
\end{equation}
with $Pr_t=1$.

\afterpage{\clearpage}

The correlation coefficient of the most models with the exact subgrid term as
it appears in the mass conservation equation (figure \ref{rdiva}) reaches a
maximum in the range $0.3$--$0.6$, is lower at the centre of the channel and
falls to or below zero near the wall.
The WALE model is here an exception as its correlation with the exact subgrid
term is very poor in the entire channel.
At the centre of the channel, the AMD and scalar AMD models have the largest
correlation coefficient. This may indicate their relevance in far-from-wall flows.
Within the influence of the wall, the most well-correlated models are the
Smagorinsky model and the Anisotropic Smagorinsky model, which is able to
slightly improve the correlation of the Smagorinsky model.
As the correlation coefficient, the regression coefficient declines from the
logarithmic layer to the wall (figure \ref{rdivb}), meaning that the investigated
subgrid-scale models fall too rapidly to zero at the wall.
The drop occurs nearer to the wall with the Anisotropic Smagorinsky model.
The Anisotropic Smagorinsky, AMD and scalar AMD models are overall in a better
agreement with the exact subgrid term (figure \ref{rdivc}).

\begin{figure}[t!]
\centerline{
\includegraphics[width=0.51\textwidth, trim={0 10 10 5}, clip]{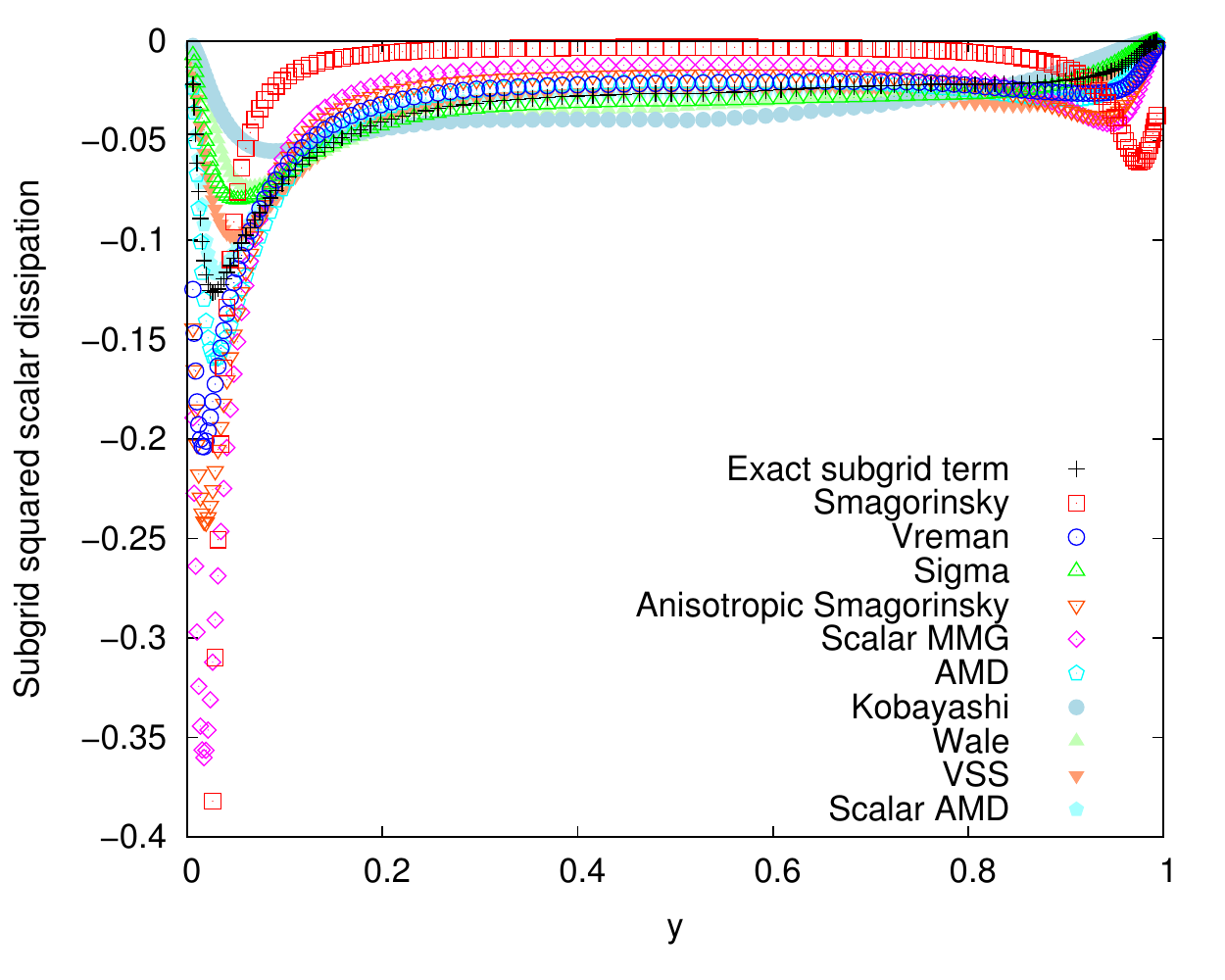}\includegraphics[width=0.51\textwidth, trim={0 10 10 5}, clip]{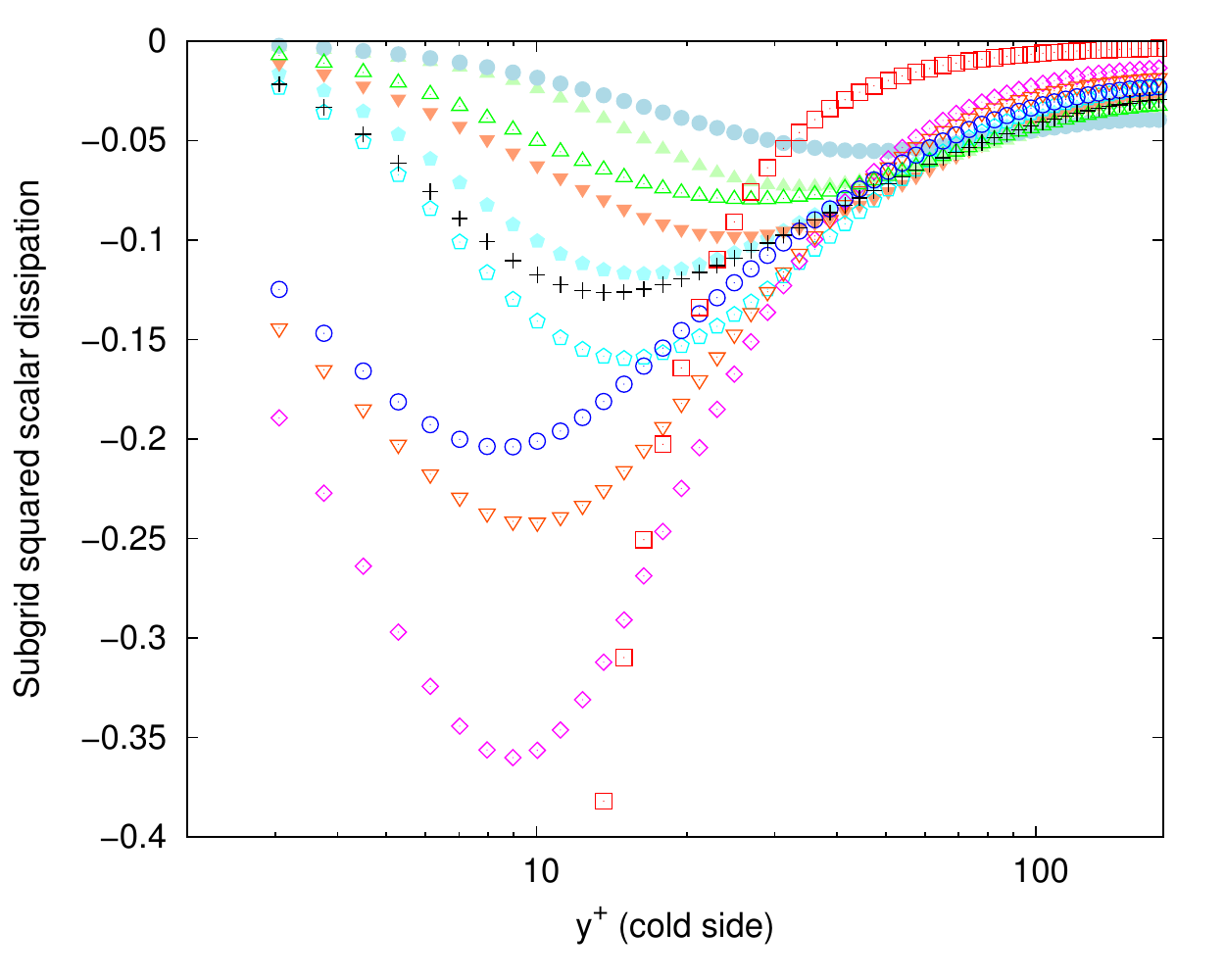}
}
\caption[Profile of the statistical average of the subgrid squared scalar dissipation of the exact density-velocity correlation subgrid term and eddy-diffusivity models.]{
Profile of the statistical average
of the subgrid squared scalar dissipation
of the exact density-velocity correlation subgrid term $F_{\rho U_j} d_j$
and eddy-diffusivity models $\pi_{j}^{\mathrm{mod}}(\vv{\f{U}}, \f{\rho}, \vv{\f{\Delta}}) d_j$.
\label{rfdrdxjd}}
\end{figure}

\begin{figure}[t!]
\centerline{
\includegraphics[width=0.51\textwidth, trim={0 10 10 5}, clip]{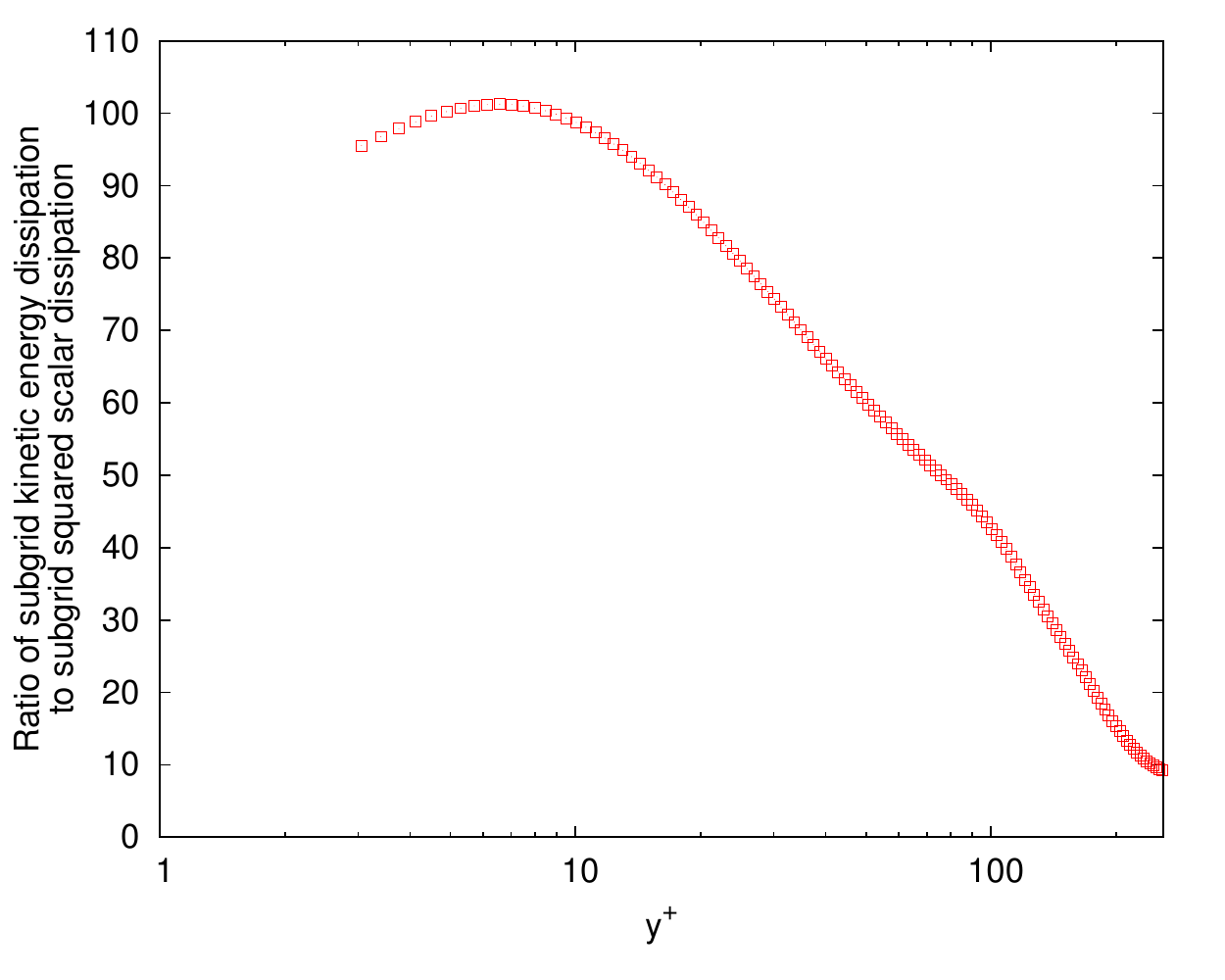}
}
\caption[Profile of the ratio of the statistical average of the subgrid kinetic energy dissipation and the subgrid squared scalar dissipation.]{
Profile of the ratio of the statistical average of the subgrid kinetic energy dissipation
and the subgrid squared scalar dissipation,
$[\f{\rho} F_{U_j U_i} S_{ij}]/[F_{\rho U_j} d_j]$.
\label{ratec}}
\end{figure}

Similarly to
eddy-viscosity models,
larger correlation coefficients are found for the subgrid squared scalar
dissipation (figure \ref{rfdrdxja}).
The AMD and scalar AMD models are clearly the models that represent the more
accurately the exact subgrid squared scalar
dissipation (figure \ref{rfdrdxjc}), with in the entire channel
a correlation coefficient over $0.8$ (figure \ref{rfdrdxja})
and a regression coefficient in the range $0.5$--$1$ (figure \ref{rfdrdxjb}).
The scalar AMD model provides an improvement compared to
the AMD model developed for the momentum convection subgrid term.
An increase of the regression coefficient of
the Smagorinsky, Vreman, Anisotropic Smagorinsky and MMG models is observed near the wall,
while the regression coefficient of
the WALE, Sigma, VSS and Kobayashi models models stabilises to a low value (figure \ref{rfdrdxjb}).
The profile of the subgrid squared scalar dissipation (figure \ref{rfdrdxjd})
shows that 
the Smagorinsky, Vreman, Anisotropic Smagorinsky and MMG models
are overdissipative in the near-wall
region and underdissipative at the centre of the channel
compared to the exact subgrid term,
and conversely for the WALE, Sigma, VSS and Kobayashi models.
These results are identical to the results obtained for the subgrid kinetic energy dissipation.
The profile of the ratio of the subgrid kinetic energy dissipation
and the subgrid squared scalar dissipation (figure \ref{ratec}) shows
that they have the same near-wall order.
The results are thus consistent with our theoretical analysis of the
asymptotic near-wall behaviour of the subgrid terms.

\begin{figure}[t!]
\centerline{
\includegraphics[width=0.67\textwidth, trim={0 0 0 0}, clip]{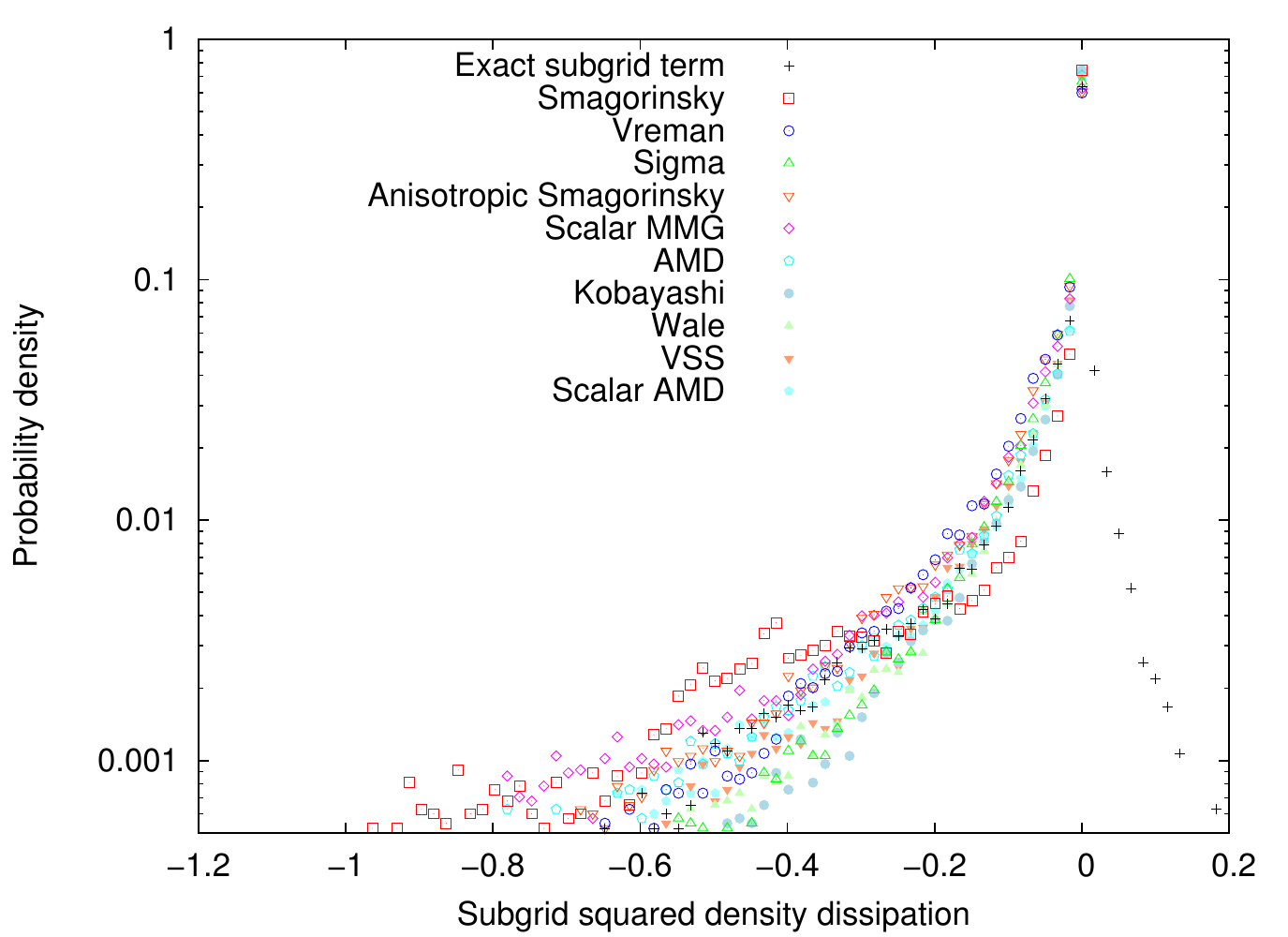}
}
\caption[Probability density function of the subgrid squared scalar dissipation of the exact density-velocity correlation subgrid term and eddy-diffusivity models.]{
Probability density function of the subgrid squared scalar dissipation
of the exact density-velocity correlation subgrid term $F_{\rho U_j} d_j$
and eddy-diffusivity models $\pi_{j}^{\mathrm{mod}}(\vv{\f{U}}, \f{\rho}, \vv{\f{\Delta}}) d_j$.
\label{rpedif}}
\end{figure}

The eddy-diffusivity assumption is as appropriate as the eddy-viscosity
assumption, in the sense the same amount of backscatter is observed for the
subgrid squared scalar dissipation than for the subgrid kinetic energy dissipation, as can be seen
in the probability density function of the subgrid squared scalar
dissipation (figure \ref{rpedif}).
However, it may be argued that the behaviour of the subgrid squared scalar
dissipation is less critical than the subgrid kinetic energy dissipation for
the numerical stability of a numerical simulation, suggesting that more
emphasis should be placed on the relevance of the model as it appears in
the mass conservation equation.

Overall, the models in better agreement with the exact subgrid term are
the AMD and scalar AMD models, followed by the Vreman, Anisotropic Smagorinsky and MMG models (figures \ref{rdivc}, \ref{rfdrdxjc}).
They are the same models than for the subgrid term associated with momentum convection.

\section{Conclusion}

The filtering of the low Mach number equations with the unweighted classical
filter or the density-weighted Favre filter leads to specific subgrid terms.
The two most significant subgrid terms are the subgrid terms associated with
the momentum convection and the density-velocity correlation.
They are compared to subgrid-scale models using the flow field from direct
numerical simulations of a strongly anisothermal turbulent channel flow.
Classical and Favre filter are found to have no influence on the performance
of the models.
Eddy-viscosity and eddy-diffusivity models are shown to be in better
agreement with the subgrid kinetic energy dissipation and the subgrid squared scalar dissipation
respectively than with the contribution of the subgrid terms in the
filtered low Mach number equations. However, eddy-viscosity and
eddy-diffusivity models are not able to account for backscatter,
present in a fifth of the points in the domain.
The AMD and scalar AMD models perform better than the other investigated models
with regard to the correlation coefficient, regression coefficient and concordance
correlation coefficient with the exact subgrid term.
This may be attributed to the strong link between the AMD and scalar AMD models
and the gradient model.
The AMD and scalar AMD inherit from the gradient model a similarity with the
exact subgrid term but, unlike the gradient model, are purely dissipative and
should not lead to numerical stability issues.

\section*{Acknowledgment}

The authors gratefully acknowledge the CEA for the development of the TRUST
platform. This work was granted access to the HPC resources of CINES under the
allocations 2017-A0022A05099 and 2018-A0042A05099 made by GENCI.